\documentclass[reprint,superscriptaddress,amsmath,amssymb,longbibliography,aps,prb,floatfix,british]{revtex4-2}

\usepackage[dvipsnames]{xcolor}
\usepackage{physics,mathtools,amsmath,amssymb,amsbsy,amsfonts,amscd,amsthm,color,colortbl,bbm,bm,multirow}
\usepackage[inline]{enumitem}
\usepackage{xspace}
\usepackage{graphicx,svg,float}
\usepackage{booktabs}
\usepackage{hyperref}
\hypersetup{
    colorlinks=true,
    citecolor=Red,
    urlcolor=Blue,
    linkcolor=Cyan,
    filecolor=magenta,
    breaklinks=true,
}
\usepackage{bibunits}

\newlength{\onecolwidth}
\newlength{\twocolwidth}
\newcommand{\angs}{~\mathring{\mathrm{A}}}
\newcommand{\nsangs}{\mathring{\mathrm{A}}}      
\newcommand{\nm}{~\mathrm{nm}}

\newcommand{\tids}{TiS\textsubscript{2}\xspace}
\newcommand{\zrds}{ZrS\textsubscript{2}\xspace}
\newcommand{\mods}{MoS\textsubscript{2}\xspace}
\newcommand{\sinfoplain}{Supplemental Materials\xspace}  
\newcommand{\sinfo}{Supplemental Materials\xspace}  

\begin{document}

\title{Long-Range Machine Learning of Electron Density for Twisted Bilayer Moir\'e Materials}
\author{Zekun Lou}
\affiliation{MPI for the Structure and Dynamics of Matter, Luruper Chaussee 149, 22761 Hamburg, Germany}
\author{Alan M. Lewis}
\affiliation{Department of Chemistry, University of York, York YO10 5DD, U.K.}
\author{Mariana Rossi}
\email{mariana.rossi@mpsd.mpg.de}
\affiliation{MPI for the Structure and Dynamics of Matter, Luruper Chaussee 149, 22761 Hamburg, Germany}
\affiliation{Yusuf Hamied Department of Chemistry, Lensfield Road, Cambridge CB2 1EW, UK}

\begin{abstract}

Moir\'e superlattices in two-dimensional (2D) materials exhibit rich quantum phenomena, but \textit{ab initio} modelling of these systems remains computationally prohibitive.
Existing machine learning methods for accelerating density-functional theory (DFT) can target the prediction of different quantities and often rely on the locality assumption.
Here we train a Gaussian process regression SALTED model exclusively on the electron densities of small displaced bilayer structures and then extrapolate electron density prediction to the large supercells required to describe small twist angles between these bilayers.
We show the necessity of long-range descriptors to yield reliable band structures and electrostatic properties of large twisted bilayer structures, when these are derived from predicted densities.
We demonstrate that the choice of descriptor determines the distribution of residual density errors, which in turn affects the downstream electronic properties.
We apply our models to twisted bilayer graphene, hexagonal boron nitride, and transition metal dichalcogenides, focusing on the model's capacity to predict complex phenomena, including flat band formation, bandwidth narrowing, domain-wall electric fields, and spin-orbit coupling effects.
Beyond moir\'e materials, this approach provides a general methodology for electronic structure prediction in large-scale systems with substantial long-range phenomena related to non-local geometric information.

\end{abstract}

\maketitle

\begin{bibunit}[apsrev4-2]  

\section{Introduction\label{sec:introduction}}
The electronic structure of two-dimensional (2D) materials exhibits significant quantum confinement effects, making them different from their bulk counterparts.
When two or more monolayers are stacked with a small twist angle, large moir\'e superlattices emerge with extended periodic interlayer modulation, inducing dramatic changes in electronic properties.
Such engineered structures have unveiled a rich landscape of quantum phases, including unconventional superconductivity, correlated insulating states, ferroelectricity, ferromagnetism, and non-trivial topology, making moir\'e materials a highly tunable platform for exploring new physics and potential applications in quantum devices~\cite{caoUnconventionalSuperconductivityMagicangle2018,xianMultiflatBandsStrong2019,sharpeEmergentFerromagnetismThreequarters2019,yasudaStackingengineeredFerroelectricityBilayer2021,kennesMoireHeterostructuresCondensedmatter2021,claassenUltrastrongSpinOrbit2022,wuCoupledFerroelectricityCorrelated2023,tsangPolarQuasicrystalVortex2024}.

While density functional theory (DFT) remains the cornerstone approach for electronic structure calculations, its computational cost limits direct simulations of large moir\'e-scale systems.
Continuum models and tight-binding approaches that fit parameters to DFT data have been developed to capture the essential low-energy physics at significantly reduced cost~\cite{lopesdossantosGrapheneBilayerTwist2007,bistritzerMoireBandsTwisted2011,suarezmorellFlatBandsSlightly2010,pathakAccurateTightbindingModel2022}.
However, their parametrisation and fixed functional form limit transferability.
For example, for twisted bilayer transition metal dichalcogenides (TMDCs), continuum model parameters vary substantially across works targeting different twist angles and physical phenomena~\cite{ahnNonAbelianFractionalQuantum2024a,xuMultipleChernBands2025}, restricting their predictive power for unexplored configurations or physics beyond the fitted data.
Tight-binding models also often require refitting when applied to different stacking configurations or materials~\cite{guineaContinuumModelsTwisted2019,pathakAccurateTightbindingModel2022}.
These limitations highlight the advantages that an \textit{ab initio}-based efficient method for accurate large-scale moir\'e simulation could have.

Machine learning (ML) techniques which predict a system's electronic structure offer a promising path forward by learning directly from complete \textit{ab initio} data and predicting unseen structures, avoiding the fixed analytical forms and the fitting limitations of traditional models, while maintaining access to the full set of ground state electronic properties.
For DFT-based methods (ML-DFT), two main ML approaches have been explored: Hamiltonian-based methods that directly predict real-space Hamiltonian matrix elements~\cite{schuttUnifyingMachineLearning2019,unkeSE3equivariantPredictionMolecular2021a,liDeeplearningDensityFunctional2022,gongGeneralFrameworkE3equivariant2023,zhongTransferableEquivariantGraph2023,tangDeepEquivariantNeural2024,maTransferableMachineLearning2025,zhangAdvancingNonadiabaticMolecular2025,zhongUniversalSpinOrbit2026}, and density-based methods that predict the electron density distribution and derive electronic properties~\cite{lewisLearningElectronDensities2021,jorgensenEquivariantGraphNeural2022,grisafiElectronicStructurePropertiesAtomCentered2023,fiedlerPredictingElectronicStructures2023,lvDeepChargeDeep2023,rackersRecipeCrackingQuantum2023,focassioLinearJacobiLegendreExpansion2023,acharMachineLearningElectron2023,leeConvolutionalNetworkLearning2024,usheninLAGNetBetterElectron2025,liImageSuperresolutionInspired2025}.

State-of-the-art Hamiltonian-based methods~\cite{liDeeplearningDensityFunctional2022,gongGeneralFrameworkE3equivariant2023} achieved impressive sub-meV accuracy for Hamiltonian matrix elements in training datasets of 2D materials and medium-sized twisted bilayers~\cite{baoDeepLearningDatabaseDensity2024}.
They exploit the nearsightedness principle of electronic matter~\cite{kohnDensityFunctionalDensity1996} to allow efficient sub-quadratic scaling with system sizes.
This assumption can limit their applicability to systems with important long-range interactions, such as systems with weak electronic screening, significant Coulomb interactions, and charge rearrangements~\cite{behlerFourGenerationsHighDimensional2021}.
Direct prediction of the electron density offers an alternative ML approach, since the density uniquely determines all ground-state properties in the DFT formalism \cite{hohenbergInhomogeneousElectronGas1964}.
In particular, using the density fitting (DF) technique \cite{renResolutionofidentityApproachHartree2012,ihrigAccurateLocalizedResolution2015} to describe the electron density, as opposed to representing it on a grid, has been shown to yield a storage-efficient and transferable model with excellent accuracy in representing both core and bonding regions of real space~\cite{lewisLearningElectronDensities2021,grisafiElectronicStructurePropertiesAtomCentered2023}.
Still, the challenge of the locality assumption for representing atomic environments persists, and the density-fitting approach can introduce noise which further complicates the intricate balance between density accuracy in training and the accuracy of downstream electronic properties.

Our work addresses these challenges by extending the SALTED (\textbf{S}ymmetry-\textbf{A}dapted \textbf{L}earning of \textbf{T}hree-dimensional \textbf{E}lectron \textbf{D}ensities) model~\cite{lewisLearningElectronDensities2021,grisafiElectronicStructurePropertiesAtomCentered2023}, which leverages a density fitting technique \cite{renResolutionofidentityApproachHartree2012}, an equivariant kernel method \cite{grisafiSymmetryAdaptedMachineLearning2018,grisafiTransferableMachineLearningModel2019}, and optimised Gaussian process regression with prediction cost linear in training set size \cite{grisafiElectronicStructurePropertiesAtomCentered2023}.
We introduce
(1) long-range descriptors to capture non-local structural features,
(2) a numerical stabilisation technique to reduce DF noise,
and (3) explicit validation of downstream electronic properties,
which together ensure reliable extrapolation of the electronic structure of 2D twisted-bilayer (TB) materials, even down to small twist angles.

By predicting the electron densities of five TB-2D materials (graphene, hexagonal boron nitride (hBN), \tids, \zrds, \mods), we demonstrate that local descriptors fail to capture the long-range physics of moir\'e systems, while long-range descriptors extending beyond the locality assumption enable accurate extrapolation of band-structures and electrostatic properties to large twisted superlattices.
We achieve robust predictions, with low-energy band error below \(5~\mathrm{meV}\) on twisted bilayer structures containing more than 1000 atoms.
The resulting SALTED models successfully predict flat-band formation, spin-orbit coupling effects, and real-space observables, while achieving speedup between one and two orders of magnitude over fully converged DFT calculations.
This framework thus provides a generalisable approach for ML-accelerated electronic structure prediction in systems dominated by long-range interactions.
Our approach is promising for accelerating the design of quantum materials and expanding the scope of first-principles simulations for complex quantum phenomena.

\section{Results\label{sec:results}}
\subsection{Extrapolation to Moir\'e Superlattices}
\label{sec:extrapolation}

We first evaluated the ability of SALTED models to predict the electronic structure of 2D materials by applying them to validation sets of aligned (displaced) bilayer structures for each 2D bilayer material (graphene, hBN, \tids, \zrds, and \mods).
To enable controlled comparisons, we fixed all descriptor hyperparameters across experiments, varying only the GPR regularisation \(\eta\) (see \autoref{sec:appendix2}). The descriptors considered are described in more detail in \autoref{sec:long_range_representation}, and comprise the short-range SOAP, and the long-range LODE and LOVV descriptors. We refer the reader to Section~\ref{sec:long_range_representation} for an explanation of the chosen acronyms for these descriptors.
\autoref{tab:experiment_results_all_materials_descriptors} summarises the accuracy of SALTED models for the electron density (direct output) and band structure (derived output). Definitions of the error metrics used can be found in \autoref{sec:appendix}.

\begin{table}[htbp]
    \centering
    \caption{
        The error in the density and band structure predictions averaged across the validation set for each of the five 2D materials, using each of the three descriptors. The error metrics are defined in \autoref{sec:appendix}.
        Hyperparameters are optimised for density prediction accuracy; see \sinfoplain~S2 for details.
    }
    \label{tab:experiment_results_all_materials_descriptors}
    \begin{tabular}{c|ccc|ccc}
        \hline
        \multirow{2}{*}{Material} & \multicolumn{3}{c|}{Density (\%)} & \multicolumn{3}{c}{Band Structure (meV)} \\
        & LOVV & LODE & SOAP & LOVV & LODE & SOAP \\
        \hline
        Graphene & 3.63 & 0.77 & 0.48 & 7.1 & 2.5 & 1.86 \\
        hBN & 2.12 & 0.95 & 0.65 & 4.0 & 2.1 & 1.51 \\
        \hline
        \tids & 2.12 & 2.20 & 1.71 & 9.6 & 16.4 & 49 \\
        \zrds & 0.45 & 0.46 & 0.40 & 6.6 & 13.4 & 27 \\
        \mods & 0.04 & 0.04 & 0.09 & 10.2 & 52 & 330 \\
        \hline
    \end{tabular}
\end{table}

Consistent with previous work by some of us~\cite{lewisLearningElectronDensities2021,lewisPredictingElectronicDensity2023}, we observed that the density error does not reliably correlate with the error in the prediction of derived observables, in this case the band structure.
For graphene and hBN, the errors in both the density and band structure follow the more common relation: Descriptors that yield larger errors for density also result in larger errors on the band structure.
However, for \tids, \zrds, and \mods, SOAP descriptors consistently achieve low density error, yet fail catastrophically in band structure predictions, while LOVV performs very well. The contrast is especially clear for \zrds, where SOAP achieves the lowest density error yet produces the largest band error, while LOVV with comparable density accuracy yields smallest band error. For \mods, LOVV and LODE are almost identical in density accuracy yet differ for five times in band error. For \tids, a larger error in the density is also seen when using LODE, in comparison to SOAP, but smaller errors in the band structure are observed.

Such behaviour reveals a nuanced relationship between the electron density and electronic structure fidelity.
Perfect density prediction guarantees accurate band structures.
However, when density predictions contain errors, the resulting electronic structure accuracy also depends on the spatial distribution of errors and their coupling to the physics governing the observables at hand. Therefore, while this initial validation of the models provides strong evidence that SALTED can be effective in predicting both densities and band structures, further tests are necessary to identify the descriptors that will produce the most accurate band structures when extrapolating to large supercells of twisted 2D materials.

\begin{figure*}[htbp]
    \centering
    \includegraphics[width=\twocolwidth]{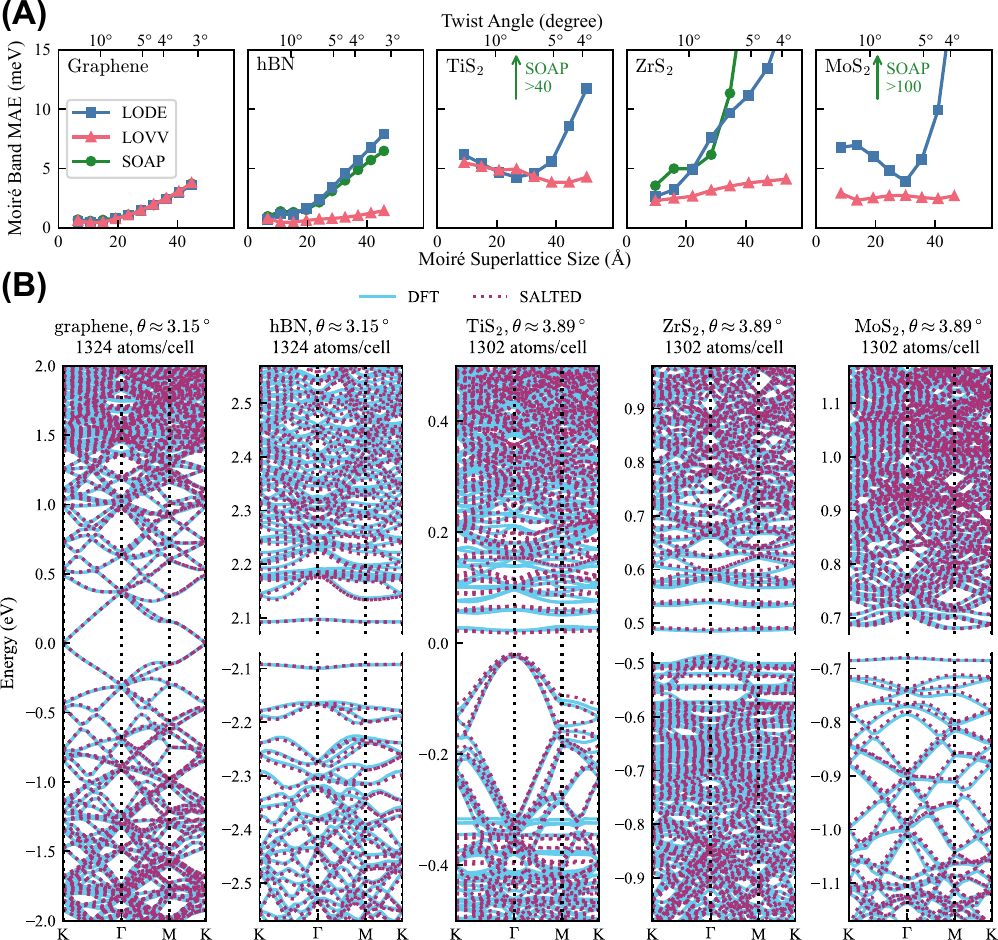}
    \caption{
        \textbf{Moir\'e band predictions across diverse twisted bilayer materials.}
        (A) Low-energy moir\'e band error vs system size for twisted bilayers.
        SALTED models used here are those with the best band structure prediction performance for each material and descriptor, obtained by following the model optimisation workflow, see \autoref{sec:train-val} and \sinfoplain~S2.
        SOAP fails catastrophically for \tids (errors \(> 40 \mathrm{meV}\), off-scale) and \mods (errors \(> 100 \mathrm{meV}\), off-scale), thus not shown here.
        (B) Comparisons between DFT-converged and SALTED-predicted moir\'e band structure for graphene, hBN, \tids, \zrds, and \mods, respectively.
        These five representative twisted bilayer systems have small twist angles \(<4^\circ\) and contain \(> 1300\) atoms per superlattice.
        Only the graphene prediction is based on SOAP; the others are based on LOVV.
        Energy alignment: band gap centred at zero for all materials, with gap central region omitted for hBN, \zrds, and \mods.
        The predictions accurately match DFT band structures across all materials, demonstrating the framework's capability to handle various 2D systems.
    }
    \label{fig:one}
\end{figure*}

We thus directly assess the framework's ability to systematically extrapolate to small twist angles (i.e., large moir\'e length scales), a regime where conventional DFT calculations become computationally prohibitive because the unit cells contain thousands of atoms.
We consider an independent test set containing twisted bilayer systems.
For graphene and hBN, structures with ten different twist angles are included in the test dataset, corresponding to twist angles from \(21.79^\circ\) to \(3.15^\circ\).
To describe these small twist angles, supercells with lattice vectors ranging in size from \(6\) to \(45\angs\) were required, which contain up to 1324 atoms at the smallest twist angles.
For the TMDCs, structures with eight different twist angles were generated, ranging from \(21.78^\circ\) to \(3.89^\circ\), which correspond to supercells with lattice vectors of \(9\) to \(50\angs\), with 1302 atoms in the largest structure.
For every material other than graphene, structures with both parallel and anti-parallel stacking symmetries were generated at each twist angle, with reported band structure errors for a given twist angle (or equivalently supercell size) being the average of the errors for these two structures.

\autoref{fig:one}(A) illustrates the dependence of the accuracy of the extrapolated band structures on both the choice of descriptor and the system size.
In \autoref{fig:one}(A), we plot the low-energy moir\'e band error against the superlattice size (equivalently, decreasing twist angle) for each material.
For graphene, all descriptors maintain reasonable accuracy (\(< 5~\mathrm{meV}\)) even at large system sizes.
However, the other materials display descriptor-dependent performance:
SOAP and LODE exhibit degraded accuracy with increasing moir\'e length scales, with SOAP failing completely for \tids and \mods;
only LOVV maintains robust extrapolation with errors \(\lesssim 5~\mathrm{meV}\) across all materials and system sizes exceeding 1000 atoms and superlattice size of \(\sim 50 \angs\).
For a better visual comparison of the prediction accuracy, we also show the DFT band structures of the smallest twist-angle systems in the test set for each material in \autoref{fig:one}(B), along with the optimal SALTED prediction of the band structure.

This material- and descriptor-dependent performance reveals systematic differences in moir\'e extrapolation capability.
SOAP uses only the local atomic density within its cutoff radius (\(6\angs\)), which proves sufficient for graphene but fails for other materials at large superlattice sizes.
LODE combines the local atomic density and an electrostatic representation,
extending the effective range beyond that of SOAP and
showing moderate extrapolation for some materials (\tids, \mods), but failing for others (\zrds, hBN) as moir\'e length scales increase.
LOVV has a different structure, using purely long-range representations, and is shown to achieve consistent accuracy across all materials and system sizes tested.
The physical and methodological origins of these performance differences are discussed in \autoref{sec:discussions}. 

Having established and validated the SALTED framework's accuracy and identified LOVV's superior extrapolation capability for moir\'e superlattices, we proceed to demonstrate its practical utility for investigating phenomena where conventional DFT is prohibitively expensive.
These case studies span three classes of problems:
(1) systematic trend mapping towards small twist angles, requiring many large structures;
(2) incorporating other physical effects at minimal additional cost;
and (3) predicting real-space observables not directly accessible to Hamiltonian-based methods.

\subsection{Band Width Extrapolation in hBN and TMDCs\label{sec:band_width_extrapolation}}

\begin{figure*}[htbp]
    \centering
    \includegraphics[width=\twocolwidth]{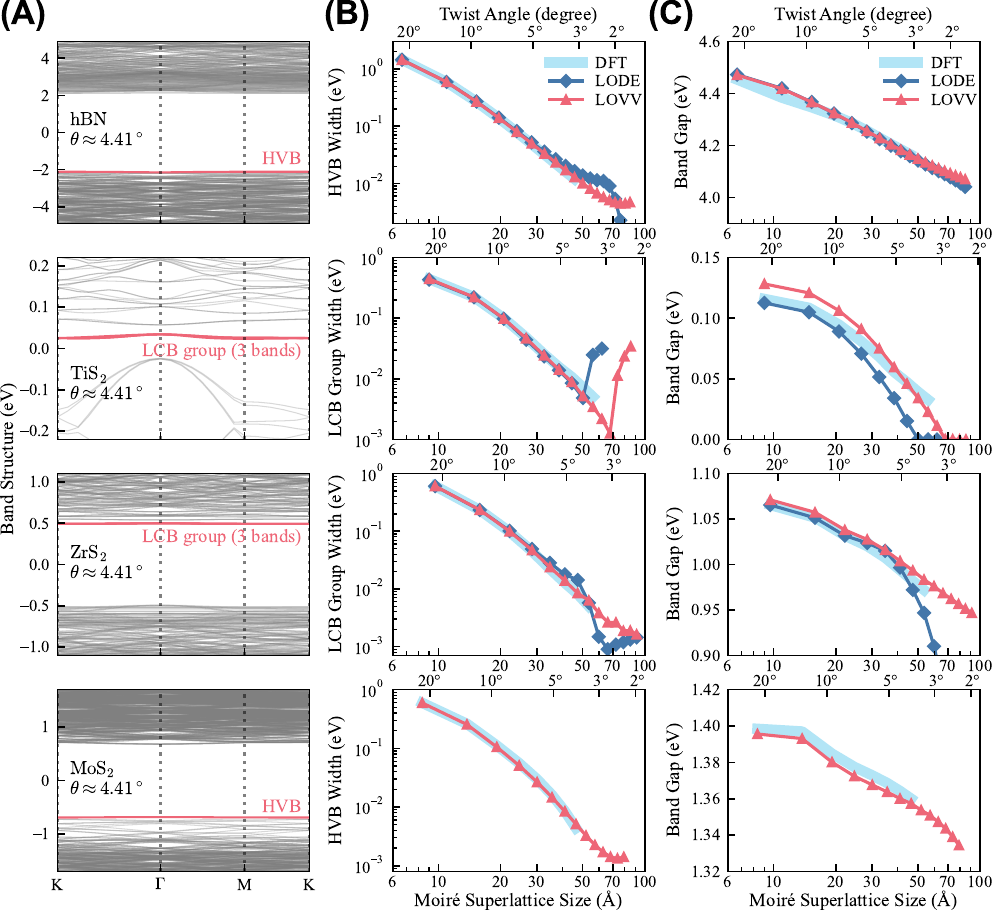}
    \caption{
        \textbf{Behaviour of band gaps and band widths with decreasing twist angle.}
        (A) Highlighted flat band(s) in the band structure, with materials' names and twist angles.
        (B) Band width vs moir\'e superlattice size (equivalently, twist angle).
        (C) Band gap vs moir\'e superlattice size.
        LODE descriptor predictions for \mods are not shown because the prediction error was too large.
        All structures use commensurate supercells with parallel unit cells in each layer.
        HVB: highest valence band. LCB: lowest conduction band.
    }
    \label{fig:flat_band_hBN_TiS2_ZrS2}
\end{figure*}

Small twist angles in 2D materials produce large-scale moir\'e superlattices that flatten electronic bands and enhance electron localisation and correlations.
Simulating twisted bilayers at very small angles (\(< 3^\circ\)) is crucial for understanding these effects.
We leveraged the efficiency of SALTED and the extrapolative power of the LOVV descriptor to map out frontier band widths of twisted-bilayer hBN, \tids, \zrds, and \mods.

\autoref{fig:flat_band_hBN_TiS2_ZrS2} shows the evolution of the bandwidth while decreasing the twist angle (and therefore increasing the number of atoms in the superlattice) for these materials.
For this dataset, the structures with the smallest twist angles contain 4564 atoms for hBN and 3768 atoms for TMDCs.
We only perform validation DFT calculations for structures with lattice constants of up to \(\approx 50\angs\) and \(\approx 1300\)~atoms per superlattice.

Both the frontier-band widths and the band gap show a decreasing behaviour with decreasing angle for each of these systems.
Panel (A) of \autoref{fig:flat_band_hBN_TiS2_ZrS2} shows representative band structures at \(\theta \approx 4.41^\circ\) with analysed bands highlighted.
For hBN and \mods, we track the highest valence band (HVB) due to their topological character, while for \tids and \zrds we focus on the group of the three lowest conduction bands (LCB) dominated by transition-metal \(d\) orbitals mixed with chalcogen \(p\) orbitals.
Panel (B) demonstrates the systematic bandwidth narrowing with decreasing twist angle: hBN's flat HVB narrows to \(\sim 5~\mathrm{meV}\), and the TMDC bands approach \(\sim 1~\mathrm{meV}\), limited by the numerical precision of DFT band structure calculations.
Panel (C) reveals the band gap evolution: hBN maintains large gap (\(\sim 4~\mathrm{eV}\)) albeit narrowing about \(0.5~\mathrm{eV}\) from \(21^\circ\) to \(1.8^\circ\), and our prediction suggests potential gap closure in \tids below \(\theta \sim 3^\circ\).

For all materials, the LOVV-based SALTED model successfully extrapolates the trend of full DFT calculations, with meV accuracy. 
In contrast, models based on LODE or SOAP, while accurate at large angles, fail as the moir\'e length increases and exceeds their training effective range, leading to unphysical predictions. We do not show the predictions with SOAP descriptors in \autoref{fig:flat_band_hBN_TiS2_ZrS2} because they fail completely for band structure predictions, especially for small twist angles, as shown in \autoref{fig:one}(A).
While models based on LOVV exhibit a diverging behaviour for \tids at twist angles below 2 degrees, where gap closure and band merging occur, it is always stable for larger system sizes than models based on LODE.
This trend highlights a limitation: SALTED may struggle to accurately predict band structures near metal-insulator transitions where the small energy gaps are sensitive to subtle density errors and minute doping of the system.
Nevertheless, this extrapolation application highlights the advantages of the LOVV descriptor in capturing long-range interactions and moir\'e physics.

\subsection{Spin-Orbit Coupling in TMDCs\label{sec:soc_in_tmdcs}}

Spin-orbit coupling (SOC) is a critical factor in the TMDCs, leading to significant band splitting that governs their topological and spintronic properties \cite{zhuGiantSpinorbitinducedSpin2011,claassenUltrastrongSpinOrbit2022}.
SOC is often included via a non-self-consistent perturbation applied after the scalar-relativistic DFT calculation has converged~\cite{huhnOnehundredthreeCompoundBandstructure2017}.
This approach integrates seamlessly with SALTED: the Kohn-Sham Hamiltonian is constructed from the predicted density, the SOC perturbation is added, and the combined Hamiltonian is diagonalised.
This is a key feature of density-based prediction: the same predicted density serves as input for SOC perturbation, electrostatic field extraction, charge analysis, without any modification to the ML model or retraining.

\begin{figure}
    \centering
    \includegraphics[width=0.48\textwidth]{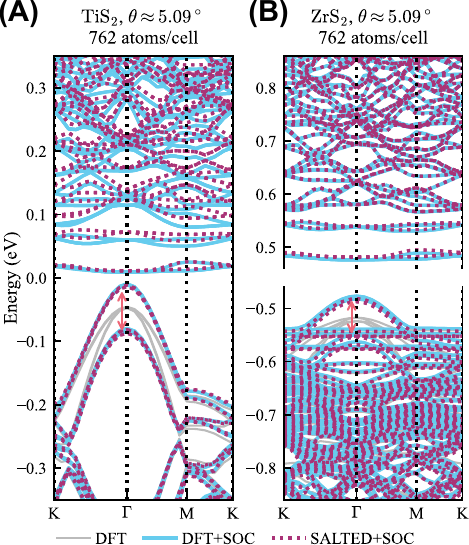}
    \caption{
        \textbf{Prediction of band structure of twisted bilayer TMDCs with spin-orbit coupling.}
        Band structures for (A) \tids and (B) \zrds at a twist angle \(\theta \approx 5.09^\circ\).
        Grey lines and cyan lines show scalar-relativistic DFT without and with perturbative SOC, respectively.
        Purple dotted lines show SALTED predictions with perturbative SOC constructed from the predicted density.
        Bands are aligned to the conduction band region to emphasise the valence band splitting indicated by the red double arrows.
        The excellent agreement between SALTED+SOC and DFT+SOC demonstrates that SALTED-predicted electron density accurately captures the orbital character necessary for perturbative SOC calculations.
    }
    \label{fig:TiS2_ZrS2_soc_band_comparison}
\end{figure}

As an illustration, we included SOC in our SALTED prediction of the band structures of twisted bilayers of \tids and \zrds, with the results shown in \autoref{fig:TiS2_ZrS2_soc_band_comparison}.
The results for \mods can be found in the \sinfo.
The predictions are in excellent agreement with full DFT calculations, accurately capturing the SOC-induced splitting and modifications to the band structure.
This confirms that the density predicted by SALTED is of sufficient quality to serve as a foundation for subsequent calculations of properties beyond the initial training.
The accurate reproduction of SOC splitting is critical for predicting spin-valley coupling, Berry curvature, and topological properties in moir\'e TMDCs, which govern their potential for valleytronic and spintronic applications.

\subsection{Structural Relaxation of Twisted Bilayer Graphene\label{sec:tbg_band_relaxation}}

We now focus on the relationship between atomic structure and electronic structure. The emergence of flat bands and correlated states near the ``magic angle'' (\(\theta \approx 1.05^\circ\)) in twisted bilayer graphene (TBG) depends critically on structural relaxation.
Interlayer van der Waals (vdW) forces induce out-of-plane and in-plane atomic displacements that significantly modify the electronic band structure compared to an idealised rigid model, especially for small twist angles around and below the magic angle~\cite{uchidaAtomicCorrugationElectron2014,carrExactContinuumModel2019, canteleStructuralRelaxationLowenergy2020}.
Obtaining these relaxed geometries is extremely challenging, since a large supercell is needed to describe such a small twist angle and high energy and force accuracy is needed to capture the buckling of the 2D sheets.

\begin{figure*}[htbp]
    \centering
    \includegraphics[width=\twocolwidth]{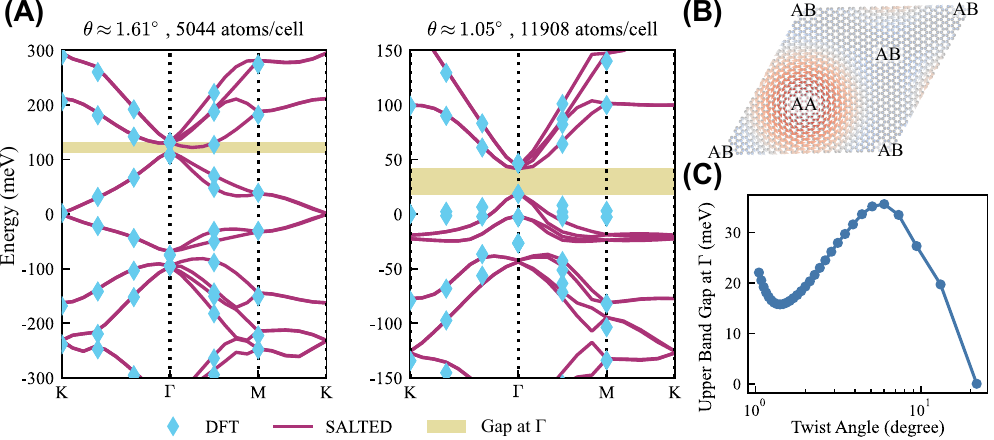}
    \caption{
        \textbf{SALTED predictions for relaxed TBG.}
        (A) SALTED predicted TBG band structure (maroon lines) at \(1.61^\circ\), and \(1.05^\circ\), compared to full DFT results (blue diamonds).
        The band structures are predicted by models based on SOAP with GPR regularisation \(\eta=10^{-3}\), guided by the band structure metric.
        (B) The relaxation effect of TBG (\(\theta \approx 2.13^\circ\), 2884 atoms/cell).
        The upper layer deregistration is shown with arrows (in-plane displacement) and colour map (out-of-plane: red for upward, blue for downward, grey for minimal).
        The energetically unfavourable AA-stacking exhibits counter-clockwise curl with upward displacement (red), while favourable AB-stacking regions show downward displacement (blue).
        These relaxations minimise energy by reducing AA-stacking area with increased interlayer distance and expanding AB-stacking area with decreased interlayer distance.
        (C) The upper band gap (between flat bands and the conduction bands above, shown in yellow in (A)) at \(\Gamma\) point against twist angles.
    }
    \label{fig:TBG_flatband}
\end{figure*}

We trained a bespoke MACE machine learning interatomic potential (MLIP) \cite{batatiaMACEHigherOrder2022} on structures spanning twist angles from \(21.79^\circ\) to \(3.48^\circ\) sampled from molecular dynamics (MD) simulations and calculated by PBE including many-body vdW corrections~\cite{tkatchenkoAccurateEfficientMethod2012} to perform these relaxations (training details in \sinfo~S3C).
The relaxation patterns, illustrated in \autoref{fig:TBG_flatband}(B) for a TBG with \(\theta=2.13^\circ\), share the same features at small twist angles:
energetically unfavourable AA-stacking regions are reduced, deforming in a curl pattern, while the areas of energetically favourable AB-stacking regions increase.
The interlayer distance shows periodic behaviour, with maximum \(\sim 3.63\angs\) at AA-stacking and minimum \(\sim 3.43\angs\) at AB-stacking, modifying local electronic interlayer hopping probabilities and ultimately changing the electronic structure~\cite{carrExactContinuumModel2019, canteleStructuralRelaxationLowenergy2020}.

When we apply SALTED to these relaxed geometries, it demonstrates robust predictive capability, benefiting from training on randomly displaced bilayers that naturally sample interlayer distances (detailed analysis in \sinfo~S3C).
\autoref{fig:TBG_flatband}(A) shows that the model accurately predicts the band structure for low twisting angles.
At the magic angle \(\theta=1.05^\circ\) with 11908 atoms per unit cell, SALTED reproduces the flat bands with a mean absolute error of approximately \(15~\mathrm{meV}\) relative to a full DFT calculation.
The predicted magic angle flat bands do exhibit residual dispersion resembling hole doping~\cite{calderonInteractions8orbitalModel2020}.
For magic angle TBG, the electronic density error amounts to about 10$^{-4}e$ per electron.
We find that normalising the deficient predicted density does not alter the band structure prediction in this case, suggesting that the observed disagreement with DFT comes rather from the distribution of errors in space.
The density errors cluster in AA-stacking domains (see \sinfo~S3C), where the density is underestimated by approximately \(2.5 \times 10^{-4}e \cdot \nsangs^{-3}\). Because flat band states sit spatially in AA regions~\cite{koshinoMaximallyLocalizedWannier2018}, the localised errors produce a hole-doping-like dispersion.
The spatial distribution of density errors relative to the electronic states determines band structure accuracy.

Nevertheless, \autoref{fig:TBG_flatband}(C) demonstrates that SALTED reliably captures the relaxation-affected band gap across the full angle range down to the magic angle, serving as a quantitative computational reference for experiments.
The MLIP+SALTED pipeline also establishes a first-principle workflow of including both structural and electronic properties in moir\'e systems, extending the reach of ab initio methods to previously inaccessible scales.

\subsection{TB-hBN Electric Field at Domain Boundaries\label{sec:tbhbn_efield}}

\begin{figure*}[htbp]
    \centering
    \includegraphics[width=\twocolwidth]{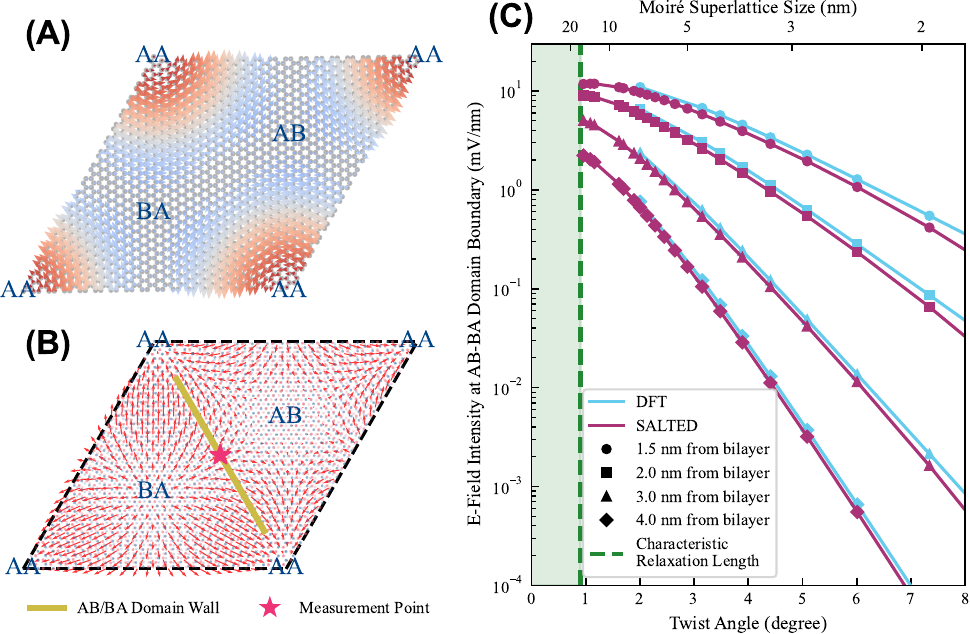}
    \caption{
        \textbf{Twisted bilayer hBN relaxation patterns and electric fields.}
        (A) Relaxed TB-hBN structure (\(\theta \approx 1.89^\circ\), 3676 atoms/cell) showing atomic position deregistration.
        The upper layer deregistration is shown with arrows (in-plane displacement) and colour map (out-of-plane: red for upward, blue for downward, grey for minimal).
        (B) Schematic view of the relaxed TB-hBN structure showing stacking regions (AA, AB, BA) and predicted in-plane electric field 1.5~nm above the bilayer centre. Blue and light red dots represent B and N nuclei respectively, with red arrows indicating the electric field direction and magnitude. The AB-BA domain wall is marked with a yellow line, and the star indicates the in-plane location of the electric field values shown in (C).
        (C) In-plane electric field intensity perpendicular to the AB-BA domain boundary as a function of twist angle, evaluated at various heights above the bilayer centre. SALTED predictions (SOAP-based model, \(\eta=10^{-6}\), optimised for density accuracy) show excellent agreement with full DFT results.
        The characteristic relaxation length represents reaching equilibrium for the domain wall.
    }
    \label{fig:TBhBN_efield_application}
\end{figure*}

Finally, we turn our attention to emerging physical properties of these systems. TB-hBN exhibits ferroelectricity \cite{bennettPolarMeronantimeronNetworks2023}, where alternating AB and BA stacking domains have opposite out-of-plane polarisations, generating in-plane electric field at the domain boundaries.
These domain wall electric fields have attracted significant interest due to their ability to modulate excitons in adjacent materials~\cite{kimElectrostaticMoirePotential2024,guQuantumConfiningExcitons2025}.
However, the spatial profile of these fields and their dependence on twist angle remain challenging to characterise experimentally.
Previous theoretical studies have predicted moir\'e electrostatic potential in TB-hBN~\cite{zhaoUniversalSuperlatticePotential2021}, but assumed rigid bilayer structures without atomic relaxation.
This approximation becomes increasingly unrealistic at small twist angles, where interlayer forces drive significant atomic deregistration and fundamentally alter the domain wall geometry and thus the electric field.

We address this challenge by combining structure relaxation using MLIPs and electronic structure prediction with SALTED.
The relaxed TB-hBN structures were obtained using a MACE model~\cite{batatiaMACEHigherOrder2022} trained on 512 displaced \(5 \times 5 \times 1\) bilayer hBN supercells with random sampling and PBE including many-body vdW corrections~\cite{tkatchenkoAccurateEfficientMethod2012} (see \sinfo ~Section S3D).
We then predicted the electron density for these relaxed structures using SALTED, computed the Hartree potential and obtained the electric fields by computing its gradient numerically on a real-space grid.

Panel (A) of \autoref{fig:TBhBN_efield_application} reveals domain-dependent structural relaxation in TB-hBN.
The energetically unfavourable AA-stacking region exhibits out-of-plane buckling and counter-clockwise in-plane displacement pattern, reducing its spatial extent while increasing interlayer separation.
Conversely, the energetically favoured AB and BA domains exhibit reduced interlayer distances and correspondingly expanded area vacated by AA regions.
These domains determine the electrostatic landscape shown in panel (B), where the in-plane electric field component is evaluated 1.5~nm above the bilayer centre.
BA and AB domains possess opposite out-of-plane polarisations, functioning as field sources and drains respectively.
The electric field intensity reaches its maximum within this plane at the centre of the AB-BA domain wall, marked as the ``Measurement Point'' in \autoref{fig:TBhBN_efield_application}(B).
The AA domain is expected to exhibit spatially uniform weak field due to zero polarisation, while the prediction exhibits residual field divergence.
This numerical artifact does not propagate to domain wall field measurements due to hexagonal symmetry (see \sinfo~S3D for detailed analysis).

Our results, presented in \autoref{fig:TBhBN_efield_application}(C), quantitatively reveal the electric field at the AB-BA domain wall.
The predicted field intensity agrees well with full DFT results, confirming SALTED accurately captures polarisation-induced electric fields.
The electric field intensity saturates as the twist angle decreases: the field strength increases sharply from large twist angles down to approximately \(\theta \approx 1^\circ\), then plateaus at smaller angles.
For example, at a probe distance of \(2 \nm\) above the bilayer centre, the field intensity saturates near \(9~\mathrm{mV/nm}\) for \(\theta < 1^\circ\).

This saturation can be understood through the formation of structural solitons~\cite{vachaspatiKinksDomainWalls2006,aldenStrainSolitonsTopological2013,niSolitonSuperlatticesTwisted2019}.
The AB-BA domain boundary is a structural soliton, which is a stable, localised deformation arising from the competition between interlayer registry energy (favouring AB or BA stacking) and elastic shearing energy (opposing sharp deformations).
As the moir\'e superlattice size \(a\) increases with decreasing twist angle \(\theta\), structural relaxation reaches an equilibrium where the domain wall adopts an equilibrium width \(w_\text{eq}\) independent of further twist angle reduction.
We extracted domain wall widths from our relaxed structures (see \sinfo~S3D), finding saturation at \(w_\text{eq} \approx 4.23 \pm 0.01~\mathrm{nm}\) for characteristic superlattice length \(a \ge 15.9 \pm 0.1~\mathrm{nm}\).
This characteristic length scale corresponds to twist angles \(\theta \approx 1^\circ\), precisely where we observe electric field saturation in \autoref{fig:TBhBN_efield_application}(C).
The electric field saturation arises because our measurement probe distance (\(z \sim 1.5 \text{--} 4~\mathrm{nm}\)) is small compared to the superlattice size at these twist angles (\(a > 15~\mathrm{nm}\)).
At this local scale, the measured electric field is dominated by the polarisation gradient at the domain wall itself.
Once the domain wall structure saturates, this local polarisation gradient stabilises.
At larger probe distances, minor contributions from the extended polarisation pattern across the larger superlattice still contribute.
The domain wall E-field is sufficiently local, being dominated by the short-range interlayer B-N charge transfer (see \sinfo~S3D), to justify using a SOAP model in this case.
These results demonstrate that structural relaxation substantially affects domain wall geometry and electric field behaviour at small twist angles, confirming that relaxation effects cannot be neglected for accurate predictions.

\section{Discussion\label{sec:discussions}}
The four case studies described above demonstrate SALTED's general applicability across the three problem classes outlined at the outset: systematic parameter mapping (bandwidth extrapolation), incorporating additional physics without retraining (SOC), and computing real-space observables (electric fields at domain walls).
The TBG and TB-hBN relaxation studies further demonstrate robustness for large systems with complex structural distortions.
The framework's computational efficiency enables systematic investigations of moir\'e physics previously inaccessible to \textit{ab initio} methods, opening new avenues for first-principles exploration and design of correlated moir\'e systems.

The necessity of long-range descriptors like LOVV for accurate electronic structure predictions in moir\'e materials is demonstrated empirically in \autoref{sec:extrapolation}.
The importance of long-range representation emerges most clearly when comparing graphene against hBN, materials with identical lattice structures but contrasting electronic properties.
As shown in \autoref{fig:one}, all three descriptors achieve comparable accuracy for graphene, while for hBN, LOVV significantly outperforms both SOAP and LODE at small twist angles.

This raises the question: Why does long-range geometric information matter for electronic structure prediction in these systems?
The most important factor is that polar bilayer systems exhibit interlayer charge rearrangement that creates out-of-plane polarisation and long-range electrostatic potentials which affect the electronic structure.
The paradigmatic example is hBN.
Despite having the same \(\text{sp}^2\) hybridisation as homoatomic graphene bilayers, hBN bilayers exhibit significant interlayer charge rearrangement between the B and N atoms at different layers, with the magnitude depending on interatomic distance.
This charge rearrangement creates out-of-plane dipole moments and long-range electrostatic fields that modulate the band edge~\cite{woodsChargepolarizedInterfacialSuperlattices2021,bennettPolarMeronantimeronNetworks2023}.
Similar polarisation effects are observed or predicted in TMDCs, where interlayer charge redistribution creates spatially varying electrostatic environments~\cite{zhangElectronicStructuresCharge2021,tsangPolarQuasicrystalVortex2024}.
In moir\'e systems, local stacking configurations vary continuously across the moir\'e supercell, creating ferroelectric domains with alternating polarisation in TB-hBN~\cite{woodsChargepolarizedInterfacialSuperlattices2021,bennettPolarMeronantimeronNetworks2023} and in-plane electric field vortices in TB-\mods~\cite{zhangElectronicStructuresCharge2021,tsangPolarQuasicrystalVortex2024}.
The electrostatic potential experienced by electrons at any site depends on the cumulative charge rearrangement and consequent polarisation from extended stacking domains spanning nanometre length scales.
The empirical comparison of band errors between graphene and hBN supports this argument.

Interestingly, the LOVV descriptor's superior performance cannot be attributed solely to its spatial extent.
Quantitative sensitivity analysis (see \sinfo Section S5A) reveals that LODE and LOVV exhibit similar distance-dependent decay characteristics, with effective ranges of approximately \(13 \angs\) and \(17 \angs\) respectively at a sensitivity threshold of \(10^{-2}\).
This small range difference contrasts with LOVV's substantial moir\'e extrapolation accuracy over LODE.
Previous work on molecular systems~\cite{grisafiMultiscaleApproachPrediction2021} demonstrated that LODE's mixed \(\ket{\rho \otimes V}\) structure provides an optimal balance between short- and long-range description for diverse molecular interactions at length scales around \(10\angs\), while our observation shows that LOVV's \(\ket{V \otimes V}\) structure performs better for extended moir\'e systems.
We propose that LOVV's \(\ket{V \otimes V}\) structure, which is capable of correlating two independent points that lie outside of the cutoff radius of the central atom, better captures the geometric features relevant to long-range physics in bilayers and moir\'e superlattice, enabling better representation of the connection between displaced bilayer training structures and moir\'e bilayer test structures.
This interpretation is supported by descriptor space coverage analysis via UMAP dimensionality reduction (see \autoref{sec:appendix2} and \sinfo Section S5B), which shows LOVV providing superior manifold overlap between displaced and twisted bilayers compared to both LODE and SOAP for certain materials, particularly TMDCs.
However, UMAP analysis shows better coverage for all descriptors in simpler systems. 
The specific mathematical structure of the descriptor representation, i.e. how it encodes correlations between atomic environments, likely plays an important role that warrants further investigation.

Regarding training costs, dominated by Hessian matrix construction~\cite{grisafiElectronicStructurePropertiesAtomCentered2023}, they are comparable across descriptors. However, prediction exhibits distinct scaling behaviour: in the current implementation, LOVV and LODE show quadratic scaling with system size, due to all-to-all electrostatic-like descriptor evaluations, whereas SOAP maintains near-linear scaling from its finite cutoff (see \sinfo~S4 for details).
Despite this scaling, LOVV-based SALTED models remain 10 to 100 times faster than fully converged DFT for moir\'e structures.
We further show in \sinfo~S4 that tuning the Gaussian kernel width \(\sigma\) can reduce LOVV prediction cost with only a small impact on band gap extrapolation accuracy.
We expect that implementation improvements, including multipole approximation for long-range descriptor contributions from distant nuclei, GPU acceleration and MPI parallelisation, will speed up the models substantially.
This computational efficiency could enable future applications such as unfolding \textit{ab initio} moir\'e band structure~\cite{quanEfficientBandStructure2026} for direct comparison with angle-resolved photoemission spectroscopy experiments~\cite{chenStrongElectronPhonon2024,chenVisualizingElectronicStructure2026}.

We further comment on our model's performance in comparison to other reports in the literature. Recent neural network methods for Hamiltonian prediction achieve sub-meV accuracy for Hamiltonian matrix elements~\cite{liDeeplearningDensityFunctional2022,zhongTransferableEquivariantGraph2023,tangDeepEquivariantNeural2024}.
These methods achieve efficiency by relying on the locality assumption and the nearsightedness principle.
Band structures can also be derived from these predictions.
For relaxed TBG at \(\theta \approx 2.0^\circ\) (3268 atoms), the pioneering work reported with the DeepH~\cite{tangDeepEquivariantNeural2024} achieves 1.3~meV low-energy band error while SALTED achieves 1.5~meV (see \sinfo~S3C), 
with both methods demonstrating \(< 0.3\%\) error over the 600~meV low-energy band range. For TBG with larger twisting angles, the band structures derived from SALTED predictions in this work can yield up to an order of magnitude lower error than DeepH results published a couple of years ago~\cite{tangDeepEquivariantNeural2024}, as summarised in the \sinfo Section~S2E.
For magic-angle TBG, DeepH produces narrower flat band width matching the expected low-energy profile more closely than SALTED's prediction.
However, graphene represents a case where the locality assumption is valid.  We expect this assumption to become limiting for other twisted bilayer systems, as discussed above. Indeed, comparing the performance of the band structure presented here for \mods and the ones derived from previously published DeepH predictions~\cite{tangDeepEquivariantNeural2024}, we observe an order of magnitude better accuracy with our model, as detailed in the \sinfo Section~S2E. Given the pioneering character of the publications using DeepH in this area, we expect that predictions with that method can also be improved by incorporating long-range interactions.
To our knowledge, published Hamiltonian prediction methods have not yet demonstrated robust band structure performance for hBN or other TMDC systems at small twist angles below \(5^\circ\)~\cite{baoDeepLearningDatabaseDensity2024}.

Many recent methods which predict the electron density directly report impressive global metrics (global density error \(\le 10^{-3} \mathrm{e \cdot \nsangs^{-3}}\) or \(\le 1\%\) if normalised by total density), but do not verify that predicted densities yield accurate downstream electronic properties~\cite{rackersRecipeCrackingQuantum2023,lvDeepChargeDeep2023,acharMachineLearningElectron2023,usheninLAGNetBetterElectron2025}.
Consistent with our previous observations~\cite{lewisLearningElectronDensities2021,lewisPredictingElectronicDensity2023}, we have found again that although accurate density prediction is necessary for accurate electronic structure prediction, the quality of the prediction of certain properties is strongly dependent on the spatial distribution of density errors.
Our results emphasise that global density error, being a single scalar quantity, is an insufficient validation metric for density prediction methods intended for electronic structure applications, and careful verification of downstream observables is essential.

We are not aware of other electronic density-prediction methods that have been employed to compute the band structure of twisted-bilayer systems.
Because such models are not straightforward to use, it has proved difficult to directly perform such a comparison in this work.
The authors of this paper nevertheless believe that a collaboration targeting widespread comparison of predicted properties among electronic-structure methods is of immense importance to the community. 

We further note the generality of SALTED.
Training separate models for each material is a pragmatic choice because of GPR's data scaling but not a constraint of SALTED: previous works~\cite{lewisLearningElectronDensities2021,grisafiElectronicStructurePropertiesAtomCentered2023} have shown multi-material SALTED models, but they need larger training datasets.
The SALTED models are not property-specific because they first predict electron densities and then derive all downstream electronic properties.
The descriptor choice between long-range and short-range is physics-driven and is analogous to neural network architectural choices such as range-separate message passing~\cite{tangDeepEquivariantNeural2024} or k-space aggregation~\cite{guoCapturingLongrangeInteractions2026a}.
The optimal regularisation \(\eta\) may differ slightly between density- and band-targeted models, but the predictions are accurate and robust across a wide range of \(\eta\) values (see \sinfo~S2B).
Using a multi-property loss function could in principle automate this balance, but the current framework works reliably without adjusting too much for each property.

We conclude by discussing perhaps the main limitation of SALTED: as it is based on Gaussian process regression, there are intrinsic limitations on the size of the training sets that can be used and, ultimately, on the model's accuracy and breadth. While active-learning strategies for the electronic density can be explored, in order to increase the model accuracy without increasing the training set size, ultimately an implementation of this model within a neural-network architecture would be desirable. Because of the numerical instabilities of SALTED discussed in this paper, related to the density representation, we expect such an implementation to face some challenges.

In summary, we established a systematic and transferable framework for predicting electronic structures in large moir\'e superlattices which overcomes the locality constraint limiting existing machine learning approaches.
We have established that the locality assumption, applied to most existing machine learning methods for electronic structure, is not universally valid for polar twisted bilayer systems.
While training only on computationally tractable displaced bilayers (\(3 \times 3\) supercells), our method achieves accurate extrapolation to twisted moir\'e structures with supercell sizes up to \(a \sim 100 \angs\) (\(\sim 4000\) atoms), at computational costs \(10 \sim 100 \times\) faster than direct DFT calculations, within the current implementation of SALTED.
This enables \textit{ab initio} exploration of quantum phenomena at twist angles and system sizes previously inaccessible due to computational constraints.

This advance builds upon three key innovations addressing distinct physical and methodological challenges.
First, we used LOVV descriptors, which are built upon atomic electrostatic-like representations and encode distant geometric information through \(1/r\) Coulomb form, to train our SALTED models.
As a result, these models were able to capture the essential physics which govern moir\'e materials: interlayer charge transfer and polarisation and moir\'e-scale modulation of electronic potentials, which require structural information far beyond limited cutoffs of local descriptors.
Second, we systematically resolved numerical instabilities from ill-conditioned density fitting auxiliary basis sets in heavy-element systems through singular value truncation, suppressing noise-dominated subspaces while preserving density fitting accuracy.
This finding is transferable to other density-fitting-based machine learning methods, which faced difficulties with heavy-elements or larger density fitting basis sets.
Third, we demonstrated that spatial density prediction accuracy alone is insufficient for reliable electronic structure modelling: the spatial distribution of density errors determines their coupling to target properties.
This necessitates explicit validation on downstream observables (band structures, real-space fields) rather than relying solely on global density metrics.

We validated the framework systematically across five diverse 2D bilayer materials (graphene, hBN, \tids, \zrds, \mods) with different electronic characters. The framework's practical utility was demonstrated through diverse applications, previously only barely accessible to \textit{ab initio} methods.
We revealed low-energy isolated band width narrowing trends with smaller twist angles, predicted geometrically relaxed magic-angle twisted bilayer graphene band structures, and computed real-space electric field profiles at ferroelectric domain walls in relaxed twisted bilayer hBN, unravelling the reason behind the plateau value of in-plane electric fields at low twisting angles.
The characterization of the impact of long-range descriptors, the singular value truncation for training stabilisation, and the demonstration that spatial distribution of density errors determines downstream electronic property accuracy enable a robust first-principles machine learning framework for polar moir\'e materials, supporting future research in this promising area.

These capabilities enable systematic materials screening, accelerating the discovery of correlated quantum phases, and supporting inverse design of moir\'e structures with targeted electronic properties. This work provides both conceptual foundation and practical tools for accurate, efficient, and transferable electronic structure prediction in moir\'e materials and beyond.

\section{Methods\label{sec:methods}}
\subsection{SALTED Foundations}

SALTED predicts electron densities using symmetry-adapted Gaussian process regression~\cite{lewisLearningElectronDensities2021,grisafiElectronicStructurePropertiesAtomCentered2023}.
The electron density is represented using the density fitting (DF) technique, where the density is expanded in atom-centred auxiliary basis functions:
\begin{equation}
\rho (\mathbf{r}) \approx \rho^\text{DF} (\mathbf{r})
= \sum_{i,n\lambda\mu,\mathbf{T}} c_i^{n\lambda\mu} \phi_i^{n\lambda\mu} (\mathbf{r} - \mathbf{R}_{i} + \mathbf{T})
\label{eq:density_fitting}
\end{equation}
where \(\mathbf{r}\) is the spatial coordinate,
\(\mathbf{R}_i\) is the position of \(i\)th atom in the unit cell,
\(\mathbf{T}\) is the lattice translation vector,
\(\phi_i^{n\lambda\mu}\) are density fitting auxiliary basis functions with radial index \(n\), angular momentum \(\lambda\), and magnetic quantum number \(\mu\),
and \(c_i^{n\lambda\mu}\) are the corresponding expansion coefficients (collectively denoted as \(\mathbf{c}^\text{DF}\)).

SALTED employs symmetry-adapted kernels~\cite{grisafiSymmetryAdaptedMachineLearning2018} based on representations of atomic environments \(A_i\). These are most commonly the Smooth Overlap of Atomic Positions (SOAP)~\cite{bartokRepresentingChemicalEnvironments2013} descriptors, which result in:
\begin{equation}
\begin{aligned}
\mathbf{k}^{\lambda} (A_i, A_j)
& = \int\limits\limits_{\hat{R} \in \mathrm{SO(3)}} \dd \hat{R} \mathbf{D}^\lambda (\hat{R})
\left| \int \dd \mathbf{r} \rho_i (\mathbf{r}) \rho_j (\hat{R} \mathbf{r}) \right|^2,
\end{aligned}
\label{eq:lambda_kernel}
\end{equation}
where \(i, j\) index atoms in the structure,
\(\mathbf{D}^\lambda\) is the Wigner-D matrix for spherical harmonics \(\lambda\).
\(\rho_i\) is the representation of the atomic environment around atom \(i\).
This kernel can be equivalently written using Dirac notation for descriptors~\cite{willattAtomdensityRepresentationsMachine2019},
\begin{equation}
\label{eq:lambda_kernel_abstract}
\mathbf{k}^{\lambda} (A_i, A_j) = \braket{\chi_j^{(2)},\lambda}{\chi_i^{(2)},\lambda},
\end{equation}
where the general feature vector \(\ket{\chi_i^{(2)}}\) is given in this case by the tensor product of atomic density representation of the atomic environment with itself, \(\ket{\chi_i^{(2)}}=\ket{\rho_i \otimes \rho_{i}}\).
Different representations of the atomic environment yield distinct descriptors with various radial range characteristics~\cite{grisafiMultiscaleApproachPrediction2021}, which are summarised in \autoref{sec:long_range_representation}.

The GPR optimisation minimises the loss function:
\begin{equation}
\mathcal{L} (\Tilde{\mathbf{w}})
= \qty(\mathbf{c}^\text{ML} - \mathbf{c}^\text{DF})^\top \mathbf{S} \qty(\mathbf{c}^\text{ML} - \mathbf{c}^\text{DF})
+ \eta \Tilde{\mathbf{w}}^\top \Tilde{\mathbf{w}}
\label{eq:salted_loss}
\end{equation}
where the first term measures the squared spatial density error between predicted DF coefficients \(\mathbf{c}^\text{ML}\) and reference DF coefficients \(\mathbf{c}^\text{DF}\) weighted by the overlap matrix \(\mathbf{S}\) of the auxiliary basis, \(\Tilde{\mathbf{w}}\) represents the regression weights, and \(\eta\) is the regularisation parameter controlling the trade-off between fitting accuracy and model generalisation ability.

Applying SALTED to moir\'e superlattice prediction introduces specific technical challenges. First, previous SALTED implementations~\cite{lewisLearningElectronDensities2021,grisafiElectronicStructurePropertiesAtomCentered2023} employed the SOAP  descriptor when calculating symmetry-adapted kernels, as in \autoref{eq:lambda_kernel}.
This descriptor relies on the locality assumption: atomic environments are represented through exponentially-decaying Gaussian densities with a cutoff radius of a few {\AA}.
However, this locality constraint can fail 
for moir\'e superlattices with periodicities reaching tens to hundreds of {\AA}. This point is addressed in more depth in  \autoref{sec:long_range_representation}.

Second, the density fitting technique used in SALTED employs intentionally overcomplete auxiliary basis sets to ensure accurate density representation across diverse chemical environments.
However, this over-completeness results in ill-conditioned overlap matrices that introduce numerical instability and degrade model performance, particularly for heavy elements requiring large and diffuse basis sets. This challenge is addressed in \autoref{sec:low_rank_approximation}.

Third, downstream electronic structure properties depend not only on density prediction accuracy measured by density error metric, but also on the spatial distribution of density errors~\cite{lewisLearningElectronDensities2021,lewisPredictingElectronicDensity2023}.
This can create a validation gap between density-based loss function optimisation and reliable electronic property prediction. Therefore, we evaluate the model explicitly on the accuracy of these predicted electronic properties, as well as the density itself. Details of these evaluation metrics are given in \autoref{sec:appendix}.

\subsection{Beyond Locality by Long-Range Representation\label{sec:long_range_representation}}

Twisted bilayer systems exhibit electronic structure governed by moir\'e-scale interlayer interactions spanning length scales far exceeding the range of conventional local descriptor cutoffs and graph neural network message-passing distances.
Local SOAP descriptors~\cite{bartokRepresentingChemicalEnvironments2013} rely on exponentially-decaying Gaussian atomic density representations with cutoff radius limited to a few {\AA} to avoid oversmoothness. A different family of descriptors has been proposed by Grisafi and coworkers, which we employ here. This is the LODE (long-distance equivariant descriptor)~\cite{grisafiIncorporatingLongrangePhysics2019} descriptor family, which uses atomic electrostatic representations \(V_i\) to naturally encode non-local geometric information through a Coulomb potential-like descriptor. \(V_i\) is defined as
\begin{equation}
V_i (\mathbf{r}) = \int \dd \mathbf{r'} \frac{\rho_i (\mathbf{r'})}{|\mathbf{r}_i - \mathbf{r'}|}
\label{eq:electrostatic_representation}
\end{equation}
where \(\rho_i\) is the atomic density representation used in SOAP.
The \(1/r\) decay of \(V_i\) enables descriptors to capture structural information beyond local neighbourhoods without requiring prohibitively large cutoffs.

Symmetry-adapted kernels sensitive to structural features at different length scales can be constructed by choosing different representations for the atomic environments, \(\ket{\chi_i^{(2)}}\) in \autoref{eq:lambda_kernel_abstract}.
SOAP uses \(\ket{\chi_i^{(2)}} = \ket{\rho_i \otimes \rho_i}\), relying solely on the local atomic density representation.
Several different LODE descriptors can be built. The one used in most recent works~\cite{grisafiPredictingChargeDensity2023,rossiLearningElectrostaticResponse2025a},  LODE(1,1) which we refer to here simply as LODE, combines local and electrostatic representation as \(\ket{\chi_i^{(2)}}=\ket{\rho_i \otimes V_{i}}\)~\cite{grisafiIncorporatingLongrangePhysics2019,grisafiMultiscaleApproachPrediction2021}.
The descriptor we call LOVV (\(\ket{\chi_i^{(2)}} = \ket{V_i \otimes V_i}\)) in this paper corresponds to LODE(0,2) in the original nomenclature, meaning it has no density components and two potential-like components. Its utility for long-range systems was not extensively investigated previously, and we demonstrate here its crucial role in moir\'e materials prediction.
These descriptors exhibit distinct sensitivity ranges as quantified in \autoref{tab:descriptor_ranges}:
SOAP is limited by its \(6\angs\) cutoff, while LODE and LOVV maintain sensitivity up to \(13\angs\) and \(17\angs\) respectively (details in \sinfo Section S5A).
Performance comparisons across different materials and twist angles are presented in \autoref{sec:results}, with physical discussion in \autoref{sec:discussions}.

\begin{table}[htbp]
    \centering
    \caption{
        Descriptor representations and sensitivity ranges.
        Sensitivity range is defined as the interatomic distance at which descriptor changes drop below \(10^{-2}~\angs^{-1}\) upon atomic displacement (see \sinfoplain Section S5A).
        The atomic density representation \(\rho_i\) uses a \(6 \angs\) cutoff.
        \(V_i\) denotes the electrostatic representation.
    }
    \label{tab:descriptor_ranges}
    \begin{tabular}{ccc}
        \textbf{Descriptor} & \(\ket{\chi_i^{(2)}}\) & \textbf{Sensitivity Range} \\
        \hline
        SOAP~\cite{bartokRepresentingChemicalEnvironments2013} & \(\ket{\rho_i \otimes \rho_i}\) & \(6 \angs\) \\
        LODE~\cite{grisafiIncorporatingLongrangePhysics2019,grisafiMultiscaleApproachPrediction2021} & \(\ket{\rho_i \otimes V_i}\) & \(13 \angs\) \\
        LOVV~\cite{grisafiIncorporatingLongrangePhysics2019} & \(\ket{V_i \otimes V_i}\) & \(17 \angs\)
    \end{tabular}
\end{table}

\subsection{Low-Rank Approximation for Mitigating Ill-Conditioning\label{sec:low_rank_approximation}}

The DF auxiliary basis sets used in SALTED are intentionally over-complete to ensure accurate electron density representation across diverse chemical environments.
However, this over-completeness leads to significant overlap between basis functions. This can result in an ill-conditioned overlap matrix \(\mathbf{S}\), which appears in the loss function in \autoref{eq:salted_loss}.
Large condition numbers \(\kappa = \sigma_\text{max} / \sigma_\text{min}\), defined as the ratio between the maximum and minimum singular values, quantitatively indicate numerical instability in coefficient calculations due to near-null space directions (corresponding to small singular values)~\cite{golubMatrixComputations2013,deisenrothMathematicsMachineLearning2020}. This introduces noise that degrades both training accuracy and generalisation capability for DF-based ML methods.

For systems which display severe ill-conditioning, we employ a low-rank approximation of the real-symmetric positive-definite overlap matrix via singular values (SV) truncation with threshold \(\delta\):
\begin{equation}
\mathbf{S} = \mathbf{U} \mathbf{\Sigma} \mathbf{U}^\top
\approx
\Tilde{\mathbf{S}} = \mathbf{U} \Tilde{\mathbf{\Sigma}} \mathbf{U}^\top
\end{equation}
where \(\mathbf{U}\) contains the left singular vectors and \(\Tilde{\mathbf{\Sigma}}\) is the diagonal matrix of truncated singular values, retaining only those satisfying \(\sigma_i > \delta\).
The truncated overlap matrix \(\Tilde{\mathbf{S}}\) is then used in the loss function in \autoref{eq:salted_loss} during model training, and also in density error evaluation in \autoref{eq:density_error} during validation.
This truncation serves two purposes:
(1) implicit regularisation by discarding near-null subspace directions associated with basis function redundancy rather than meaningful physical variations;
and (2) numerical stability during training is improved by removing near-null space directions.
Also, since we target extrapolating to moir\'e superlattices, the low-rank approximation improves generalisation by preventing overfitting to noise-dominated components.

Considering the real, symmetric, and positive-definite nature of \(\mathbf{S}\), eigenvalue decomposition is mathematically equivalent to singular value decomposition (SVD) for this application.
In this work, we use SVD.

\subsection{Validation Metrics\label{sec:appendix}}

We quantified the prediction accuracy of the electron density using the integrated root mean square error (RMSE), normalised by the residual density:
\begin{equation}
\begin{aligned}
    \Delta\rho
    = & \qty(
        \frac{
            \sum_k \int \dd \mathbf{r} \qty| \rho_k^\mathrm{ML}(\mathbf{r}) - \rho_k^\mathrm{DF}(\mathbf{r}) |^2
        }{
            \sum_k \int \dd \mathbf{r} \qty| \rho_k^\mathrm{DF}(\mathbf{r}) - \rho_k^\mathrm{sph}(\mathbf{r}) |^2
        }
    )^{1/2} \\
    = & \qty(
        \frac{
            \sum_k \qty( \mathbf{c}_k^\mathrm{ML} - \mathbf{c}_k^{\mathrm{DF}} )^\top \mathbf{S}_{k} \qty( \mathbf{c}_k^\mathrm{ML} - \mathbf{c}_k^{\mathrm{DF}} )
        }{
            \sum_k \qty( \mathbf{c}_k^\mathrm{DF} - \mathbf{c}_k^{\mathrm{sph}} )^\top \mathbf{S}_{k} \qty( \mathbf{c}_k^\mathrm{DF} - \mathbf{c}_k^{\mathrm{sph}} )
        }
    )^{1/2}
\end{aligned}
\label{eq:density_error}
\end{equation}
where \(k\) is the geometry index in the dataset, \(\mathbf{S}_k\) is the DF auxiliary basis overlap matrix, \(\mathbf{c}_k^\text{sph}\) represents spherical components of the DF coefficients.
Normalisation by the deviation from spherical approximation makes this metric extremely sensitive: for graphene, a \(1\%\) RMSE in \autoref{eq:density_error} corresponds to less than \(0.02 \%\) error if normalised by total density; \mods has large residual densities, which suppresses its normalised density error in \autoref{tab:experiment_results_all_materials_descriptors} (see \sinfo~S2D for details).
Density fitting is expensive for large moir\'e structures, so twisted bilayer density error is reported within memory limits in \sinfo~S2D. The band structure error is reported for all test structures.

To evaluate subsequent electronic properties, we restart DFT calculation from the predicted density using FHI-aims~\cite{blumInitioMolecularSimulations2009,abbottRoadmapAdvancementsFHIaims2026shortAuthorList}.
The SALTED-predicted density is used to initialise a DFT calculation, with just one Hamiltonian diagonalisation performed to establish the electronic state and obtain energies.
With this electronic state fixed, we evaluate subsequent electronic properties, including band structures as well as real-space electronic properties.
This procedure avoids iterative convergence, while enabling direct comparison with fully converged DFT results.

Since moir\'e physics emerges from low-energy bands near the Fermi level, and these bands are most sensitive to long-range density features, we directly evaluated band structure prediction accuracy by mean absolute error (MAE):
\begin{equation}
\Delta b \qty(\{B\})
= \frac{1}{N_\mathrm{k} N_{\{B\}}} \sum_{\substack{i \in \text{k-path},\\ b \in \{B\}}}
\qty| \epsilon_{b}^\text{ML} (\mathbf{k}_i) - \epsilon_{b}^\text{DFT} (\mathbf{k}_i) + \delta_\epsilon |
\label{eq:low_energy_error}
\end{equation}
where \(N_\mathrm{k}\) is the number of k-points (\(\{\mathbf{k}_i\}\)) in the k-path (see \sinfo~S1C),
\(\{B\}\) is a set of low-energy bands close to the Fermi level (4 valence + 4 conduction bands for graphene and hBN, and 6 + 6 for \tids, \zrds, and \mods), \(N_{\{B\}}\) is the number of bands in \(\{B\}\),
\(\epsilon_{b}^\text{ML} (\mathbf{k}_i)\) and \(\epsilon_{b}^\text{DFT} (\mathbf{k}_i)\) are band energies from SALTED-predicted density and fully converged DFT respectively,
and \(\delta_\epsilon\) aligns band structures from ML and DFT by gap centres.
This metric simultaneously evaluates band gap and band dispersion accuracy in the low-energy regime governing moir\'e phenomena.

\subsection{Model Hyperparameters and Descriptor Coverage \label{sec:appendix2}}

\begin{figure}[htbp]
    \centering
    \includegraphics[width=0.75\onecolwidth]{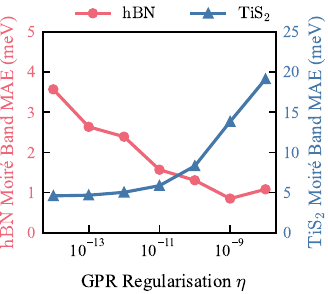}
    \caption{
        \textbf{Low-energy moir\'e band error varies by GPR regularisation \(\eta\) for hBN and \tids.}
        As \(\eta\) increases, low-energy moir\'e band error decreases for hBN but increases for \tids, showing contrary trends.
        The descriptors here are LOVV.
    }
    \label{fig:band_metric_by_regul_graphene_tis2}
\end{figure}

The optimal GPR regularisation \(\eta\) exhibits system-dependent behaviour, related to the application of the low-rank approximation to the overlap matrix.
For all materials, the validation density error decreases monotonically with decreasing \(\eta\) (see \autoref{eq:salted_loss}) until reaching a saturation value determined by the descriptor, which is expected since our loss function in \autoref{eq:salted_loss} relies on the spatial integral of the squared error in the density.
However, the error in the band structure prediction of the twisted bilayers shows contrasting dependence on \(\eta\) across material classes, as shown in \autoref{fig:band_metric_by_regul_graphene_tis2}.
For well-conditioned systems (graphene, hBN), the optimal \(\eta\) for band prediction is larger than that for density minimisation.
Stronger regularisation suppresses overfitting to local structural variations and residual noise from density fitting.
For ill-conditioned systems (\tids, \zrds, \mods), the optimal \(\eta\) for band prediction is smaller than that for the density.
GPR regularisation here stabilises against low-rank truncation artifacts rather than suppressing noise. These results suggest that a modification directly to the loss function targetting property optimisation could improve the models further.
Nevertheless, the model accuracy in practice is not highly sensitive to the precise value of \(\eta\): for graphene and hBN, a single compromise \(\eta\) between the density and band optima increases either metric by less than a factor of two; for TMDCs, both metrics share the same small-\(\eta\) plateau. The total energies error on the moir\'e test sets remains low regardless of whether \(\eta\) is optimised for band structure or density (see \sinfo~S2B for details). The workflow does not require precise per-property hyperparameter tuning.

Gaussian process regression predicts via weighted similarity in descriptor space, meaning that accurate extrapolation from aligned to twisted structures relies on the descriptors used in the prediction of extrapolated structures being represented within the descriptors associated with the training data.
We used dimensionality reduction via UMAP \cite{mcinnesUMAPUniformManifold2020} to enable a qualitative descriptor coverage assessment.

\begin{figure}[htbp]
    \centering
    \includegraphics[width=\onecolwidth]{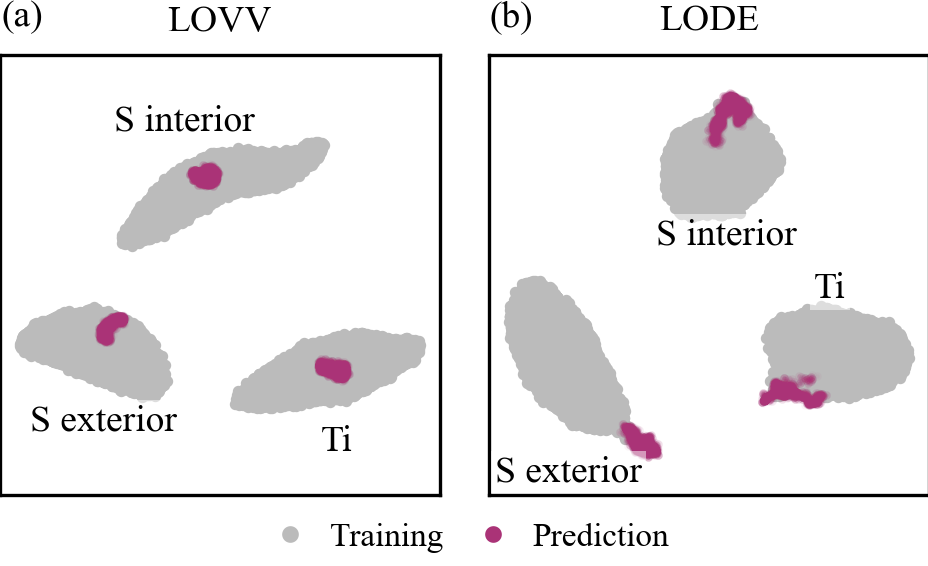}
    \caption{
        \textbf{Descriptor distribution dimension reduction by UMAP for \tids.}
        For both (a) LOVV and (b) LODE descriptors, descriptors for each atom are automatically clustered according to atomic species (Ti and S) and geometric position (interior S atoms and exterior S atoms).
        Training (aligned bilayers) and test (twisted structures) atomic descriptors are coloured in grey and purple, respectively.
        The prediction distribution can be well-covered by the training distribution in LOVV,
        while in LODE, the prediction cluster of Ti atoms is not well-covered by their training counterpart, and both interior and exterior S atom clusters lie on the edge of the training clusters.
    }
    \label{fig:descriptors_umap_results_tis2}
\end{figure}

As an example, the UMAP plots of the LOVV and LODE descriptors for \tids are shown in \autoref{fig:descriptors_umap_results_tis2}. LODE descriptors show limited overlap between the training (aligned bilayers) and extrapolated (twisted bilayers) distributions, while the LOVV descriptors of extrapolated structures all lie within the distribution of the descriptors of the training structures. Equivalent plots for the other materials are provided in the \sinfo~S5B.
These results suggest qualitatively that short-range descriptors are less likely to cover the twisted bilayer descriptor space, leading to extrapolation failures, while LOVV's long-range nature ensures sufficient coverage for accurate predictions.

\subsection{Model Training and Conditioning \label{sec:train-val}}

\begin{figure*}[htbp]
    \includegraphics[width=\twocolwidth]{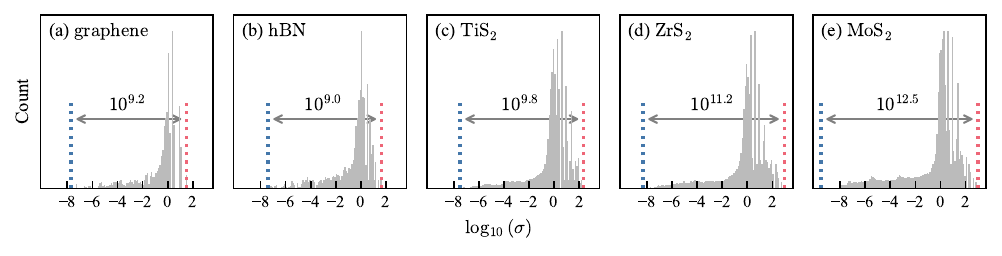}
    \caption{
        \textbf{Singular value distribution of overlap matrices.}
        (a-e) Singular value decomposition results for DF basis overlap matrices across all training structures.
        The y-axis shows counts (arbitrary units), and the x-axis shows log-scale singular values.
        Vertical green/red lines indicate mean minimum/maximum singular values, with their separation representing the average condition number.
        TMDCs, especially \mods, exhibit significantly larger condition numbers than light-element systems.
    }
    \label{fig:overlap_matrix_singular_values_histogram}
\end{figure*}

We trained SALTED models to predict the electron density and band structure of a range of 2D bilayer materials: graphene, hBN, \tids, \zrds, and \mods.
We began by compiling a dataset of 512 aligned (displaced) bilayer structures of size \(3\times3\) for each material, which were generated by randomly sampling in-plane displacements, interlayer spacings between the bilayers, and per-atom coordinate perturbations within each layer.
The range of the displacements in each one of these coordinates is described in the \sinfo~S1A. 
Each dataset was split into training and validation subsets of 400 and 112 structures respectively.
    
The electron densities used to train the machine learning models and the reference band structures of the twisted bilayer structures were calculated using the PBE functional \cite{perdewGeneralizedGradientApproximation1996}. 
All of these calculations were performed using FHI-aims~\cite{blumInitioMolecularSimulations2009,abbottRoadmapAdvancementsFHIaims2026shortAuthorList}, using light or intermediate basis sets depending on DF accuracy in reproducing band structure as described in \sinfo~S1.
Detailed specifications of dataset generation, auxiliary basis set selection, hyperparameter optimisation, and structural parameters are provided in the \sinfo~S1 and S2. Despite the shortcomings of PBE and other GGA functionals regarding the description of band-gaps, we took a pragmatic choice of employing this functional to train the model because this is the \textit{de facto} standard on previous studies of the materials reported here~\cite{fangElectronicStructureTheory2016,kimElectrostaticMoirePotential2024,liStraininducedEnhancementThermoelectric2017,claassenUltrastrongSpinOrbit2022,xiaoEffectsVanWaals2014}. Training SALTED on data from hybrid functionals also works seamlessly and will be the topic of future work.

\autoref{fig:overlap_matrix_singular_values_histogram} shows the distribution of the singular values of the overlap matrices for every structure in the training dataset of each material, along with the average condition numbers of these matrices.
The condition numbers indicate particularly severe ill-conditioning for the TMDC training datasets, likely due to the larger and more diffuse DF auxiliary basis sets required by heavier elements.
As a result, the low-rank approximation to the overlap matrix described in \autoref{sec:low_rank_approximation} was applied to the TMDC training datasets, with \(\delta=10^{-6}\) used for \tids, and \(\delta=10^{-2}\) for both \mods and \zrds.

\textbf{Workflow:} The overall workflow for predicting electron densities and band structures of 2D materials described in the preceding sections is summarised here: 

\begin{itemize}
\item We begin by generating a training dataset, ensuring a range of intra-layer and inter-layer geometries are sampled. 
\item The electron densities of this training set are computed using DFT, and the corresponding DF coefficients are obtained using an auxiliary basis set that ensures density and band structure reproducibility. 
\item The condition number of the overlap matrix of these auxiliary basis functions is calculated for each structure in the dataset, and if the average condition number exceeds our empirical threshold \((\kappa > 10^{9.5})\) the low rank approximation to the overlap matrices is applied. 
\item A descriptor is then selected to encode the training structures, with the appropriate choice depending on both the material of interest and the property being targeted. 
\item The SALTED model is then trained, optimising the regularisation hyperparameter \(\eta\), and if appropriate the descriptor hyperparameters. The model is evaluated against both error in the predicted electron density and accuracy of the predicted band structure, since low errors in the former do not guarantee low errors in the latter.
\end{itemize}

To illustrate the practical computational gains of using the SALTED model, we take \zrds as an example:
computing a twisted structure at \(\theta=2.281^\circ\) (3786 atoms per unit cell) would require about 11.5k CPU hours to converge the DFT electronic density (estimated by the scaling reported in \sinfo Figure~S28).
In comparison, the total cost of \zrds dataset generation and training of five SALTED models for \zrds is around 6.7k CPU hours, attesting to the efficiency of using such models on the study of twisted bilayer systems.

\section*{Associated Content}

\textbf{Data Availability Statement}

Datasets and models are available on Zenodo~\cite{LouZenodoDataset2026}.

\textbf{Notes}

The authors declare no competing financial interest.

\section*{Acknowledgments}

The authors thank Andrea Grisafi for discussions and clarifications regarding different descriptors and the SALTED infrastructure.
The authors also thank Lede Xian for initial discussions on twisted bilayer systems and \'Angel Rubio for insightful input on the application aspects of this framework.
Z.L. thanks Dante Kennes, Jingkai Quan, and Xinle Cheng for helpful discussions.
Z.L. was partially funded by the Max Planck Graduate Center for Quantum Materials.
Computational time was provided by the MPCDF.
The work was partially funded by the Deutsche Forschungsgemeinschaft (DFG, German Research Foundation) – Project-ID 555467911 – CRC 1772 / TP A06.

\putbib[ref_zotero_export_bibtex,customized]

\end{bibunit}

\clearpage
\onecolumngrid
\appendix
\renewcommand{\thesection}{S\arabic{section}}
\renewcommand{\thesubsection}{\Alph{subsection}}
\renewcommand{\appendixname}{}
\global\setcounter{figure}{0}
\global\setcounter{table}{0}
\global\setcounter{equation}{0}
\global\setcounter{section}{0}
\renewcommand{\theequation}{S\arabic{equation}}
\renewcommand{\thefigure}{S\arabic{figure}}
\renewcommand{\thetable}{S\arabic{table}}
\makeatletter
\renewcommand{\theHfigure}{S\arabic{figure}}
\renewcommand{\theHtable}{S\arabic{table}}
\renewcommand{\theHequation}{S\arabic{equation}}
\renewcommand{\theHsection}{S\arabic{section}}
\makeatother

\begin{center}
    {\large\textbf{Long-Range Machine Learning of Electron Density for Twisted Bilayer Moir\'e Materials}}

    {\large\textbf{SUPPLEMENTAL MATERIALS}}
    \vspace{1em}

    Zekun Lou,$^{1}$ Alan M. Lewis,$^{2}$ and Mariana Rossi$^{1,3}$
    \vspace{0.5em}

    {\small\itshape
    $^{1}$MPI for the Structure and Dynamics of Matter,\\
    Luruper Chaussee 149, 22761 Hamburg, Germany}

    {\small\itshape
    $^{2}$Department of Chemistry, University of York, York YO10 5DD, U.K.}

    {\small\itshape
    $^{3}$Yusuf Hamied Department of Chemistry, Lensfield Road, Cambridge CB2 1EW, U.K.}
\end{center}

\vspace{3em}

\begin{bibunit}[apsrev4-2]  

\section{Dataset Details\label{app:setup_summary}}

All density functional theory calculations in this work were performed using FHI-aims \cite{blumInitioMolecularSimulations2009,abbottRoadmapAdvancementsFHIaims2026shortAuthorList}.

\subsection{Geometry Parameters\label{app:materials_geometries}}

\begin{table}[htbp]
    \centering
    \caption{
        Structural parameters and computational details for bilayer primitive cells.
        \(a\): bilayer primitive cell lattice constant;
        \(d\): bilayer interlayer distance, i.e. $z$-direction separation between layer centres (superscript r: obtained by DFT relaxation calculations);
        \(\Tilde{\sigma}_\text{a}\): per-atomic random perturbation standard deviation before atomic mass normalisation;
        \(\sigma_{z}\): $z$-direction interlayer distance displacement standard deviation.
        Cost: computational cost, wall time in CPU hours for generating a training dataset.
        The k-grid refers to the converged k-grid, via \(\Gamma\)-centred k-grid sampling by converging total energy within 0.1 meV per atom on a \(3 \times 3 \times 1\) bilayer supercell in the training dataset.
    }
    \label{tab:material_parameters}
    \begin{tabular}{lcccccccl}
    \hline
    Material & \(a (\nsangs)\) & \(d (\nsangs)\)  & \(\Tilde{\sigma}_\text{a} (\nsangs\cdot\mathrm{u}^\frac{1}{2})\) & \(\sigma_z (\nsangs)\) & Cost (CPU Hours) & Basis Set & k-grid & Stacking Symmetry \\
    \hline
    Graphene & 2.4595~\cite{zakharchenkoFiniteTemperatureLattice2009} & 3.4866\(^{\mathrm{r}}\) & \(0.05\sqrt{6}\) & \(0.12\) & 625 & light & \(8 \times 8 \times 1\) & - \\
    hBN & 2.5124~\cite{jainCommentaryMaterialsProject2013} & 3.4677\(^{\mathrm{r}}\) & \(0.05\sqrt{7}\) & \(0.12\) & 661 & light & \(4 \times 4 \times 1\) & in-plane inversion~\cite{gilbertAlternativeStackingSequences2019} \\
    1T-\tids & 3.4079~\cite{krogelPerspectivesVanWaals2020} & 5.6989~\cite{krogelPerspectivesVanWaals2020} & \(0.5\) & \(0.3\) & 9125 & light & \(8 \times 8 \times 1\) & z inversion~\cite{anishaThermoelectricPerformance1TZrS22023} \\
    1T-\zrds & 3.6394\(^{\mathrm{r}}\) & 5.9835\(^{\mathrm{r}}\) & \(0.5\) & \(0.2\) & 5184 & light & \(6 \times 6 \times 1\) & z inversion~\cite{anishaThermoelectricPerformance1TZrS22023} \\
    1H-\mods & 3.1530\(^{\mathrm{r}}\) & 6.3019\(^{\mathrm{r}}\) & \(0.5\) & \(0.2\) & 34121 & intermediate & \(4 \times 4 \times 1\) & in-plane inversion~\cite{heStackingEffectsElectronic2014} \\
    \hline
    \end{tabular}
\end{table}

Our datasets consist of five materials.
\autoref{tab:material_parameters} shows detailed bilayer primitive cell geometries for all materials used in this work.
Graphene interlayer distance is obtained by relaxing twisted bilayer graphene structures with the intermediate species default and MBD@rsSCS for van der Waals (vdW) interactions~\cite{tkatchenkoAccurateEfficientMethod2012}.
hBN primitive cell data is retrieved from the Materials Project~\cite{jainCommentaryMaterialsProject2013} entry mp-984 from database version v2025.09.25. The interlayer distance is obtained by relaxing twisted bilayer hBN structures with the intermediate species default and MBD@rsSCS for vdW correction.
\zrds bilayer primitive cell size and interlayer distance are obtained by relaxing the most energy-favoured AA1 stacking bilayer structure~\cite{anishaThermoelectricPerformance1TZrS22023} with tight species defaults and MBD-NL for vdW interactions~\cite{hermannDensityFunctionalModel2020}.
\mods bilayer primitive cell size and interlayer distance are obtained by relaxing the most energy-favoured AA' stacking bilayer structure~\cite{heStackingEffectsElectronic2014} with tight species defaults and MBD-NL for vdW interactions~\cite{hermannDensityFunctionalModel2020}.

\subsection{Datasets for Training and Testing\label{app:dataset_details}}

For each material, training datasets are generated by random sampling on \(3 \times 3\) bilayer supercells.
The sampling method works in the following manner.
Starting from a rigid bilayer, the upper layer is randomly displaced in the $xy$-plane (parallel to the bilayer) following a uniform distribution, and then moved in the $z$-direction (perpendicular to the bilayer) according to a normal distribution with standard deviation \(\sigma_{z}\).
After these displacements, all atoms are further randomly displaced in all directions by a normal distribution.
This perturbative displacement uses a standard deviation defined as \(\sigma_\text{a}^{M} = \Tilde{\sigma}_\text{a}/\sqrt{M}\) (\(M\) is the atomic mass in Dalton), which is normalised by the nuclear mass to mimic phonon vibrations.
All training datasets consist of 512 structures.
All structure cells contain a vacuum region of at least \(90 \angs\) to minimise the interaction between periodic images.
For bilayer stacking symmetry (column ``Stacking Symmetry'' in \autoref{tab:material_parameters}), each symmetry-related stacking configuration is equally represented with the same share in the training dataset.

Testing datasets are generated by constructing proper supercells guided by commensurate angles and their corresponding supercell transformation matrices.
The commensurate twist angles are parametrised by integers \(m\) and \(r\) ~\cite{lopesdossantosContinuumModelTwisted2012}:
\begin{equation}
    \theta = \mathrm{arccos} \frac{3 m^2 + 3 mr + r^2 / 2}{3 m^2 + 3 mr + r^2}.
    \label{eq:twist_angle_m_r}
\end{equation}
In this work, we fix \(r=1\) and vary \(m\) from 1 to larger integers.
For graphene and hBN, the \(m\) values from 1 to 10 are included, corresponding to twist angles from \(21.79^\circ\) to \(3.15^\circ\).
For TMDCs, the \(m\) values are from 1 to 8, corresponding to twist angles from \(21.78^\circ\) to \(3.89^\circ\).
Bilayer stacking symmetries (column ``stacking symmetry'' in \autoref{tab:material_parameters}) are included in the testing datasets by including all the high-symmetry stacking configurations for each commensurate twist angle.

\subsection{Density Functional Theory Calculations\label{app:dft_details}}

All the calculations are performed with the Perdew-Burke-Ernzerhof (PBE) \cite{perdewGeneralizedGradientApproximation1996} exchange-correlation functional, atomic ZORA approximation~\cite{blumInitioMolecularSimulations2009}, without spin polarisation.
Different species default and density fitting basis functions are determined by the band structure reproducibility described below.
The converged $k$-grids for \(3 \times 3 \times 1\) bilayer supercells in each material are shown in \autoref{tab:material_parameters}.
The DFT convergence criteria and smearing scheme are set to default values of FHI-aims.

\begin{figure}[H]
    \centering
    \includegraphics[width=\twocolwidth]{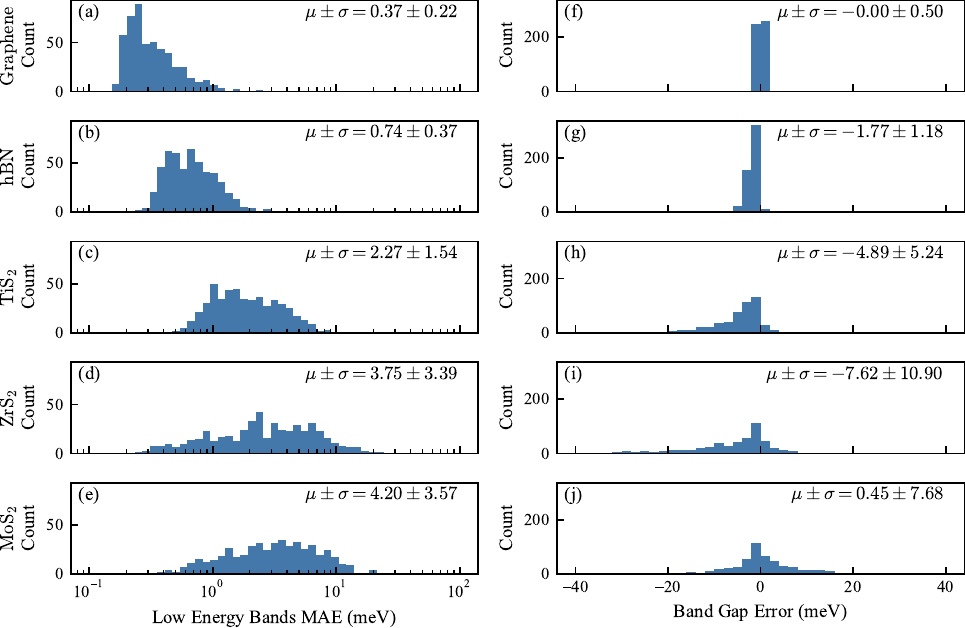}
    \caption{
        \textbf{Comparing DFT-converged bands and DF-restarted bands for the aligned datasets in training models.}
        (a-e) Low energy band error for each material. The bands considered for each material follow those in the main paper: 4 top valence bands and 4 bottom conduction bands for graphene and hBN, and 6 top valence bands and 6 bottom conduction bands for \tids, \zrds, and \mods.
        (f-j) Band gap error for each material.
        Mean values \(\mu\) and standard deviations \(\sigma\) are listed per metric and material.
        For k-path sampling, we sample 11 points on \(\mathrm{K \Gamma}\), 6 points on \(\mathrm{\Gamma M}\), and 9 points on \(\mathrm{M K}\) for all materials including both endpoints.
    }
    \label{fig:trainset_df_restart_compare_to_dft_bands}
\end{figure}

The density fitting (DF) basis is chosen to ensure bilayer band structure reproducibility.
For each material, a \(3 \times 3 \times 1\) bilayer structure is chosen from the training dataset, and DF coefficients are calculated with various basis sets; for TMDCs, extra DF basis functions are added to the transition metal atoms if necessary.
Detailed DF accuracy in terms of band structures for each material is shown in \autoref{fig:trainset_df_restart_compare_to_dft_bands}.
The DF basis sets used are listed in \autoref{tab:material_parameters}. For \zrds, an extra auxiliary basis is applied to the Zr atoms, and also for \mods, one is applied to the Mo atoms, both using the setting \verb|for_aux hydro 5 g 0.0| in FHI-aims.
The shared density fitting numerical settings in FHI-aims are:
\begin{verbatim}
    default_max_l_prodbas 6
    default_prodbas_acc 1e-4
    wave_threshold 1e-6
\end{verbatim}

Dipole correction is included by \verb|use_dipole_correction|.
For k-path sampling, we sample 11 points on \(\mathrm{K \Gamma}\), 9 points on \(\mathrm{\Gamma M}\), and 6 points on \(\mathrm{M K}\) for all the materials including both endpoints.
Any other different k-path settings will be stated.

For FHI-aims, versions after 240403 are recommended to avoid basis reordering due to spherical harmonics conventions.

\section{SALTED Model Details\label{app:validation_results}}

\subsection{Training Parameters\label{app:training_parameters}}

For all materials and descriptors, the following descriptor parameters suggested in SALTED are used: cutoff radius \(r_\text{cut}=6\angs\), number of radial basis functions \(n_\text{rad} = 6\), number of angular momentum channels \(n_\text{ang}=6\), and Gaussian width \(\sigma=0.3 \angs\).
The kernel trick parameter \(z=2\) and RKHS sparsification \(M_\text{env}=200\) are shared across all experiments.

\begin{figure}[H]
    \centering
    \includegraphics[width=\twocolwidth]{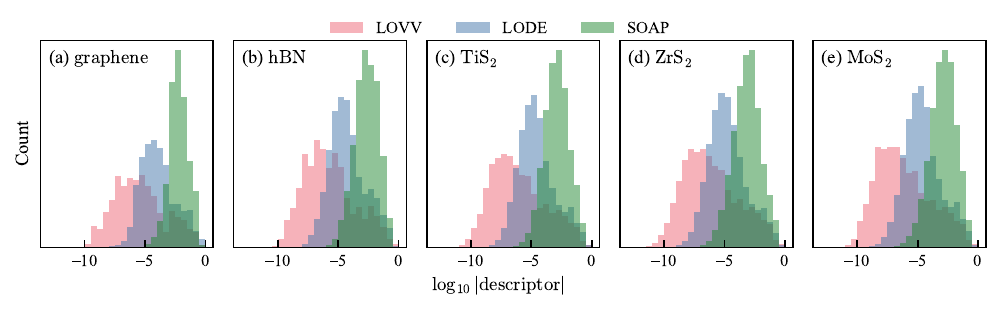}
    \caption{
        \textbf{Descriptor magnitude distribution for \(\lambda=0\) component.}
        Histograms of absolute descriptor values (\(\log_{10}\) scale) in each training dataset.
        For each material-descriptor combination, the \(\lambda=0\) component of \(\lambda\)-descriptors~\cite{grisafiSymmetryAdaptedMachineLearning2018} is extracted for every atom, element-wise absolute values are computed, and the distribution is plotted.
        The systematic ordering of descriptor magnitudes (LOVV \(<\) LODE \(<\) SOAP) directly determines optimal GPR regularisation requirements, with higher-magnitude descriptors needing stronger regularisation.
    }
    \label{fig:2507ppr_dcpt_histogram}
\end{figure}

Different from neural network methods, where the regularisation parameters have a broad range of acceptable values, Gaussian process regression (GPR) regularisation in SALTED must be optimised on a per-descriptor basis.
This descriptor dependence arises from the reproducing kernel Hilbert space (RKHS) framework employed in SALTED \cite{grisafiElectronicStructurePropertiesAtomCentered2023}, where the GPR solution is obtained by inverting the Hessian matrix:
\begin{equation}
H = \Psi^\top S \Psi + \eta \mathbb{I},
\end{equation}
where \(S\) is the overlap matrix of the DF auxiliary basis functions, and \(\Psi\) is the RKHS vector whose magnitude scales by the descriptor magnitude proportionally.
Since the first term scales quadratically with \(\Psi\), larger descriptor magnitudes yield larger Hessian eigenvalues, necessitating larger \(\eta\) to maintain numerical stability and optimal generalisation.
\autoref{fig:2507ppr_dcpt_histogram} reveals systematic differences in descriptor magnitude distributions.
The histograms show the \(\lambda=0\) component~\cite{grisafiIncorporatingLongrangePhysics2019} of each descriptor for each material.
A consistent ordering across materials exists: SOAP exhibits the largest magnitudes, followed by LODE, then LOVV, with peak positions differing by several orders of magnitude.
Descriptor components with higher \(\lambda\) are constructed from \(\lambda=0\) with Wigner matrices, so such magnitude pattern is shared across all \(\lambda\)s.

While this ordering establishes the general hierarchy of regularisation requirements, the specific optimal \(\eta\) value also depends on other factors, including the material's electronic structure, the DF basis conditioning, and the geometric diversity of the training dataset.
Consequently, \(\eta\) should be optimised for each material-descriptor combination, as detailed in \autoref{tab:salted_hyperparameters_for_band_predictions} and \autoref{tab:salted_hyperparameters_for_density_predictions}, while explicit optimal regularisation depends on more aspects.

\subsection{Hyperparameters\label{app:full_experiment_results}}

\begin{figure}[H]
    \centering
    \includegraphics[width=\onecolwidth]{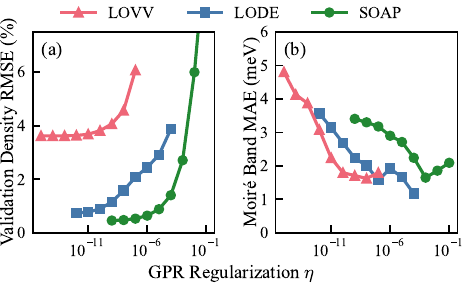}
    \caption{
        \textbf{GPR regularisation effect on graphene prediction.}
        Left: validation density error on displaced bilayer structures.
        Right: low-energy moir\'e band error on twisted bilayer test structures, averaged over \(\theta = 21.8^\circ \text{\textendash} 3.15^\circ\) (10 commensurate structures).
    }
    \label{fig:graphene_density_and_band_metric_by_regul}
\end{figure}

\begin{figure}[H]
    \centering
    \includegraphics[width=\onecolwidth]{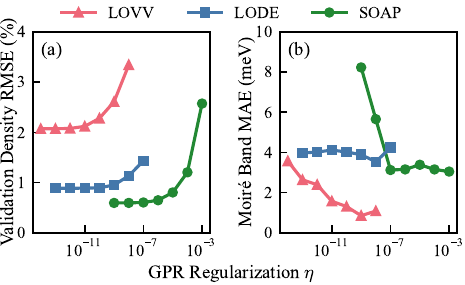}
    \caption{
        \textbf{GPR regularisation effect on hBN.}
        Left: validation density error on displaced bilayer structures.
        Right: low-energy moir\'e band error on twisted bilayer test structures, averaged over \(\theta = 21.8^\circ \text{\textendash} 3.15^\circ\) (\(10 \times 2\) commensurate structures due to stacking symmetry).
    }
    \label{fig:hBN_density_and_band_metric_by_regul}
\end{figure}

\begin{figure}[H]
    \centering
    \includegraphics[width=\onecolwidth]{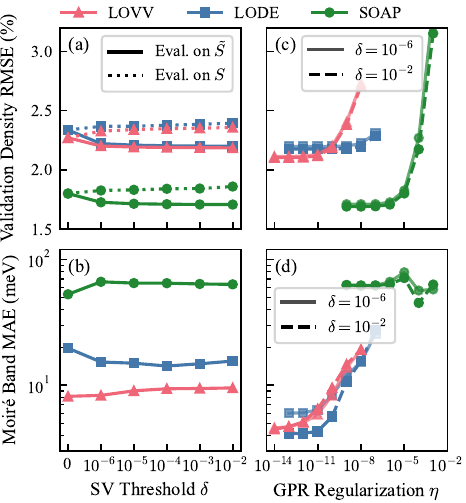}
    \caption{
        \textbf{GPR regularisation and SV threshold effects on \tids prediction.}
        Left column: effect of SV threshold \(\delta\) with fixed GPR regularization \(\eta\) (LOVV: \(\eta=10^{-10}\), LODE: \(\eta=10^{-8}\), SOAP: \(\eta=10^{-6}\)).
        (Solid/dotted lines in top left: evaluation using truncated \(\Tilde{S}\) / original \(S\) overlap matrices.)
        Right column: effect of GPR regularization \(\eta\) with fixed SV threshold \(\delta=10^{-6}\) or \(10^{-2}\).
        Top row: validation density error on displaced bilayer structures.
        Bottom row: low-energy moir\'e band error on twisted bilayer test structures, averaged over \(\theta = 21.8^\circ \text{\textendash} 3.89^\circ\) (\(8 \times 2\) commensurate structures due to stacking symmetry).
        The validation density error varies with \(\delta\) only in the range \(0 \sim 10^{-6}\).
        However, since the SALTED training workflow recommends starting the hyperparameter search at \(\delta=10^{-2}\), we benchmark both \(\delta=10^{-6}\) and \(10^{-2}\) in the right column when evaluating the effect of \(\eta\).
        Results show that \(\delta=10^{-2}\) achieves similar performance to \(\delta=10^{-6}\) when \(\eta\) is properly optimised.
        In both cases, smaller \(\eta\) values are preferred for better performance.
        The plateau behaviour in validation density error suggests satisfactory noise filtering at \(\delta=10^{-6}\).
        Therefore, we recommend \(\delta=10^{-6}\) for \tids to preserve more information in DF overlap matrices, rather than the more aggressive threshold of \(\delta=10^{-2}\) without too much benefit in performance.
    }
    \label{fig:tis2_density_and_band_metric_by_svt_regul_all_svt_at_regul}
\end{figure}

\begin{figure}[H]
    \centering
    \includegraphics[width=\onecolwidth]{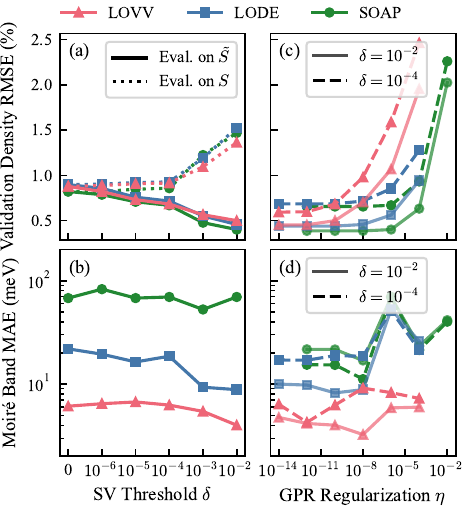}
    \caption{
        \textbf{GPR regularisation and SV threshold effects on \zrds prediction.}
        Left column: effect of SV threshold \(\delta\) with fixed GPR regularization \(\eta\) (LOVV: \(\eta=10^{-10}\), LODE: \(\eta=10^{-8}\), SOAP: \(\eta=10^{-6}\)).
        Right column: effect of GPR regularization \(\eta\) with fixed SV threshold \(\delta=10^{-4}\) or \(10^{-2}\).
        Top row: validation density error on displaced bilayer structures.
        Solid/dotted lines: evaluation using truncated \(\Tilde{S}\) / original \(S\) overlap matrices.
        Bottom row: low-energy moir\'e band error on twisted bilayer test structures, averaged over \(\theta = 21.8^\circ \text{\textendash} 3.89^\circ\) (16 commensurate structures).
        The validation density error evaluated with original overlap matrices shows a turning point at \(\delta=10^{-4}\).
        Since the SALTED training workflow recommends starting at \(\delta=10^{-2}\), we compare both \(\delta=10^{-4}\) and \(\delta=10^{-2}\) in the right column.
        Results show that \(\delta=10^{-2}\) achieves slightly better overall performance than \(\delta=10^{-4}\) when \(\eta\) is optimised. Both thresholds suggest that smaller \(\eta\) values yield better performance.
        Since the validation density error evaluated with truncated overlap matrices decreases almost monotonically from \(\delta=0\) to \(\delta=10^{-2}\) for LODE and LOVV, we recommend \(\delta=10^{-2}\) for \zrds as it effectively filters noise properly in the DF basis band and benefits overall model performance.
    }
    \label{fig:zrs2_density_and_band_metric_by_svt_regul_all_svt_at_regul}
\end{figure}

\begin{figure}[H]
    \centering
    \includegraphics[width=\onecolwidth]{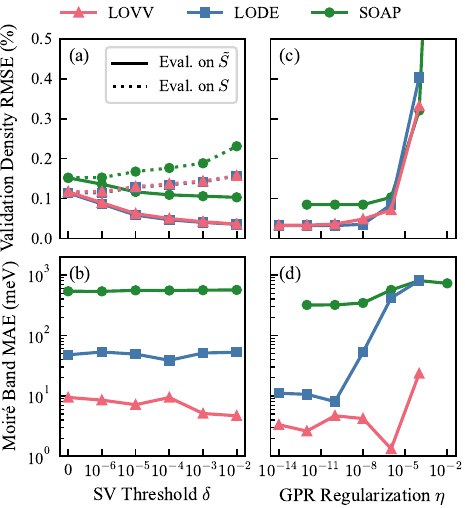}
    \caption{
        \textbf{GPR regularisation and SV threshold effects on \mods prediction.}
        Left column: effect of SV threshold \(\delta\) with fixed GPR regularization \(\eta\) (LOVV: \(\eta=10^{-10}\), LODE: \(\eta=10^{-8}\), SOAP: \(\eta=10^{-6}\)).
        Right column: effect of GPR regularization \(\eta\) with fixed SV threshold \(\delta=10^{-2}\).
        Top row: validation density error on displaced bilayer structures.
        Solid/dotted lines: evaluation using truncated \(\Tilde{S}\) / original \(S\) overlap matrices.
        Bottom row: low-energy moir\'e band error on twisted bilayer test structures, averaged over \(\theta = 21.8^\circ \sim 3.89^\circ\) (\(8 \times 2\) commensurate structures due to stacking symmetry).
    }
    \label{fig:mos2_density_and_band_metric_by_svt_regul_all}
\end{figure}

\autoref{fig:graphene_density_and_band_metric_by_regul}, \autoref{fig:hBN_density_and_band_metric_by_regul}, \autoref{fig:tis2_density_and_band_metric_by_svt_regul_all_svt_at_regul}, \autoref{fig:zrs2_density_and_band_metric_by_svt_regul_all_svt_at_regul}, and \autoref{fig:mos2_density_and_band_metric_by_svt_regul_all} show the effects of GPR regularisation \(\eta\) and singular value (SV) threshold \(\delta\) on prediction accuracy for all five materials.
These figures are used in determining the recommended hyperparameters for different descriptors and materials listed in \autoref{tab:salted_hyperparameters_for_density_predictions} and \autoref{tab:salted_hyperparameters_for_band_predictions} for density prediction and band prediction respectively, following the model training and conditioning methods described in Section Methods in the main text.

The model is not very sensitive to the precise value of \(\eta\).
For graphene and hBN, choosing a single value of \(\eta\) as a compromise between the two values optimised for density and band prediction only increases either density or band structure error by less than a factor of two relative to their respective optima (see \autoref{fig:graphene_density_and_band_metric_by_regul} and \autoref{fig:hBN_density_and_band_metric_by_regul}).
For TMDCs, both metrics follow the same trend and their optima lie within a small-\(\eta\) plateau where performance is insensitive to moderate deviations (see \autoref{fig:tis2_density_and_band_metric_by_svt_regul_all_svt_at_regul}, \autoref{fig:zrs2_density_and_band_metric_by_svt_regul_all_svt_at_regul}, and \autoref{fig:mos2_density_and_band_metric_by_svt_regul_all}).
\autoref{tab:energy_robustness} assesses the robustness of the total energy error on the moir\'e test sets when \(\eta\) is optimised for band structure accuracy versus electronic density accuracy. Errors remain stable between the two \(\eta\) choices for each descriptor and material, confirming that the model is not sensitive to \(\eta\) within this range. Large errors for \tids and \mods with SOAP reflect descriptor failure, also observed in band structure prediction.

\begin{table}[H]
    \centering
    \caption{
        Recommended hyperparameters for moir\'e band structure prediction for different descriptors and materials following the SALTED framework guidelines.
        The notation (f) indicates that the descriptor failed to produce reasonable band structure predictions for that material, with errors comparable to the material's band gap.
        Suggested \(\eta\) values when applying SALTED to new 2D materials for band structure prediction: LOVV \(10^{-9}\), LODE \(10^{-7}\), SOAP \(10^{-4}\), for cases without low-rank approximation; LOVV \(10^{-12}\), LODE \(10^{-10}\), SOAP \(10^{-8}\), for cases with low-rank approximation.
    }
    \label{tab:salted_hyperparameters_for_band_predictions}
    \begin{tabular}{lccccc}
    \hline
    \multirow{2}{*}{Material} & \multirow{2}{*}{SV Threshold \(\delta\)} & \multicolumn{3}{c}{Regularisation Parameter \(\eta\)} \\
    \cline{3-5}
     & & LOVV & LODE & SOAP \\
    \hline
    Graphene & \(0\) & \(10^{-8}\) & \(10^{-7}\) & \(10^{-3}\) \\
    hBN & \(0\) & \(10^{-9}\) & \(10^{-8}\) & \(10^{-6}\) \\
    \tids & \(10^{-6}\) & \(10^{-14}\) & \(10^{-11}\) & \(10^{-7}\)(f) \\
    \zrds & \(10^{-2}\) & \(10^{-8}\) & \(10^{-8}\) & \(10^{-8}\) \\
    \mods & \(10^{-2}\) & \(10^{-12}\) & \(10^{-10}\) & \(10^{-8}\)(f) \\
    \hline
    \end{tabular}
\end{table}

\begin{table}[H]
    \centering
    \caption{
        Recommended hyperparameters for density prediction for different descriptors and materials following the SALTED framework guidelines.
        Suggested \(\eta\) values when applying SALTED to new 2D materials for electronic density prediction: LOVV \(10^{-11}\), LODE \(10^{-9}\), SOAP \(10^{-7}\).
    }
    \label{tab:salted_hyperparameters_for_density_predictions}
    \begin{tabular}{lccccc}
    \hline
    \multirow{2}{*}{Material} & \multirow{2}{*}{SV Threshold \(\delta\)} & \multicolumn{3}{c}{Regularisation Parameter \(\eta\)} \\
    \cline{3-5}
     & & LOVV & LODE & SOAP \\
    \hline
    Graphene & \(0\) & \(10^{-12}\) & \(10^{-11}\) & \(10^{-8}\) \\
    hBN & \(0\) & \(10^{-11}\) & \(10^{-9}\) & \(10^{-6}\) \\
    \tids & \(10^{-6}\) & \(10^{-12}\) & \(10^{-9}\) & \(10^{-7}\) \\
    \zrds & \(10^{-2}\) & \(10^{-12}\) & \(10^{-8}\) & \(10^{-6}\) \\
    \mods & \(10^{-2}\) & \(10^{-10}\) & \(10^{-8}\) & \(10^{-8}\) \\
    \hline
    \end{tabular}
\end{table}

\begin{table}[H]
    \centering
    \caption{
        \textbf{Robustness of total energy predictions against property-specific regularisation.}
        The total energy MAE per atom for the moir\'e test sets, for models where GPR regularisation hyperparameter \(\eta\) was optimised either for band structure accuracy or electronic density accuracy (parameters are listed in \autoref{tab:salted_hyperparameters_for_band_predictions} and \autoref{tab:salted_hyperparameters_for_density_predictions}).
        The total energy error remains stable between density- and band-targeted \(\eta\) choices for each descriptor and material combination, confirming that the model is not so sensitive to the value of \(\eta\) in this range.
        The large errors for \tids and \mods with the SOAP descriptor reflect descriptor failure for these systems, also observed in band structure prediction.
    }
    \label{tab:energy_robustness}
    \begin{tabular}{cc cc}
        \toprule
        & & \multicolumn{2}{c}{\textbf{Total Energy MAE per Atom} (meV/atom)} \\
        \cmidrule(lr){3-4}
        \textbf{Material} & \textbf{Descriptor} & Optimal ``Bands $\eta$"  & Optimal ``Density $\eta$"  \\
        \midrule
        \multirow{3}{*}{Graphene} 
        & LOVV & 0.08 & 0.09 \\
        & LODE & 0.06 & 0.11 \\
        & SOAP & 0.27 & 0.03 \\
        \midrule
        \multirow{3}{*}{hBN} 
        & LOVV & 0.00 & 0.00 \\
        & LODE & 0.05 & 0.11 \\
        & SOAP & 0.09 & 0.09 \\
        \midrule
        \multirow{3}{*}{TiS\textsubscript{2}} 
        & LOVV & 0.01 & 0.03 \\
        & LODE & 2.07 & 1.89 \\
        & SOAP & 29.48 & 29.48 \\
        \midrule
        \multirow{3}{*}{ZrS\textsubscript{2}} 
        & LOVV & 4.26 & 5.45 \\
        & LODE & 3.38 & 3.38 \\
        & SOAP & 5.57 & 3.44 \\
        \midrule
        \multirow{3}{*}{MoS\textsubscript{2}} 
        & LOVV & 5.02 & 4.86 \\
        & LODE & 3.86 & 2.70 \\
        & SOAP & 119.20 & 119.20 \\
        \bottomrule
    \end{tabular}
\end{table}

\subsection{Learning Curves\label{app:learning_curves}}

For each material, the displaced bilayer training dataset is split into training and validation sets containing 400 and 112 structures respectively (for dataset details, see \autoref{app:dataset_details}).
Learning curves are constructed by training models on increasingly large subsets of the training set while keeping the validation set fixed.
Hyperparameters follow \autoref{tab:salted_hyperparameters_for_density_predictions}, optimised for density prediction accuracy.
Most materials and descriptors converge at approximately 200 training structures.
The LOVV learning curves for graphene and hBN show larger fluctuations, likely due to statistical variation arising from the single random seed (42, the SALTED default) used for the train-validation split, which could result in the training dataset being more diverse than the validation set.
Averaging over multiple random seeds would yield smoother convergence behaviour.

\begin{figure}[H]
    \centering
    \includegraphics[width=0.75\twocolwidth]{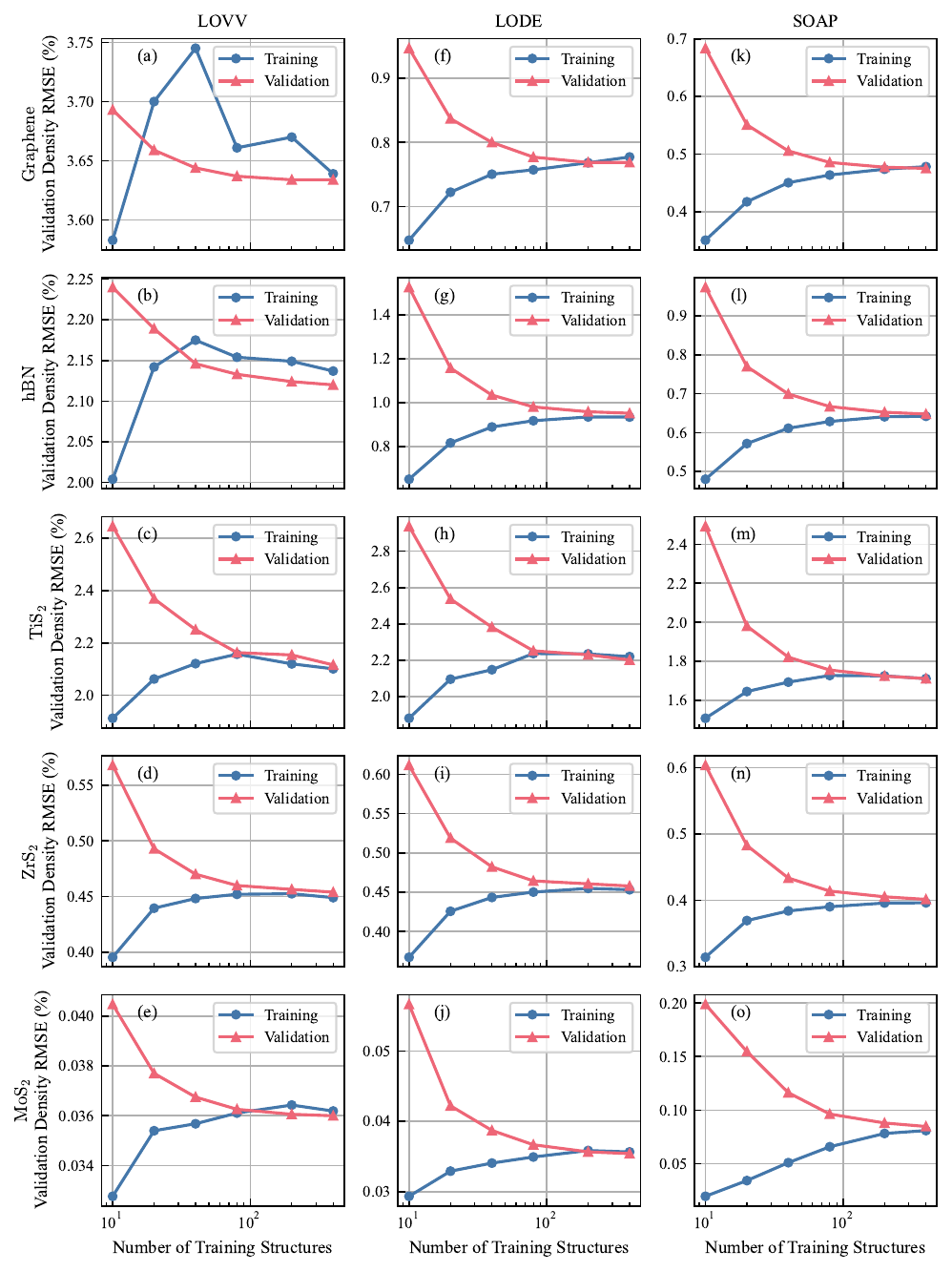}
    \caption{
        \textbf{SALTED's learning curve for different descriptors and different materials training dataset.}
    }
    \label{fig:learning_curve_all_materials}
\end{figure}

\subsection{Accuracy and Generalisation of Electron Density Error}

\begin{figure}[H]
    \centering
    \includegraphics[width=\onecolwidth]{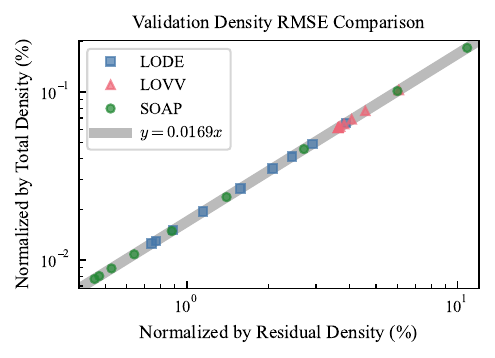}
    \caption{
        \textbf{SALTED validation density error normalised by residual v.s. total density.}
        The example shown uses graphene from this work (the left subfigure in \autoref{fig:graphene_density_and_band_metric_by_regul}).
        The relationship shows good linearity, with a slope (ratio between the two normalisation schemes) of approximately 60.
        For TMDCs, the ratio is expected to be even larger, because the core density is larger for heavy elements, leading to a relatively smaller residual density compared to the total density.
    }
    \label{fig:valid_density_ratio}
\end{figure}

In this work, the electron density error is normalised by the residual density, defined as the total density minus spherical component per atom (computed as the average of all l=0 channels across all structures for each element).
Compared to renormalising by total density, the residual density is much smaller in magnitude, since the inactive core electron density dominates the total density.
As a result, normalisation by residual density makes the metric extremely sensitive to deviations from the targeting density.
\autoref{fig:valid_density_ratio} shows the ratio between the density error normalised by the total density and the residual density.

Since the density metric is normalised by residual density as defined in Section IV.D, we list the root-mean-square residual density \(\rho^\text{res}\) for each dataset in \autoref{tab:residual_density_per_material}, defined as:
\begin{equation}
\rho^\text{res} = \sqrt{ \frac{1}{N} \sum_k \qty( \mathbf{c}_k^\mathrm{DF} - \mathbf{c}_k^{\mathrm{sph}} )^\top \mathbf{S}_{k} \qty( \mathbf{c}_k^\mathrm{DF} - \mathbf{c}_k^{\mathrm{sph}} ) }.
\label{eq:residual_density}
\end{equation}
\mods has \(\rho^\text{res} = 72.2\), nearly two orders of magnitude larger than all other materials.
This large normalisation denominator suppresses the normalised density error, explaining the small validation density errors of \mods reported in the main text.

\begin{table}[H]
    \centering
    \caption{
        Root-mean-square residual density \(\rho^\text{res}\) for each material's training dataset as in Equation~\ref{eq:residual_density}.
    }
    \label{tab:residual_density_per_material}
    \begin{tabular}{cc}
        \toprule
        Material & \(\rho^\text{res}\) \\
        \midrule
        Graphene & 0.57 \\
        hBN      & 0.46 \\
        \tids    & 1.42 \\
        \zrds    & 3.06 \\
        \mods    & 72.2 \\
        \bottomrule
    \end{tabular}
\end{table}

\begin{figure}[H]
    \centering
    \includegraphics[width=0.75\onecolwidth]{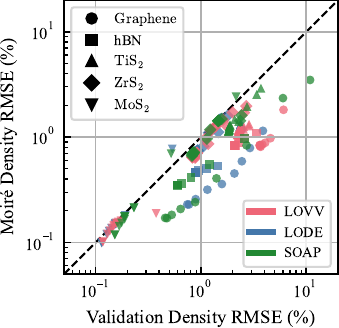}
    \caption{
        \textbf{Comparing model predicted density RMSE for validation dataset displaced structures and test dataset twisted bilayers} (TBs).
        The moir\'e density RMSE are evaluated against DFT-converged density fitting coefficients.
        Due to machine memory limit, we computed 6 twist angles (\(21.786^\circ \sim 5.086^\circ\)) for graphene (6 TBs in total) and hBN (12 TBs in total), 2 twist angles (\(21.787^\circ, 13.174^\circ\)) for \tids (4 TBs in total) and one twist angle (\(21.787^\circ\)) for \zrds and \mods (2 TBs in total for each).
        In all cases, the twisted bilayers' density RMSE is smaller or comparable to displaced bilayers'.
    }
    \label{fig:density_rmse_valid_vs_moire}
\end{figure}

\autoref{fig:density_rmse_valid_vs_moire} compares validation (displaced bilayers) and test (twisted bilayers) density RMSE for all the models tested in \autoref{app:full_experiment_results}.
In all cases, the test density RMSE for moir\'e structures is smaller than or comparable to the displaced bilayer validation error, confirming that the model generalises stably from displaced bilayers to twisted bilayers. 

\subsection{Comparison with Hamiltonian-based Machine Learning Methods\label{app:compare_deeph}}

Table~\ref{tab:deeph_comparison} compares moir\'e band accuracy between SALTED and DeepH~\cite{tangDeepEquivariantNeural2024} for TBG and \mods at larger twist angles, where identical \(k\)-paths for DeepH and DFT in the original publication enable direct MAE computation.
We note that this DeepH model is fitted to hybrid functional, but we are only comparing the errors within each methodology. This particular DeepH model employs an extended two-cutoff architecture to improve long-range accuracy beyond standard DeepH model, yet still performs worse than SALTED.
We further compared a TBG structure at \(\theta \approx 2.0^\circ\) in \autoref{app:tbg_band_relaxation} and \autoref{fig:2507ppr_TBG_deeph_vs_dft_bands}, where DeepH shows a low-energy band error of 1.34~meV and SALTED shows a low-energy band error of 1.50~meV.

\begin{table}[H]
    \centering
    \caption{
        \textbf{Comparison of moir\'e band MAE between DeepH-hybrid and SALTED.}
        The mean absolute error (MAE) is calculated for the low-energy moir\'e bands across different twist angles for twisted bilayer graphene (TBG) and \mods.
        DeepH data corresponds to Figure~3(b,c) and Figure~4(c,e) from Tang et al.~\cite{tangDeepEquivariantNeural2024}.
    }
    \label{tab:deeph_comparison}
    \begin{tabular}{ccccc}
        \toprule
        & \multicolumn{2}{c}{\textbf{Moir\'e Band MAE} (meV)} & \multicolumn{2}{c}{\textbf{Band Gap Error} (meV)} \\
        \cmidrule(lr){2-3} \cmidrule(lr){4-5}
        \textbf{Structure} & DeepH & SALTED & DeepH & SALTED \\
        \midrule
        TBG, $\theta=21.79^\circ$ & 5.71 & 0.70 & --- & --- \\
        TBG, $\theta=13.17^\circ$ & 4.07 & 0.59 & --- & --- \\
        \midrule
        \mods, $\theta=21.79^\circ$ & 20.41 & 2.91 & 35.25 & 2.85 \\
        \mods, $\theta=13.17^\circ$ & 16.77 & 2.34 & 33.89 & 3.89 \\
        \bottomrule
    \end{tabular}
\end{table}

\section{Application Details\label{app:applications_details}}

\subsection{Band Width Extrapolation in hBN and TMDCs\label{app:band_width_extrapolation}}

Twisted bilayers with parallel symmetry are predicted in this section.
These structures consist of parallel monolayers with twist angles smaller than \(30^\circ\).
The supercells are constructed from primitive cells in \autoref{app:setup_summary} using commensurate twist configurations with \(r=1\) and varying \(m\) from 1 to large integers described in \autoref{eq:twist_angle_m_r}.
For band structure predictions with DFT references (see \autoref{app:dataset_details}), a dense \(\mathrm{K-\Gamma-M-K}\) k-path (see \autoref{app:dft_details}) is used, consistent with full DFT calculation settings to ensure comparison on the same footing.
For the rest structures that have smaller twist angles, a coarse four-k-point \(\mathrm{\Gamma-\text{mid}(\Gamma K)-K-M}\) path (where \(\text{mid}(\mathrm{\Gamma K})\) denotes the midpoint between \(\mathrm{\Gamma}\) and \(\mathrm{K}\)) is used to sample all the high symmetry points and estimate the band width.
Given that the low-energy band is narrow for small twist angle structures, this k-path is sufficient to capture the band width and band gaps.
All SALTED parameters are listed in \autoref{tab:salted_hyperparameters_for_band_predictions}.

\begin{figure}[H]
    \centering
    \includegraphics[width=\twocolwidth]{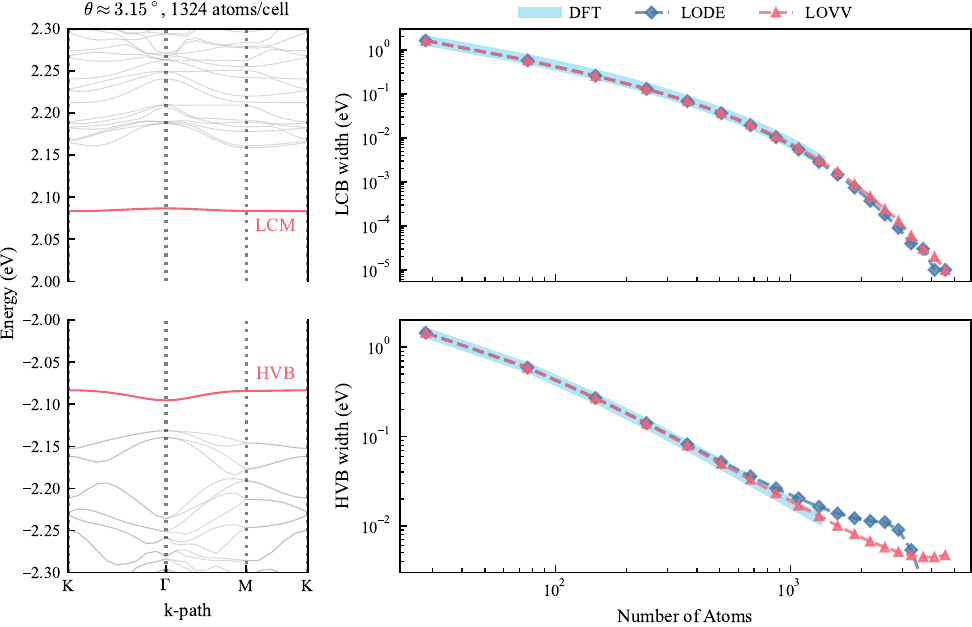}
    \caption{
        \textbf{TB-hBN flat band extrapolation by twist angle.}
        Left: TB-hBN (\(\theta \approx 3.15^\circ\), 1324 atoms/cell) band structure, with two flat bands at the VBM and CBM.
        Top right and bottom right: CBM and VBM flat band widths as a function of twist angle.
        SOAP and LODE show large deviation starting from 1000 atoms, and LOVV provides the best match to full SCF results and good extrapolation to smaller twist angles.
    }
    \label{fig:TBhBN_flat_band}
\end{figure}

\begin{figure}[H]
    \centering
    \includegraphics[width=\twocolwidth]{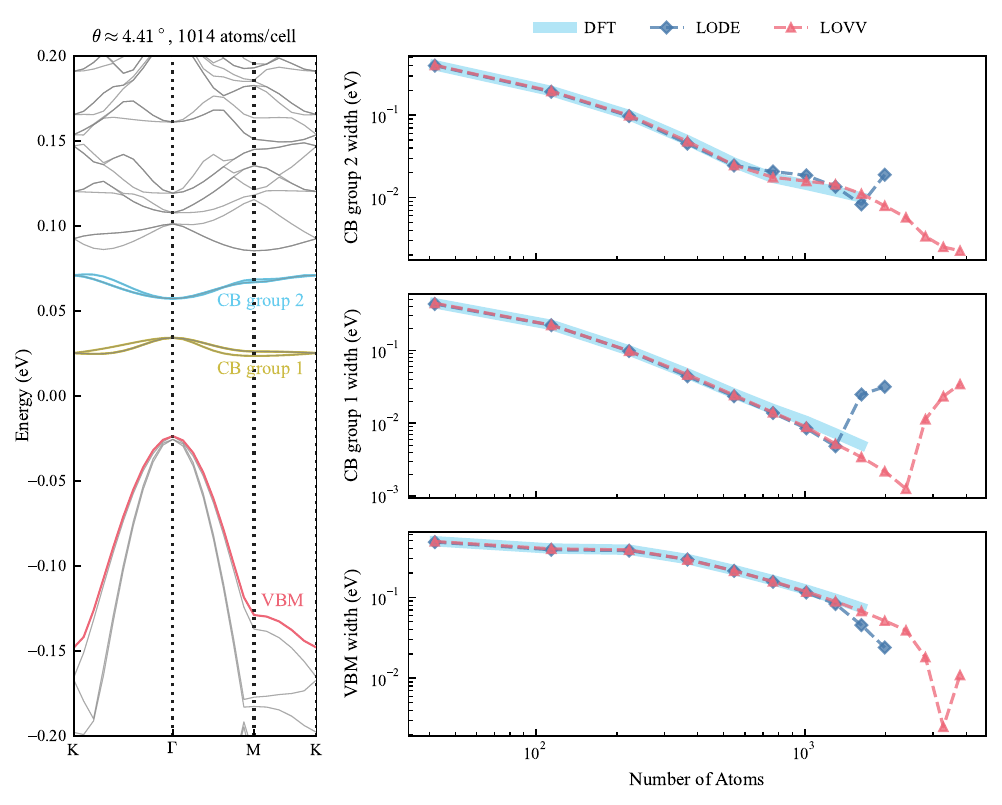}
    \caption{
        \textbf{TB-TiS2 flat band extrapolation by twist angle.}
        Left: TB-\tids (\(\theta \approx 4.41^\circ\), 1014 atoms/cell) band structure showing two flat band groups at the conduction minimum, each consisting of 3 bands; the VBM is not flat.
        Right: Band width predictions of conduction band groups and VBM as a function of twist angle.
        Both LODE and LOVV show increases after 1000 atoms, corresponding to the gap closure discussed in the main text.
    }
    \label{fig:TBTiS2_flat_band}
\end{figure}

\autoref{fig:TBhBN_flat_band} shows the band width extrapolation of the conduction band minimum (CBM) and valence band maximum (VBM) flat bands of TB-hBN as a function of number of atoms in the superlattice.
For 1T-\tids and 1T-\zrds (\autoref{fig:TBTiS2_flat_band} and \autoref{fig:TBZrS2_flat_band} respectively), the band width of the first and second group of lowest conduction bands (each containing 3 bands \cite{tanwarFirstprinciplesStudyStructural2023}) and the VBM are calculated.
For 1H-\mods results in \autoref{fig:TBMoS2_flat_band}, the band width of the lowest conduction bands (containing 6 bands) and the VBM are calculated.
The LOVV descriptor consistently performs the best, accurately capturing both the decrease in band width with increasing number of atoms (equivalent to decreasing twist angles).

\begin{figure}[H]
    \centering
    \includegraphics[width=0.5\linewidth]{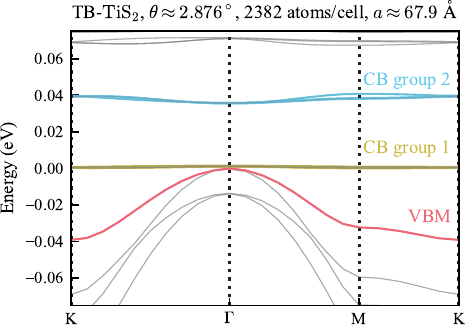}
    \caption{
        \textbf{\tids band structure from LOVV prediction.}
        Low-energy band dispersion remains physically reasonable despite predicted gap closure.
    }
    \label{fig:TiS2_geom_idx_11_band_gap_close}
\end{figure}

\begin{figure}[H]
    \centering
    \includegraphics[width=\twocolwidth]{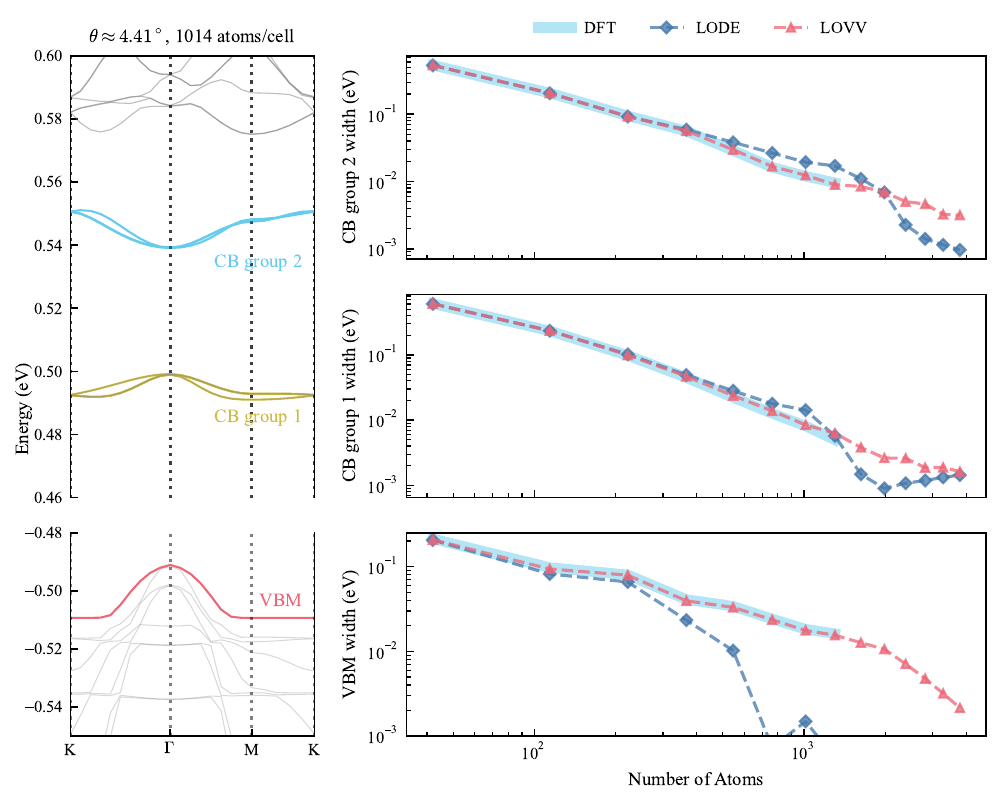}
    \caption{
        \textbf{TB-ZrS2 flat band extrapolation by twist angle.}
        Left: TB-\zrds (\(\theta \approx 4.41^\circ\), 1014 atoms / cell) band structure showing two flat band groups at the conduction minimum, each consisting 3 bands; the VBM is not flat.
        Right: Band width predictions for the conduction band groups and VBM as a function of twist angle.
        LOVV correctly extrapolates the decreasing trend in band width with decreasing twist angle.
    }
    \label{fig:TBZrS2_flat_band}
\end{figure}

\begin{figure}[H]
    \centering
    \includegraphics[width=\twocolwidth]{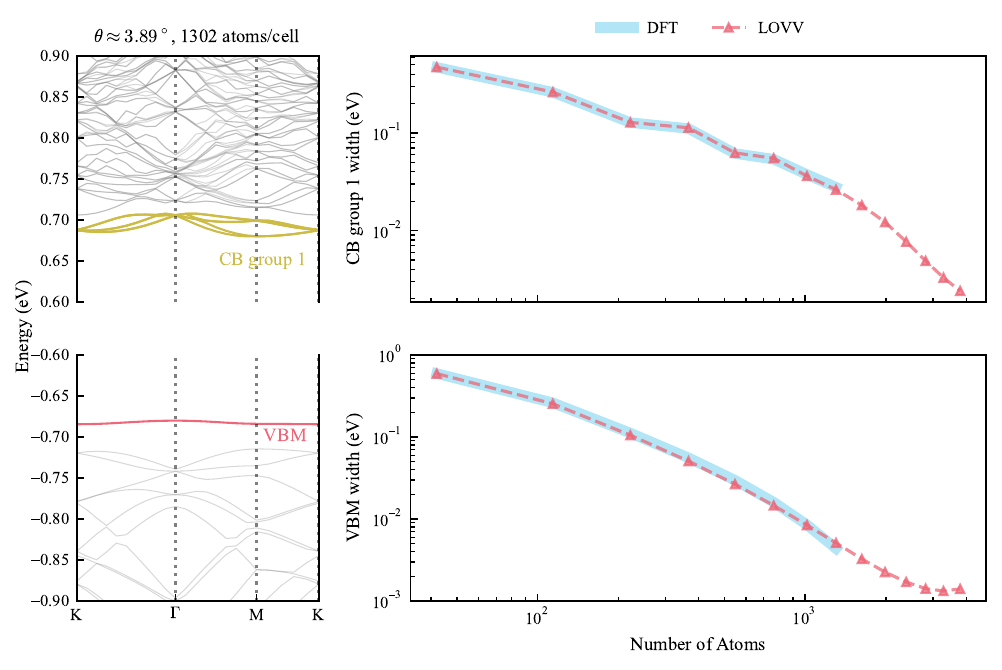}
    \caption{
        \textbf{TB-MoS2 flat band extrapolation by twist angle.}
        Left: TB-\mods (\(\theta \approx 3.89^\circ\), 1302 atoms/cell) band structure showing one flat band group at the conduction minimum consisting of 6 bands, and a flat VBM.
        Right: Band width predictions of the conduction band group and VBM as a function of twist angle.
        LOVV correctly extrapolates the decreasing trend in band width with decreasing twist angle.
    }
    \label{fig:TBMoS2_flat_band}
\end{figure}

As discussed in the main text, the band gap of TB-\tids gets smaller as the twist angle reduces, even below \(50~\mathrm{meV}\) at \(\theta=3.481^\circ\).
\autoref{fig:TiS2_geom_idx_11_band_gap_close} shows the LOVV-predicted band structure at \(\theta = 2.876^\circ\) (2382 atoms/cell), where we predict gap closure.
Despite this, the low-energy band dispersion and band widths (top and middle panels of \autoref{fig:TBTiS2_flat_band}) remain on the correct extrapolation trend, confirming that LOVV describes the moir\'e electronic structure correctly.
The gap closure is a near-degeneracy artefact: near-metallic levels are sensitive to small residual density errors, in the same way that near-metallic DFT calculations are slow to converge. It does not signal a breakdown of the descriptor or model.

\subsection{Spin-Orbit Coupling in TMDCs\label{app:soc_in_tmdcs}}

\begin{figure}[H]
    \centering
    \includegraphics[width=\onecolwidth]{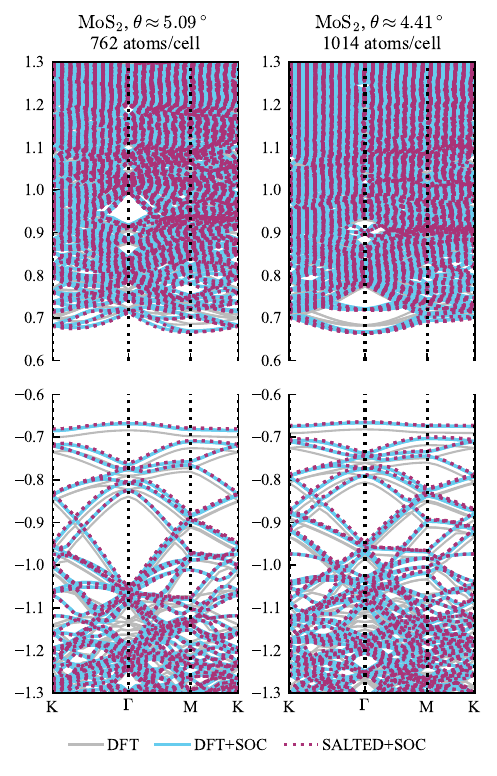}
    \caption{
        \textbf{TB-MoS2 with SOC predicted by SALTED.}
        SOC effects in TB-\mods with AA-series stacking, where two monolayers are almost identical except for the small twist angle.
        SOC reduces the band gap and modifies the deep valence bands (around \(-1.14~\mathrm{eV}\)).
        These changes are captured by SALTED using LOVV descriptor.
    }
    \label{fig:TBMoS2_soc_band_comparison}
\end{figure}

In FHI-aims, spin-orbit coupling (SOC) is included via a non-self-consistent perturbation applied after the scalar-relativistic DFT calculation has converged \cite{huhnOnehundredthreeCompoundBandstructure2017}.
The corresponding keyword in FHI-aims is \verb|include_spin_orbit|.
In our work, SOC band structure prediction for \tids and \zrds are presented in the main text due to their significant low-energy band-splitting effect.
In \autoref{fig:TBMoS2_soc_band_comparison}, analogous results for \mods are presented, where SOC effect modifies deep valence bands.

\subsection{TBG Band Structure with Structural Relaxation\label{app:tbg_band_relaxation}}

A MACE model \cite{batatiaMACEHigherOrder2022} is trained on a TBG dataset consisting of 900 structures with twist angles ranging from \(21.79^\circ\) to \(3.48^\circ\) (train-valid-test split ratio 8:1:1).
These structures are sampled from MLIP-accelerated molecular dynamics (MD) simulations at temperatures from \(10 \mathrm{K}\) to \(300 \mathrm{K}\), where the MLIP is trained on another bilayer graphene dataset.
The TBG dataset is constructed using FHI-aims with intermediate species default, the PBE functional, and MBD@rsSCS for vdW correction~\cite{tkatchenkoAccurateEfficientMethod2012}.
The MACE model employs default settings except for hidden irreps \verb|32x0e+32x1o| and the number of interactions set to 2.
It achieves force MAE \(2.6~\mathrm{meV/\angs}\) and relative force MAE \(0.5 \%\) in the test set.
The model is then used to relax TBG structures (constructed from the primitive cells described in \autoref{tab:material_parameters}) to a maximum force threshold of \(10^{-1}~\mathrm{meV}/\nsangs\) using FIRE algorithm implemented in ASE~\cite{bitzekStructuralRelaxationMade2006,hjorthlarsenAtomicSimulationEnvironment2017}.

Relaxed structures are predicted by the SALTED model using SOAP and \(\eta=10^{-3}\), optimised for graphene's band structure accuracy (parameters listed in \autoref{tab:salted_hyperparameters_for_band_predictions}).
Band structures are obtained through a half SCF cycle in FHI-aims using the SALTED-predicted density as initialisation.

\begin{figure}[htbp]
    \centering
    \includegraphics[width=\onecolwidth]{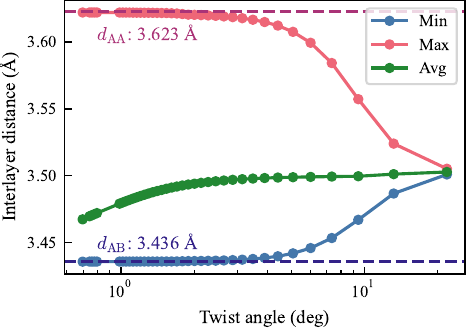}
    \caption{
        \textbf{Interlayer distance variation in relaxed TBG structures.}
        Local interlayer separation as a function of twist angle, showing minimum (blue), maximum (red), and average (green) values across the moir\'e supercell.
        Horizontal dashed lines indicate the equilibrium distances for AA stacking (\(d_\text{AA} = 3.623~\angs\)) and AB stacking (\(d_\text{AB} = 3.436~\angs\)) obtained by relaxing AA and AB bilayer primitive cells using the MACE model.
        As twist angle decreases, structural relaxation induces pronounced z-direction buckling: the maximum separation saturates near \(d_\text{AA}\) in AA-stacked regions while the minimum approaches \(d_\text{AB}\) in AB-stacked regions.
        Due to the symmetric z-direction buckling in relaxed TBG structures, interlayer distances are extracted as follows:
        The structure is first centred such that the bilayer centre of mass lies at \(z=0\). Due to mirror symmetry against \(z=0\) xy plane, where each upper-layer carbon atom at position \(z_i\) has a corresponding close-by lower-layer atom at \(-z_i\), the local interlayer distance is \(2 z_i\).
        The minimum, maximum, and average separations are obtained by taking the minimum, maximum, and mean values of \(2z_i\) over all upper-layer carbon atoms, respectively.
    }
    \label{fig:TBG_relaxed_v2_separation}
\end{figure}

Previous studies have studied how buckling affects TBG band structure, especially the flat band width and shape~\cite{uchidaAtomicCorrugationElectron2014,carrExactContinuumModel2019,canteleStructuralRelaxationLowenergy2020}, showing that z-direction buckling plays an important role.
\autoref{fig:TBG_relaxed_v2_separation} shows relaxation-introduced deregistration in the z-direction for different twist angles.
Starting from the largest twist angle, where the interlayer distance is almost uniform, the maximum and the minimum of interlayer distances changes as twist angle decreases and gradually saturate to those of AA- and AB-stacking bilayer primitive cells, showing significant z-direction buckling in large TBGs.
Also, the average interlayer distance reduces, since the AA-stacking area shrinks.
Although such shrinking is found not to be important for electronic properties in TBG, it might be pronounced in other structures, like in TB-hBN with ferroelectricity characters, and continuum models that assume in-plane rigidity cannot manage such geometric features.

\begin{figure}[H]
    \centering
    \includegraphics[width=\twocolwidth]{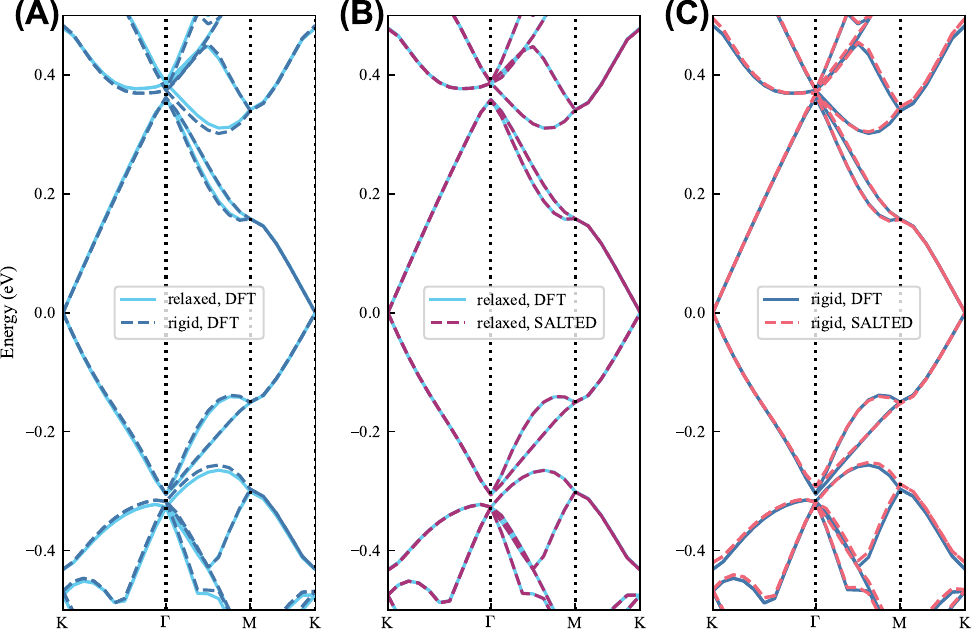}
    \caption{
        \textbf{Relaxation affects TBG band structure, results from converged DFT and SALTED predictions.}
        Band structures for a TBG structure with \(\theta=3.15^\circ\) and 1324 atoms, either calculated by converged DFT or SALTED prediction.
        (A) Band structure before and after relaxation, calculated by converged DFT. Relaxation introduces larger band gaps around the low-energy bands at \(\Gamma\) around \(0.35~\mathrm{eV}\) and \(-0.30~\mathrm{eV}\) in the TBG band structure.
        (B) Comparing SALTED prediction and converged DFT for relaxed TBG.
        (C) Comparing SALTED prediction and converged DFT for rigid TBG.
    }
    \label{fig:TBG_compare_geom_idx_10_relaxed_rigid_dft_salted}
\end{figure}

\autoref{fig:TBG_compare_geom_idx_10_relaxed_rigid_dft_salted} demonstrates the fact that, for band structure prediction, SALTED predicts relaxed structures better than rigid structures.
For a representative structure at \(\theta = 3.15^\circ\) (1324 atoms), panel (A) compares the band structure between relaxed and rigid TBG using converged DFT results.
Panels (B) and (C) compare SALTED predictions against full DFT for both relaxed and rigid structures, and SALTED predicts band structure more accurately for the relaxed structure.

The accuracy difference can be explained via the geometry space feature manifold of the training dataset.
The random sampling strategy discussed in \autoref{app:materials_geometries} spans a high-dimensional manifold where per-atom coordinates vary independently, with interlayer distance following a normal distribution covering the \(\sim 0.2~\angs\) range shown in \autoref{fig:TBG_relaxed_v2_separation}.
Relaxed structures, with their spatially varying interlayer distances and local atomic displacements, naturally locate within this high-dimensional manifold of the training dataset.
In contrast, rigid twisted bilayers that have uniform interlayer distances form a low-dimensional submanifold that is under-represented in the training set.
Since GPR prediction accuracy depends on descriptor overlap between test and training configurations, relaxed structures benefit from greater similarity to the training manifold, while rigid structures suffer from limited interpolation.
To improve rigid structure predictions, the training dataset should be augmented with structures sampled specifically along the low-dimensional rigid manifold, i.e. only applying interlayer distance variation and in-plane sliding and without per-atom coordinate random perturbation.

\begin{figure}[H]
    \centering
    \includegraphics[width=\twocolwidth]{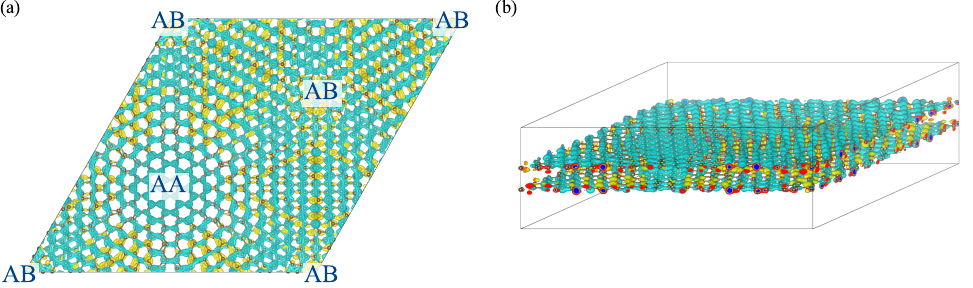}
    \caption{
        \textbf{Spatial distribution of density prediction errors in relaxed TBG.}
        The relaxed TBG structure has a twist angle \(\theta=3.15^\circ\) and contains 1324 atoms per superlattice.
        (a) Top view and (b) side view show the total density difference between SALTED prediction (after half an SCF cycle) and fully converged DFT, with isosurface value \(2.5 \times 10^{-4}~\mathrm{e} \cdot \nsangs^{-3}\).
        Yellow (positive) and blue (negative) isosurfaces represent regions where SALTED over- and under-predicts the density, respectively.
        Negative errors (density depletion) concentrate around AA-stacking centres in both layers, while positive errors (density excess) appear predominantly in AB-stacking regions.
        Visualisation generated by VESTA.
    }
    \label{fig:graphene_TBG_cube_visualization_concatenated}
\end{figure}

\begin{figure}[H]
    \centering
    \includegraphics[width=\twocolwidth]{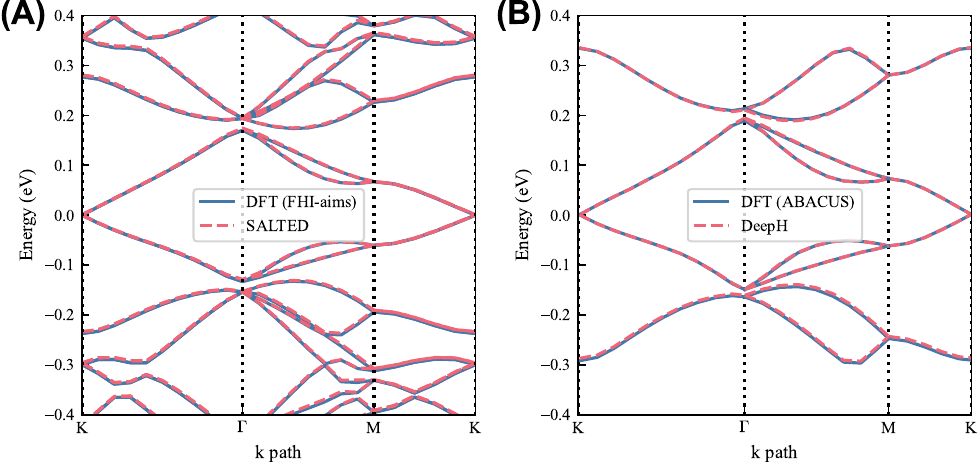}
    \caption{
        \textbf{Comparison of SALTED and DeepH predictions for TBG band structure.}
        The TBG structure has twist angle \(\theta=2.00^\circ\) with 3268 atoms per superlattice, providing a direct benchmark between SALTED and DeepH~\cite{tangDeepEquivariantNeural2024}.
        (A) SALTED prediction compared to converged DFT from FHI-aims, with low-energy band MAE 1.50 meV.
        (B) DeepH prediction compared to converged DFT from ABACUS, extracted from Tang et al.~\cite{tangDeepEquivariantNeural2024}, with low-energy band MAE 1.34 meV.
        Both structures are relaxed respectively.
        Low-energy band MAE is calculated over 4 highest valence bands and 4 lowest conduction bands.
        Both methods achieve excellent agreement with DFT over the low-energy band structure spanning around 600~meV, with errors \(< 1\%\) of the total band energy range.
    }
    \label{fig:2507ppr_TBG_deeph_vs_dft_bands}
\end{figure}

To enable quantitative comparison with the neural network-based Hamiltonian prediction method of Tang et al.~\cite{tangDeepEquivariantNeural2024}, we extracted band structure data from their Figure~3(d) for a relaxed TBG at \(\theta \approx 2.00^\circ\) using WebPlotDigitizer since \(k\)-paths for DFT and DeepH don't match in published band structure data.
We extracted both the DFT reference bands and their DeepH-predicted bands, focusing on the 4 highest valence bands and 4 lowest conduction bands to match our low-energy band error metric defined in our paper, shown in \autoref{fig:2507ppr_TBG_deeph_vs_dft_bands}(B).
The band structure alignment followed that in the figure.
Calculating the low-energy band error, we obtained 1.34~meV for DeepH.
For direct comparison, our SALTED method achieves an error of 1.50~meV, demonstrating comparable performance to the neural network approach for this structure.

\subsection{TBhBN Electric Field at Domain Boundaries\label{app:tbhbn_efield}}

A MACE model \cite{batatiaMACEHigherOrder2022} is trained on a displaced bilayer hBN dataset consisting of 512 \(5 \times 5 \times 1\) bilayers (train-valid-test split 8:1:1), sampled using the same random sampling procedure as the SALTED training dataset (described in \autoref{app:materials_geometries}).
The dataset is calculated by FHI-aims using intermediate species default, PBE functional, and MBD@rsSCS for vdW correction \cite{tkatchenkoAccurateEfficientMethod2012}.
The model has default MACE settings except hidden irreps \verb|64x0e+64x1o+64x2e|, and it reaches force RMSE \(2.5~\mathrm{meV/\angs}\) and relative force RMSE \(0.13 \%\) in the test set.
The model is used to relax bilayer hBN structures to a maximum atomic force of \(10^{-1}~\mathrm{meV}/\nsangs\) using FIRE algorithm implemented in ASE \cite{bitzekStructuralRelaxationMade2006,hjorthlarsenAtomicSimulationEnvironment2017}.
The SALTED model with SOAP and \(\eta=10^{-6}\) (according to \autoref{tab:salted_hyperparameters_for_density_predictions}, as the Hartree potential calculation is more sensitive to density error) predicts the relaxed structures.
Hartree potential cube files are obtained through a half SCF cycle in FHI-aims using the SALTED-predicted density as initialisation.
The cube file settings ensure a spatial cube grid in-plane spacing of \(0.1\angs\) and z-direction spacing \(1\angs\).
The cube grids form a uniform equilateral triangular lattice in the xy plane.
The Hartree potential along a straight path through AA-AB-BA-AA stacking is extracted and fitted using trigonometric functions (cosine and sine) up to the fourth Fourier harmonics.
The derivative at the AB-BA domain boundary is then calculated analytically from the fitted functions to obtain the electric field.

The in-plane electric field presented in the main text is obtained through the following procedure:
(1) interpolating in-plane Hartree potential with third-order spline with periodic boundary conditions using \verb|scipy.ndimage.map_coordinates|;
(2) calculating the first-order derivative using central difference method with spacing of \(10^{-4}\) (approximately \(10^{-5}\angs\)) on the cube file's local grid coordinates;
and (3) converting the derivative from cube coordinates to Cartesian coordinates using the Jacobian matrix:
\begin{equation}
\begin{bmatrix}
    x \\
    y
\end{bmatrix}
= L
\begin{bmatrix}
    1 & \cos(60^\circ) \\
    0 & \sin(60^\circ)
\end{bmatrix}
\begin{bmatrix}
    a \\
    b
\end{bmatrix}
\quad
\begin{bmatrix}
    \frac{\partial}{\partial x} \\
    \frac{\partial}{\partial y}
\end{bmatrix}
=
\begin{bmatrix}
    \frac{\partial a}{\partial x} & \frac{\partial b}{\partial x} \\
    \frac{\partial a}{\partial y} & \frac{\partial b}{\partial y}
\end{bmatrix}
\begin{bmatrix}
    \frac{\partial}{\partial a} \\
    \frac{\partial}{\partial b}
\end{bmatrix}
\end{equation}
where \((x,y)\) are the Cartesian coordinates, \((a,b)\) are the cube file's local grid coordinates in the in-plane directions, and \(L\) is the cube grid in-plane spacing.
The cube grid has an integer \((a,b)\) coordinate.

\begin{figure}[H]
    \centering
    \includegraphics[width=\twocolwidth]{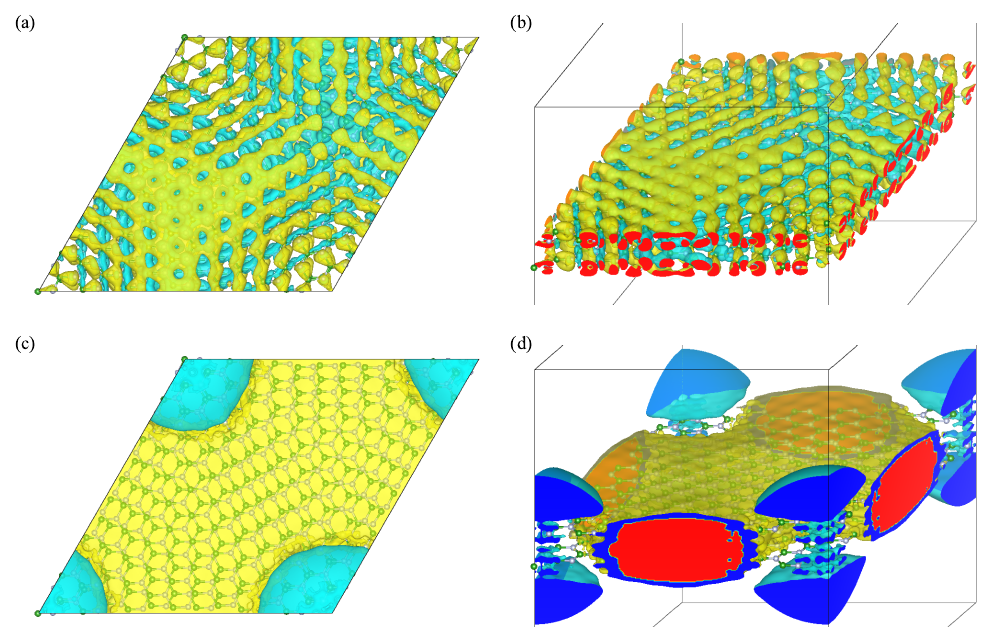}
    \caption{
        \textbf{Density and Hartree potential spatial error distribution.}
        The example TBhBN structure has twist angle \(\theta=5.09^\circ\) and has 508 atoms per superlattice.
        AA stacking is at the four vertices, and BA and AB stacking are at 1/3 and 2/3 along the diagonal line from lower left to upper right, respectively.
        (a) Vertical view and (b) side view of total electron density error between SALTED prediction after half an SCF cycle and converged DFT, with isosurface value \(10^{-4}~e/\nsangs^3\).
        (c) Vertical view and (d) side view of Hartree potential error between SALTED prediction after half an SCF cycle and converged DFT, with isosurface value \(5\times 10^{-4}~\mathrm{Hartree/e} = 13.6~\mathrm{mV}\).
        Visualisation by VESTA, with yellow and blue for positive and negative errors, respectively.
    }
    \label{fig:hbn_spatial_error_distribution}
\end{figure}

\autoref{fig:hbn_spatial_error_distribution} demonstrates the spatial error distributions of density and Hartree potential.
While the density prediction error is not significant, the Hartree potential exhibits an AA-localised deviation.
Despite this pattern, the in-plane electric field at the AB-BA domain wall is not significantly affected due to cancellation between symmetric AA-localised deviation pairs on both sides of the domain wall centre.

To explain the saturation behaviour of the AB-BA domain wall electric field at small twist angles, we obtained the domain wall saturation width both theoretically and from our relaxed structures, and the latter analysis further provides insight into its scaling behaviour against moir\'e superlattice size.

\begin{figure}[H]
    \centering
    \includegraphics[width=\onecolwidth]{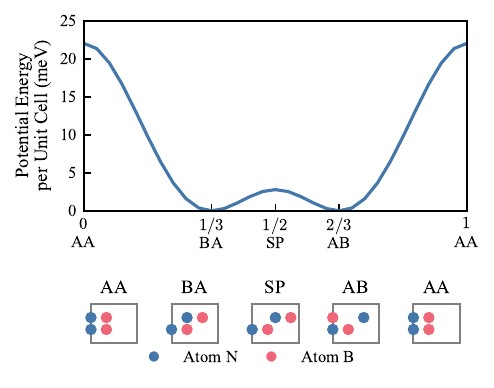}
    \caption{
        \textbf{Potential energy per unit cell for bilayer hBN as a function of in-plane interlayer displacement along path AA--BA--SP--AB--AA.}
        The subfigures below show side views demonstrating the local stacking pattern in these regions.
        The SP region exhibits slightly higher energy than both AB and BA stacking configurations, while AA stacking has the highest energy.
    }
    \label{fig:hbn_registry_energy_vs_displacement}
\end{figure}

Following the continuum elasticity theory \cite{aldenStrainSolitonsTopological2013}, the equilibrium domain wall full width at half maximum (FWHM) is:
\begin{equation}
w_\text{eq} = \frac{a_\text{BN}}{2} \sqrt{\frac{k}{V_\text{sp}}}
\label{eq:domain_wall_width_by_shear_strength_and_saddle_point_energy}
\end{equation}
where \(a_\text{BN} = 1.451~\angs\) is the B-N bond length~\cite{jainCommentaryMaterialsProject2013}, \(k\) is the bilayer shear stiffness, and \(V_\text{sp}\) is the saddle point (SP) energy per unit area, where the two layers are stacked as the relative position in AB-BA domain wall centre.
The shear stiffness \(k = 36.7~\mathrm{N/m}\) has been calculated via controlled in-plane shear deformations to AA and AB stacked bilayers~\cite{zhaoMechanicalVibrationalBehaviors2024}.
The saddle point energy obtained by our MACE model using the bilayer hBN geometry in \autoref{tab:material_parameters} is \(2.79~\mathrm{meV}\) per primitive cell (see \autoref{fig:hbn_registry_energy_vs_displacement}), yielding \(V_\text{sp} = 8.17 \times 10^{-3}\mathrm{N/m}\) where the primitive cell area is \(5.47~\angs^2\).
Substituting these values in \autoref{eq:domain_wall_width_by_shear_strength_and_saddle_point_energy}, we obtain equilibrium domain wall width \(w_\text{eq} = 4.85 \nm\).

\begin{figure}[H]
    \centering
    \includegraphics[width=\twocolwidth]{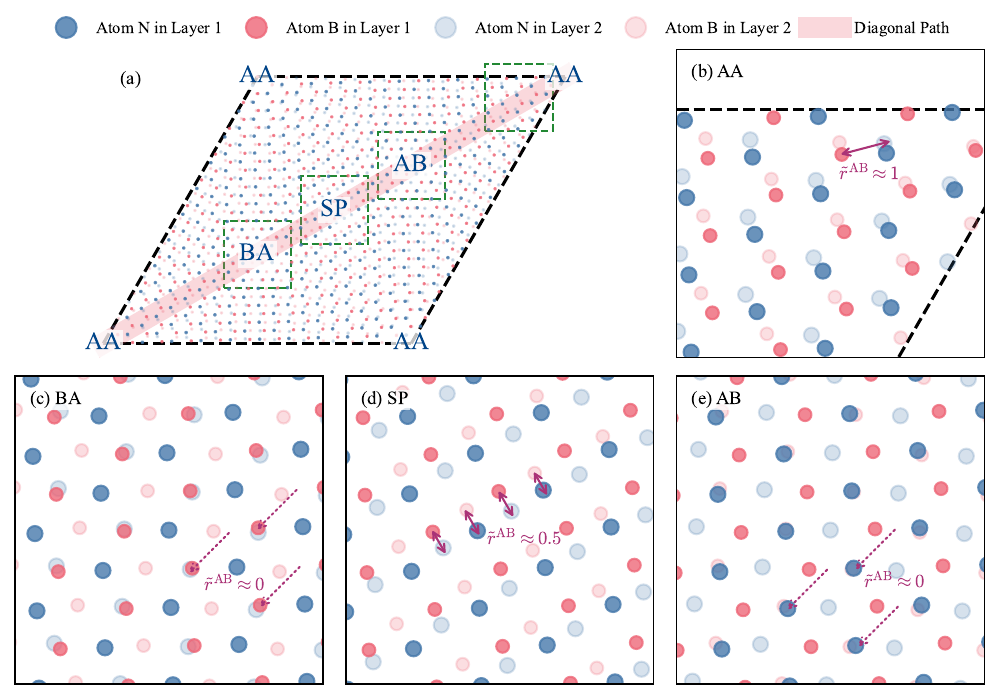}
    \caption{
        \textbf{The normalised interlayer homoatomic in-plane atomic distance / AB-centred normalised registry \(\Tilde{r}^\text{AB}\) in each region in a moir\'e superlattice.}
        (a) Relaxed TBhBN structure with twist angle \(\theta \approx 3.15^\circ\) (1324 atoms, lattice size \(45.70~\angs\)). Stacking centres (AA, AB, BA) and the SP region are labelled.
        (b-e) Region AA, BA, SP, and AB, with arrows indicating \(\Tilde{r}^\text{AB}\). In (b,d) the \(\Tilde{r}^\text{AB}\)s are labelled by double arrows with solid line; in (c,e) the atom-pairs are marked by a single arrow with dashed line, because \(\Tilde{r}^\text{AB}\) are close to 0 around BA and AB stacking regions.
    }
    \label{fig:hbn_local_atomic_distance_illustration}
\end{figure}

\begin{figure}[H]
    \centering
    \includegraphics[width=\twocolwidth]{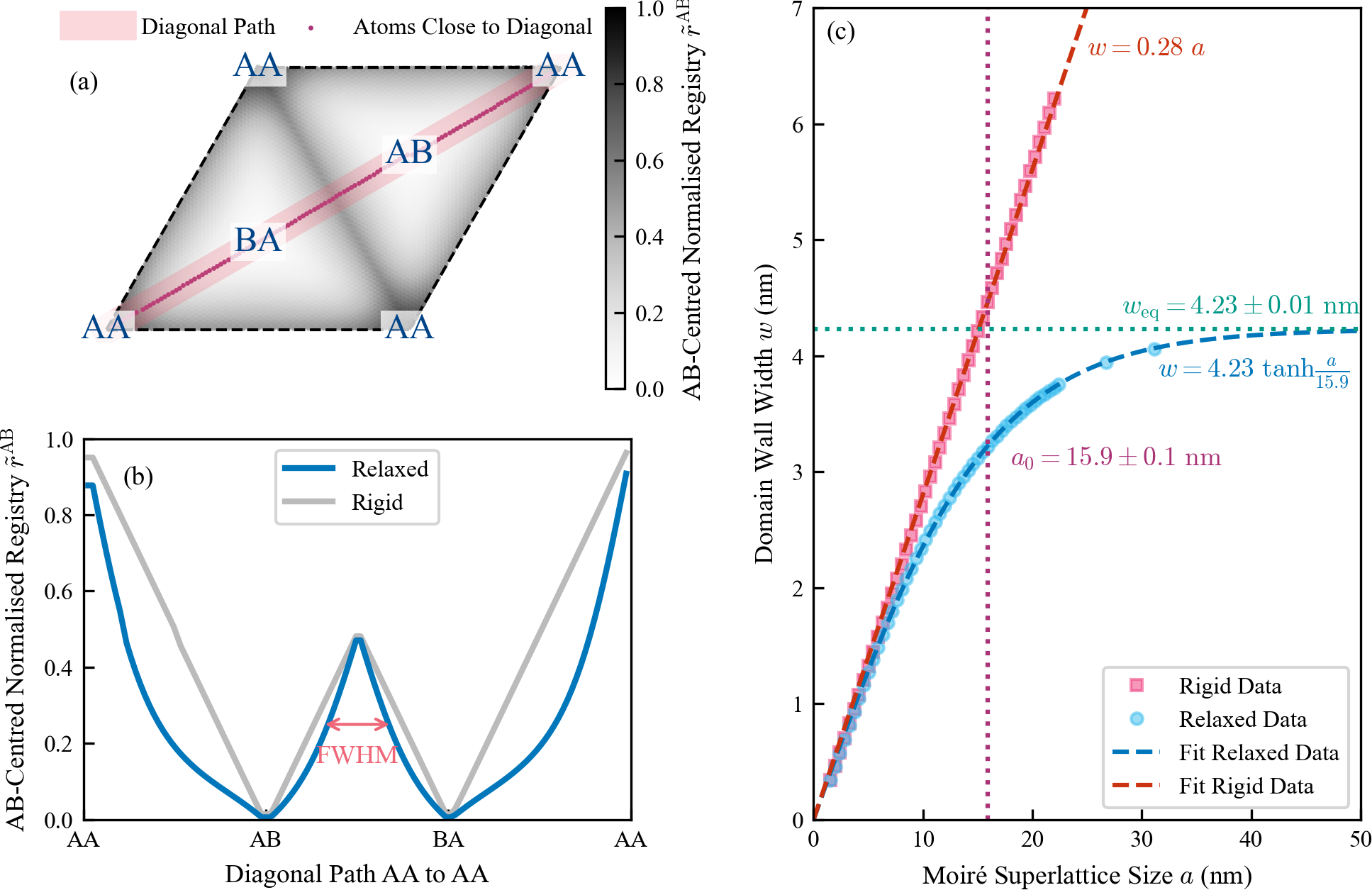}
    \caption{
        \textbf{Modelling AB-BA domain wall width.}
        (a) Relaxed TBhBN structure at twist angle \(\theta=0.82^\circ\) (19684 atoms, lattice size \(17.63 \nm\)).
        Grey colour shows AB-centred normalised registry \(\Tilde{r}^{\mathrm{AB}}\), where smaller values indicate more AB-like stacking.
        The diagonal path (pink line) crosses AA-BA-AB-AA regions perpendicular to the domain wall.
        (b) Normalised interlayer atomic distance for atoms along the diagonal path for relaxed and rigid structures.
        The full width at half maximum (FWHM) of the peak between AB and BA stacking represents the domain wall width \(w\).
        (c) Domain wall FWHM \(w\) extracted from relaxed and rigid structures, fitted to hyperbolic tangent saturation and linear functions respectively.
        Relaxed structures show equilibrium domain wall width \(w_\text{eq} = 4.23 \pm 0.01 \nm\) and characteristic relaxation length \(a_0 = 15.9 \pm 0.1 \nm\).
        The hyperbolic tangent form \(w = w_\text{eq} \tanh(a/a_0)\) is characteristic of soliton kink solutions.
    }
    \label{fig:hbn_domain_wall}
\end{figure}

We extracted the domain wall width from our relaxed TBhBN structures, and extrapolated the scaling behaviour with moir\'e superlattice size and obtained the equilibrium domain wall width, as shown in \autoref{fig:hbn_local_atomic_distance_illustration} and \autoref{fig:hbn_domain_wall}.
Since the domain wall is formed due to the competition between shear stiffness and neighbouring AB and BA stacking registry, where the local stacking gradually changes from standard BA stacking to standard AB stacking via the intermediate SP (see \autoref{fig:hbn_local_atomic_distance_illustration}), we define a local stacking metric to characterise local stacking configurations: the nearest interlayer hetero-atomic distance normalised by the equilibrium B-N bond length in each monolayer primitive cell, or AB-centred normalised registry, symbolled as \(\Tilde{r}^\text{AB}\).
At AB and BA stacking centres, \(\Tilde{r}^\text{AB}\) is minimised to 0 since the hetero-atoms from two layers are perfectly aligned (see \autoref{fig:hbn_local_atomic_distance_illustration}~(c) and (e)), while at AA stacking \(\Tilde{r}^\text{AB}\) is maximised to approximately 1 due to aligned interlayer atoms having the same atomic species (see \autoref{fig:hbn_local_atomic_distance_illustration}~(b)).
The SP stacking is demonstrated in \autoref{fig:hbn_local_atomic_distance_illustration}~(d), where \(\Tilde{r}^\text{AB}\) is around \(0.5\).
We extract the values for each atom in the rigid or relaxed TBhBN structures, shown in~\autoref{fig:hbn_domain_wall}~(a).
To accurately picture the domain wall width, we select those atoms that are closest to the diagonal AA-BA-SP-AB-AA line in each primitive cell in one of the two layers, and plot their \(\Tilde{r}^\text{AB}\) against their projected positions along the diagonal path, shown in~\autoref{fig:hbn_domain_wall}~(b).
This diagonal path is perpendicular to the domain wall.
The rigid structure shows straight lines, while relaxed structure shows curved lines, meaning that AB and BA region areas are larger after relaxation since they are energy-favoured.
We extract the domain wall width \(w\) by FWHM of the central peak~\cite{aldenStrainSolitonsTopological2013}, then fit the scaling behaviour with superlattice size as shown in~\autoref{fig:hbn_domain_wall}~(c).
For rigid structures, \(w\) increases linearly with moir\'e superlattice size (\(w = 0.28a\)), reflecting the geometric constraint in rigid structures that the domain wall must span the fixed fraction of the supercell determined by the twist angle.
For relaxed structures, they exhibit saturation behaviour well-described by the hyperbolic tangent function:
\begin{equation}
w = w_\text{eq} \tanh \qty( \frac{a}{a_0} )
\end{equation}
where \(w_\text{eq} = 4.23 \pm 0.01~\mathrm{nm}\) is the equilibrium domain wall width and \(a_0 = 15.9 \pm 0.1~\mathrm{nm}\) is the characteristic superlattice size for relaxation.
The hyperbolic tangent analytical form is the standard solution for the spatial characteristics of solitons and kinks~\cite{vachaspatiKinksDomainWalls2006}.
For \(a \gtrsim a_0\), the domain wall width stabilises at \(w_\text{eq}\), and the domain wall width converges to the equilibrium value.
The fitted equilibrium width \(w_\text{eq} = 4.23~\mathrm{nm}\) is not far from the continuum elasticity prediction of \(4.85~\mathrm{nm}\) in \autoref{eq:domain_wall_width_by_shear_strength_and_saddle_point_energy}. However, it highlights limitations of such simplified models, as a difference of about 6~\AA~ is observed.

\begin{figure}[H]
    \centering
    \includegraphics[width=0.8\twocolwidth]{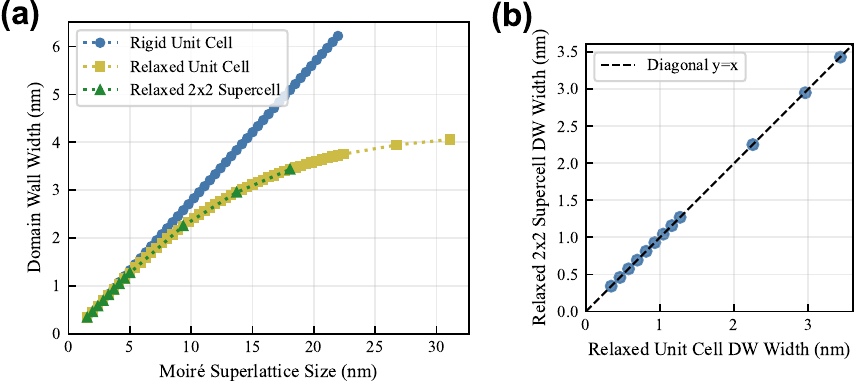}
    \caption{
        \textbf{Comparing domain wall width between relaxed unit cell and relaxed \(2 \times 2\) supercell}.
        We relaxed \(2 \times 2\) TBhBN supercells with different twist angles, extracted their AB-BA domain wall (DW) width, and compare with relaxed unit cell data presented in the paper.
        (a) Comparing rigid DW with relaxed DW width in unit cell and in \(2 \times 2\) supercell at different moir\'e superlattice size. For \(2 \times 2\) supercell, the moir\'e superlattice size is half of its simulated cell size.
        (b) Comparing relaxed DW widths in unit cells and in \(2 \times 2\) supercells.
    }
    \label{fig:TBhBN_domain_wall_size_compare_unitcell_2x2_supercell}
\end{figure}

Since structural relaxation could be different when considering more than a single Moir\'e unit cell, we further relaxed several \(2 \times 2\) TB-hBN supercells using the same MACE model and relaxation workflow and compared their AB-BA domain wall widths to those extracted from unit cells, as shown in \autoref{fig:TBhBN_domain_wall_size_compare_unitcell_2x2_supercell}.
The domain wall widths from unit cells and \(2 \times 2\) supercells are in agreement with each other, confirming that structural finite size effects are negligible within our setup. 

The in-plane electric field measurements are performed at fixed heights \(z = 1.5\text{--}4.0~\mathrm{nm}\) above the domain wall centre.
For large superlattice sizes (\(a > 15~\mathrm{nm}\) at \(\theta < 1^\circ\)), this measurement distance becomes small compared to the superlattice size.
In this regime, the measured electric field is dominated by the local polarisation gradient at the domain wall.
Once the domain wall structure saturates gradually (\(a > a_0\)), the local polarisation gradient stabilises, thus the measured electric field intensity plateaus.

Weak far-field contributions from the extended moir\'e polarisation pattern continue to vary with superlattice size, explaining why different measurement heights show slightly different saturation behaviour: for measurements further away from the domain wall, their electric field has less contribution from the domain wall polarisation gradient, and thus the saturation superlattice size will increase as the measurement distance increases.
Such effect is secondary compared to the stabilised domain wall contribution.

\begin{figure}[H]
    \centering
    \includegraphics[width=\onecolwidth]{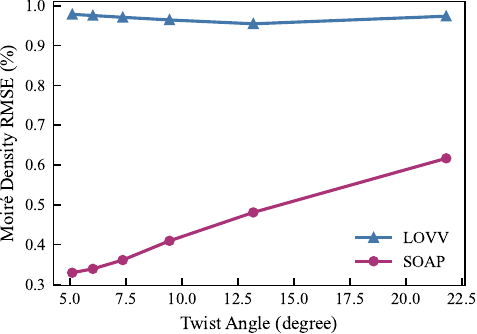}
    \caption{
        \textbf{TB-hBN density RMSE as a function of twist angle}.
        Electron density RMSE evaluated on rigid twisted bilayer hBN structure at twist angles ranging from \(\theta \approx 21.787^\circ\) to \(5.086^\circ\), predicted by the SOAP-based or LOVV-based SALTED models used for TB-hBN E-field extrapolation in Figure~\ref{fig:TBhBN_e_field_compare_soap_lovv}.
        SOAP's RMSE decreases monotonically with decreasing twist angle; LOVV's RMSE remains stable near \(0.97\%\).
    }
    \label{fig:TBhBN_density_error}
\end{figure}

\begin{figure}[H]
    \centering
    \includegraphics[width=\onecolwidth]{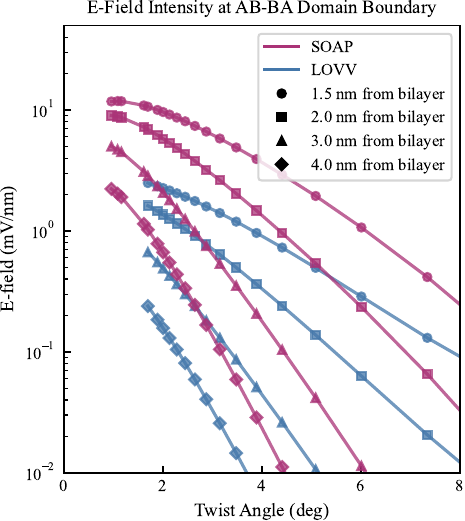}
    \caption{
        \textbf{E-field intensity at the AB-BA domain wall in twisted bilayer hBN, comparing SOAP and LOVV predictions.}
        In-plane electric field intensity perpendicular to the AB-BA domain boundary as a function of twist angle, evaluated at heights of 1.5--4.0~nm above the bilayer centre, predicted by SALTED models using SOAP (in purple) and LOVV (in blue) descriptors.
        SOAP achieves quantitative agreement with DFT across all twist angles and measurement heights.
        LOVV correctly reproduces the qualitative trend towards small twist angle but underestimates the absolute field magnitude, consistent with its long-range representation for long-wavelength density features rather than the short-range atomic-scale dipoles that generate the domain wall E-field.
    }
    \label{fig:TBhBN_e_field_compare_soap_lovv}
\end{figure}

Although LOVV is the preferred descriptor for accurate band structure predictions across materials, the domain wall E-field in TB-hBN arises primarily from short-range interlayer B-N charge-transfer dipoles at bond length scales rather than long-range electrostatics. Because this dipolar field decays rapidly in space, its accurate prediction primarily depends on capturing these short-range density features.
SOAP already resolves these local dipoles and achieves lower density RMSE than LOVV (Table~1 in the main text: \(0.65\%\) vs \(2.12\%\)).
Figure~\ref{fig:TBhBN_density_error} shows that the SOAP density RMSE decreases monotonically with decreasing twist angle while LOVV's error remains stable. Figure~\ref{fig:TBhBN_e_field_compare_soap_lovv} confirms that compared to SOAP, which achieves quantitative E-field agreement with DFT (Figure~5 in the main text), LOVV captures the qualitative trend and underestimates the field magnitude.
Additionally, LOVV's quadratic scaling with system size makes it more expensive for this application.

\section{Computational Cost Comparison\label{app:computation_cost}}

\begin{figure}[htbp]
    \centering
    \includegraphics[width=\twocolwidth]{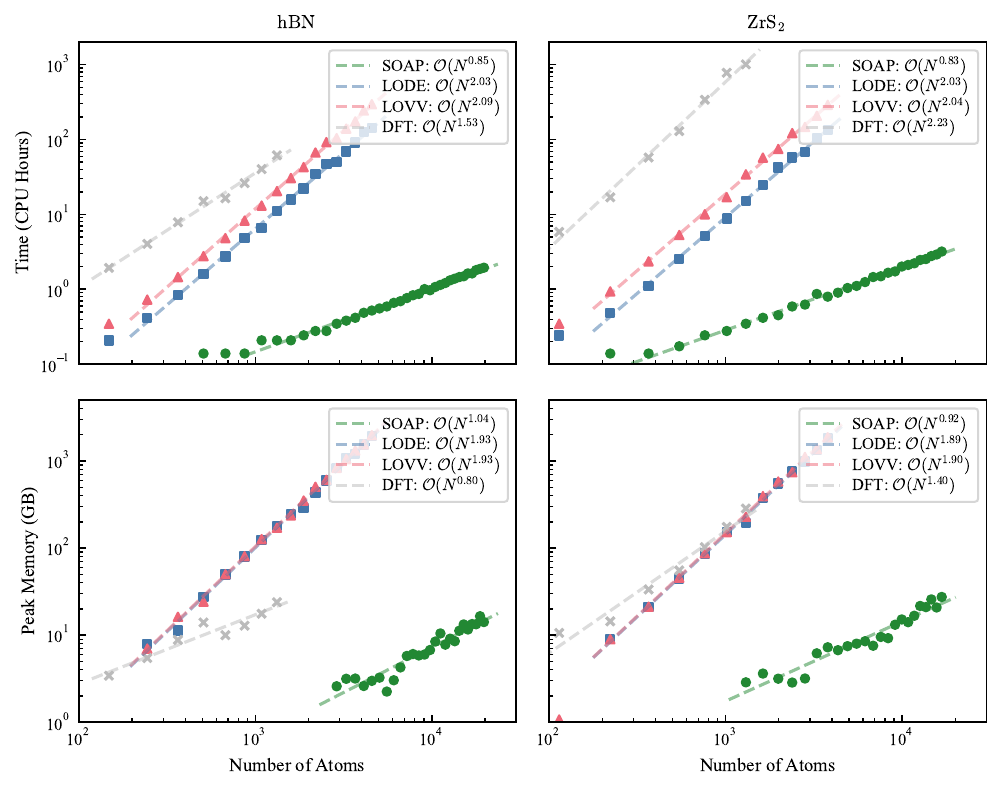}
    \caption{
        \textbf{Computational efficiency of SALTED framework.}
        Wall time and peak memory scaling with system size for SALTED predictions using different descriptors (SOAP, LODE, LOVV) compared to fully converged DFT calculations for hBN and \zrds moir\'e structures.
        The computational cost for SALTED predictions is extracted from slurm output logs for the SALTED prediction steps.
        For DFT calculations, the computational costs are those spent for converging the DFT, without band structure output.
        Some low-value data points are omitted because slurm experiences noise in statistics due to the small computational cost.
        Fitted scaling exponents are shown in the legends.
        Training a LOVV-based model requires 8.87 CPU hours for hBN and 248 CPU hours for \zrds.
    }
    \label{fig:computational_efficiency}
\end{figure}

To assess the practical efficiency of SALTED for large-scale moir\'e systems, we benchmark the computational cost of predictions using different descriptors against full DFT calculations performed with FHI-aims.
\autoref{fig:computational_efficiency} compares the wall time and peak memory scaling for systems ranging from \(\sim 100\) to \(>10,000\) atoms for hBN and \zrds moir\'e structures.
The results demonstrate that SALTED achieves substantial speedup, varying from \(\sim 10\times\) to \(\sim 100\times\) depending on material, system size, and DFT configurations, while maintaining the predictive accuracy shown in the main text.

The computational scaling exhibits distinct behaviour depending on the descriptor choice.
SOAP shows near-linear scaling in time and memory, reflecting the locality character of the atomic density representations with a finite cutoff.
In contrast, both LOVV and LODE display quadratic scaling in time and memory, consistent with their all-to-all long-range nature.
DFT calculations show intermediate time scaling between linear and quadratic.
For memory scaling, FHI-aims benefits from its sparse matrix representations, achieving sub-linear memory scaling in hBN and near-linear scaling in \zrds.

Despite the quadratic scaling of long-range descriptors, SALTED predictions remain significantly faster than DFT across the entire system size range studied.
For a system with 1000 atoms, SALTED predictions using LOVV achieve \(\sim 4\times\) speedup for hBN, and \(\sim 20\times\) speedup for \zrds.
These speedups can be doubled, as the current LOVV implementation redundantly calculates the same atomic electrostatic representations twice (the major cost in prediction).

Although the memory scaling remains quadratic for LOVV and LODE descriptors, and shows higher memory usage than DFT for small and light-element systems using smaller basis sets, these descriptors can predict large heavy-element systems with manageable memory requirements, e.g. a \zrds moir\'e systems containing up to 2000 atoms can be predicted within 1~TB of memory.
Future optimisations, such as efficient approximations via Ewald summation for electrostatic representations, could potentially reduce the scaling to sub-quadratic.

The computational advantage shall be weighed against the one-time training cost.
Training a SALTED model with LOVV descriptors requires 8.87 CPU hours for hBN and 248 CPU hours for \zrds, comparable to a single DFT calculation for systems with 300 atoms (hBN) or 700 atoms (\zrds), respectively.
Since hyperparameter tuning for GPR regularisation typically requires training \(\sim 5\) models, and the trained model may be used for tens or hundreds of predictions, the training cost becomes negligible compared to the cumulative computational savings from predictions.
This makes SALTED highly cost-effective for exploring configurational spaces (e.g., scanning twist angles and stacking configurations) or conducting \textit{ab initio} molecular dynamics simulations where thousands of DFT calculations on large structures are required.

\begin{figure}[H]
    \centering
    \includegraphics[width=\twocolwidth]{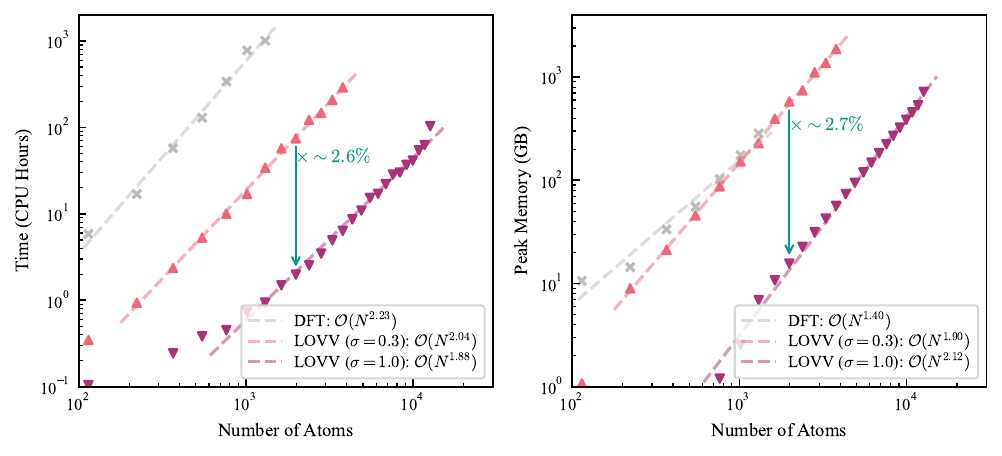}
    \caption{
        \textbf{Time and memory benchmark for DFT and LOVV with different \(\sigma\) settings, for TB-ZrS\textsubscript{2} prediction.}
        By changing descriptor hyperparameter from \(\sigma=0.3\angs\) to \(\sigma=1.0\angs\), one obtains \(\sim 37 \times\) reduction in both time and memory cost.
        The SALTED model with LOVV \(\sigma=1.0\angs\) predicts \(\sim 1000 \times\) faster than converging a DFT calculation.
    }
    \label{fig:computational_cost_benchmark_zrs2_lovv_sigma10_ref_B}
\end{figure}

\begin{figure}[H]
    \centering
    \includegraphics[width=0.5\linewidth]{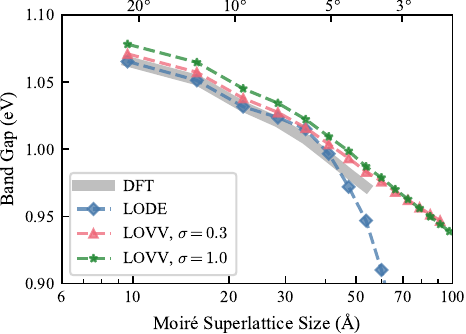}
    \caption{
        \textbf{Band gap comparison between DFT and LOVV prediction, with different \(\sigma\) parameters.}
        LOVV prediction with \(\sigma=0.3\angs\) (red) and \(\sigma=1.0\angs\) (green) both correctly extrapolate the band gap decreasing trend following DFT (in grey).
    }
    \label{fig:ZrS2_band_gap_sigma10}
\end{figure}

To reduce LOVV's cost, we tested a larger descriptor Gaussian width \(\sigma=1.0~\angs\) for \zrds against the default \(\sigma=0.3~\angs\).
The larger \(\sigma\) reduces both time and memory by \(37\times\) (\autoref{fig:computational_cost_benchmark_zrs2_lovv_sigma10_ref_B}) with no significant change in band gap extrapolation accuracy (\autoref{fig:ZrS2_band_gap_sigma10}).

All benchmarks were performed on CPU-based computing clusters with various Intel Xeon / Skylake and AMD EPYC processors.
While specific software versions and hardware configurations may affect absolute computational costs, the relative speedups should remain comparable across modern HPC systems, as server CPU performance has not improved dramatically in recent years.

\section{Descriptor Analysis\label{app:descriptor_analysis}}

\subsection{Descriptor Sensitivity Analysis\label{app:descriptor_sensitivity}}

To quantify how descriptor sensitivity decays by interatomic separation, we evaluated how descriptors' response to atomic coordinate perturbation varies by interatomic distance.

Two atoms (A and B) are placed in a \(200 \times 100 \times 100~\angs^3\) periodic cell along the x-axis with varying interatomic distance \(r\) (identical y and z coordinates).
For each configuration, atom B is displaced by \(\Delta x = 0.1~\angs\) in the x-direction, and we calculate the resulting change in the \(\lambda=0\) descriptor component at atom A's position.
The descriptor sensitivity is defined as:
\begin{equation}
S(r) = \frac{\| \chi_A(r + \Delta x) - \chi_A(r) \|}{\Delta x}
\end{equation}
where \(\|\cdot\|\) denotes the \(\ell_2\)-norm.
Displacement perpendicular to the interatomic axis (y or z directions) produces negligible descriptor changes and is not considered.
Notice that descriptor \(\chi_A\) is normalised to 1 in SALTED.

\begin{figure}[htbp]
    \centering
    \includegraphics[width=\onecolwidth]{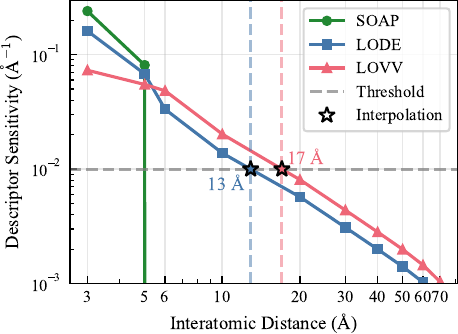}
    \caption{
        \textbf{Descriptor sensitivity as a function of interatomic distance.}
        Sensitivity is defined as the \(\ell_2\)-norm of descriptor change at atom A when atom B is displaced by unit length.
        SOAP vanishes beyond its \(6\angs\) cutoff.
        LODE and LOVV exhibit similar long-range decay behaviour for \(r > 6 \angs\), reaching sensitivity ranges of \(13 \angs\) and \(17 \angs\) respectively at a sensitivity threshold of \(10^{-2}\angs^{-1}\) (horizontal dashed grey line).
    }
    \label{fig:descriptor_sensitivity_vs_distance}
\end{figure}

\autoref{fig:descriptor_sensitivity_vs_distance} shows the distance-dependent sensitivity for all three descriptors.
SOAP is limited by its cutoff radius \(r_\text{cut} = 6.0\angs\), confirming its locality character.
LODE and LOVV maintain sensitivity at extended distance, and have identical decay behaviour for \(r > r_\text{cut}\).
This similarity reflects their shared electrostatic \(1/r\) character in the long-range regime.
We interpolate the log-log plot linearly to obtain the interatomic ranges corresponding to a sensitivity threshold of \(10^{-2}~\angs^{-1}\), which characterise the descriptors' sensitivity range: LODE reaches \(13\angs\), and LOVV reaches \(17\angs\).

Despite this small difference in spatial extent, LOVV significantly outperforms LODE for moir\'e superlattice prediction for many materials discussed in this work.
This discrepancy indicates that although the electrostatic representation plays a vital role in representing distance geometries, LOVV's superior performance cannot be attributed only to this long-range representation.
We hypothesise that the \(\ket{V \otimes V}\) structure in LOVV encodes the geometric relationships in displaced bilayers in a way that better overlaps with moir\'e bilayer geometries in descriptor space, compared to the \(\ket{\rho \otimes V}\) structure in LODE.
This hypothesis is supported by descriptor manifold coverage analysis in \autoref{app:descriptor_distribution}, which shows LOVV providing superior overlap between displaced bilayer training structures and moir\'e bilayer test structures compared to LODE.

\subsection{Descriptor Distribution Analysis\label{app:descriptor_distribution}}

For kernel-based machine learning methods, prediction accuracy on unseen configurations depends critically on whether the descriptor space of the prediction set is covered by the training set.
This is particularly important for our study, where training data consist of displaced bilayers (broad spatial sampling) while prediction targets are twisted bilayers (moir\'e configurations).
To assess whether descriptors can adequately represent both regimes within a unified descriptor space, we employ dimensionality reduction visualisation using Uniform Manifold Approximation and Projection (UMAP) \cite{mcinnesUMAPUniformManifold2020}.

UMAP reduces high-dimensional descriptor vectors to low-dimensional projections while preserving local neighbourhood structure, quantified by trustworthiness (TW) metrics \cite{vennaNeighborhoodPreservationNonlinear2001}.
TW values exceeding 0.9 indicate reliable dimensionality reduction \cite{vennaLocalMultidimensionalScaling2006,maatenLearningParametricEmbedding2009}.
However, UMAP assumes approximately uniform data distribution on the underlying manifold \cite{mcinnesUMAPUniformManifold2020}.
When training and prediction sets have significantly different densities, UMAP can produce ``condensed core'' artifacts where high-density regions appear artificially separated from low-density regions in 2D projection, despite actual overlap in the original high-dimensional space.

\begin{figure}[H]
    \centering
    \includegraphics[width=\twocolwidth]{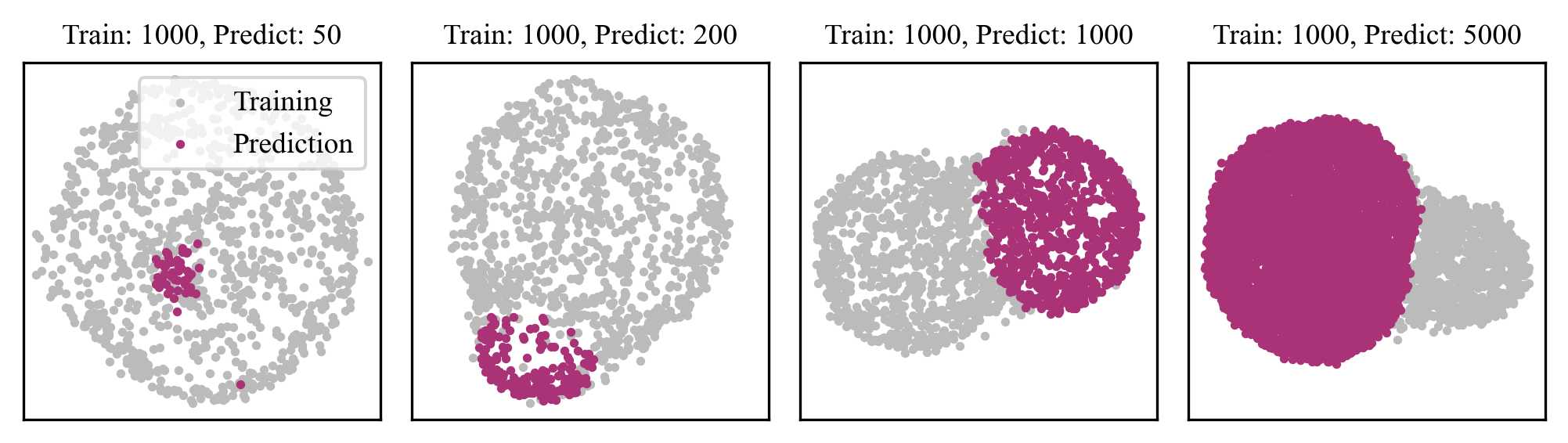}
    \caption{
        \textbf{UMAP artifact demonstration with synthetic datasets.}
        Training data: 1000 samples from a Gaussian distribution (\(\mu=100, \sigma=10\)).
        Prediction data: varying sample size from a narrower Gaussian distribution (\(\mu=100, \sigma=1\)).
        All samples are normalised to unit length in the 10D space.
        UMAP parameter: \texttt{n\_neighbors=40}, \texttt{min\_dist=0.5}.
        Titles of subplots indicate the number of training and prediction samples.
        Despite the prediction distribution residing entirely within the training distribution in high-dimensional space, increasing prediction set density (left to right) creates apparent separation in the 2D UMAP projection.
        This artifact necessitates density-balanced subsampling for reasonable UMAP visualisation.
}
    \label{fig:eg_density_umap_results}
\end{figure}

\autoref{fig:eg_density_umap_results} demonstrates this artifact using synthetic Gaussian data.
One can observe that, although the prediction set distribution is fully contained within the training set in high-dimensional space, the 2D UMAP projection shows increasing separation as the prediction set density increases.
To compensate for this artifact and enable fair comparison across materials and descriptors, we apply density-based subsampling in \autoref{tab:umap_dataset_info} while keeping UMAP parameters constant, including number of neighbours (\verb|n_neighbors=40|) and minimum distance (\verb|min_dist=0.5|).
The number of samples in subsets from training and prediction sets is kept consistent within each material across all three descriptors, and also consistent for materials with the same lattice type (e.g. graphene and hBN both use 368 training samples and 53 prediction samples).
This controlled subsampling maintains computational efficiency while preserving the essential distribution structure for meaningful visualisation and reasonable comparison.

\begin{table}[H]
    \centering
    \begin{tabular}{c|cccc}
        \hline
        Material & Train Ratio & Train Count & Predict Ratio & Predict Count \\
        \hline
        Graphene & 2.0\% & 368 & 1.0\% & 53 \\
        hBN & 2.0\% & 368 & 0.5\% & 53 \\
        1T-\tids & 10.0\% & 2764 & 1.0\% & 159 \\
        1T-\zrds & 10.0\% & 2764 & 1.0\% & 159 \\
        1H-\mods & 20.0\% & 5229 & 1.0\% & 159 \\
        \hline
    \end{tabular}
    \caption{
        \textbf{Density-based subsampling parameters for UMAP visualisation.}
        Ratios are consistent within each material across all descriptors, and consistent for materials of the same lattice type for controlled comparison.
    }
    \label{tab:umap_dataset_info}
\end{table}

\begin{figure}[H]
    \centering
    \includegraphics[width=0.7\twocolwidth]{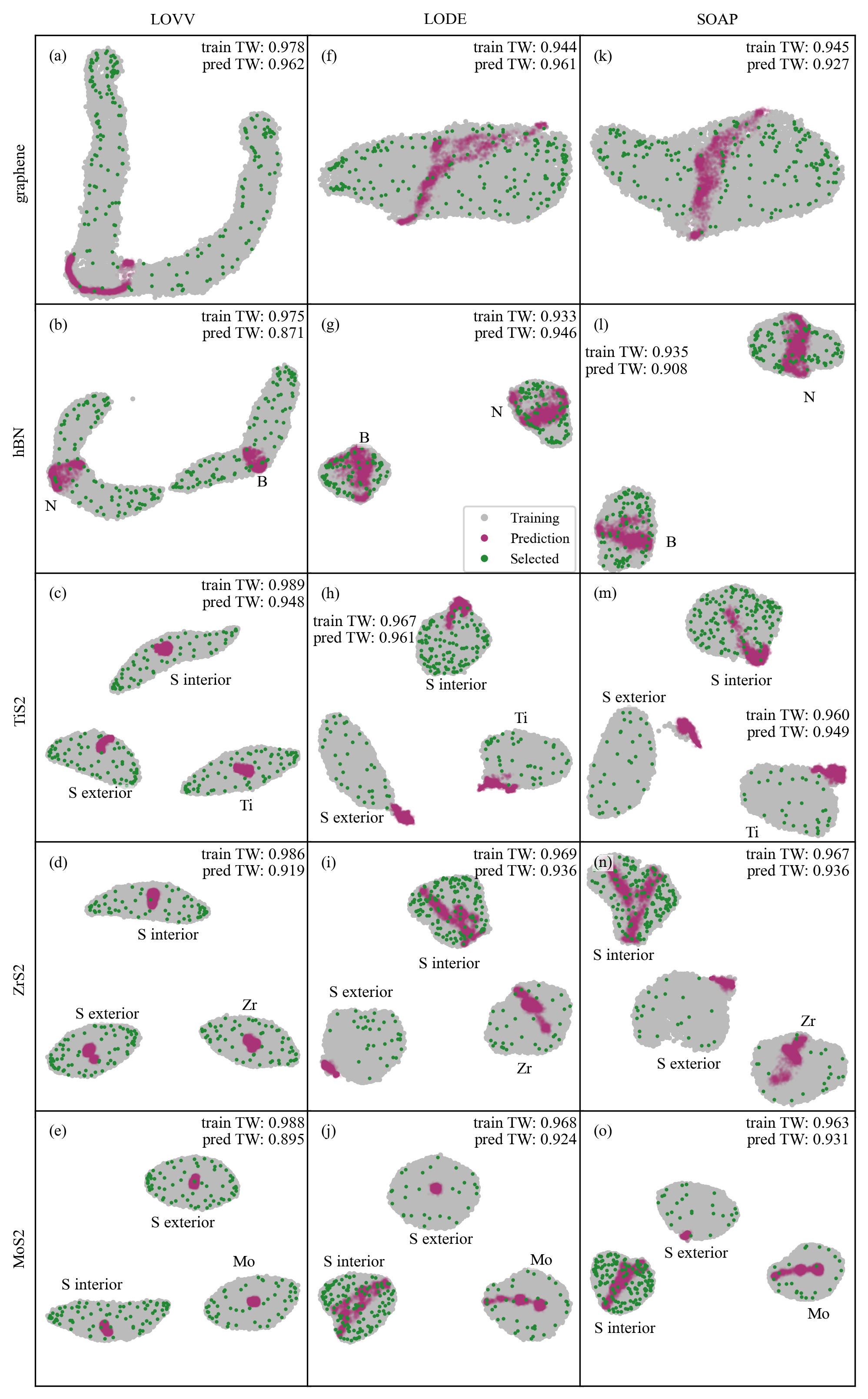}
    \caption{
        \textbf{UMAP visualisation reveals descriptor-dependent extrapolation capability.}
        Training (displaced bilayers, grey) and prediction (twisted bilayers, purple) descriptor distributions for all five materials and three descriptors.
        Green points indicate selected atomic environments for RKHS representation \cite{grisafiElectronicStructurePropertiesAtomCentered2023}.
        Trustworthiness (TW) scores quantify neighbourhood preservation quality.
        All reductions achieve TW \(>0.87\), confirming reliable local structure preservation.
        All UMAP projections use consistent parameters (\texttt{n\_neighbors=40}, \texttt{min\_dist=0.5}) and material-specific subsampling ratios (\autoref{tab:umap_dataset_info}).
        Atomic species naturally cluster by chemical species and structural position (interior vs. exterior for sulfur atoms in TMDCs).
    }
    \label{fig:descriptors_umap_for_all_materials}
\end{figure}

\autoref{fig:descriptors_umap_for_all_materials} presents UMAP projections for all five materials and three descriptors.
The following observations can be drawn:
\begin{itemize}
\item Graphene and hBN: All three descriptors show comparable UMAP coverage, despite the clear performance differences for band structure prediction.
\item TMDCs: Descriptor performance shows material-dependent degradation patterns.
For \tids, LOVV has better train-prediction overlap than LODE and SOAP.
For \zrds, LODE and SOAP prediction clusters of exterior S atoms lie at the boundaries of training distributions rather than within the core, indicating marginal coverage, while LOVV maintains central overlap.
For \mods, SOAP shows weak peripheral overlap for exterior S atoms, while both LODE and LOVV maintain substantial central overlap, correlating with their improved prediction accuracy.
\end{itemize}
This discrepancy indicates that descriptor space coverage is necessary but not sufficient.
While UMAP effectively identifies inadequate coverage (SOAP for TMDCs), it cannot fully explain performance differences when coverage appears similar.
The non-uniform density distribution between training and prediction sets further complicates UMAP interpretation, as this violates the method's uniform sampling assumption and necessitates careful parameter selection to avoid density-related visualisation artifacts~\cite{mcinnesUMAPUniformManifold2020}.
While we employed systematic density-balanced subsampling and consistent hyperparameters across materials to ensure fair comparisons, these methodological requirements highlight that UMAP, while useful for qualitative insight, cannot serve as a definitive validation metric for descriptor performance in predicting long-range phenomena in moir\'e materials.

Understanding why LOVV's \(\ket{V \otimes V}\) structure outperforms LODE's \(\ket{\rho \otimes V}\) beyond simple coverage, likely through better encoding of interlayer electrostatic correlations, requires systematic kernel analysis and descriptor decomposition, which we leave for future work.

\putbib[ref_zotero_export_bibtex,customized]

\end{bibunit}

\end{document}